\pdfoutput=1
\documentclass[11pt,a4paper]{article}

\usepackage[margin=2cm]{geometry}
\usepackage[utf8]{inputenc}
\usepackage[T1]{fontenc}
\usepackage{authblk}
\usepackage{hyperref}

\usepackage[numbers,sort&compress]{natbib}
\usepackage{graphicx}%
\usepackage{multirow}%
\usepackage{amsmath,amssymb,amsfonts}%
\usepackage{amsthm}%
\usepackage{mathrsfs}%
\usepackage[title]{appendix}%
\usepackage{xcolor}%
\usepackage{textcomp}%
\usepackage{manyfoot}%
\usepackage{booktabs}%
\usepackage{algorithm}%
\usepackage{algorithmicx}%
\usepackage{algpseudocode}%
\usepackage{listings}%
\usepackage{subcaption}
\usepackage{gensymb}
\graphicspath{{figs/}}

\title{Magnetic pendulum under external excitation with state-dependent phase control}

\author[1]{Krystian Polczy\'nski\thanks{Corresponding author. E-mail: \texttt{krystian.polczynski@p.lodz.pl} (ORCID: 0000-0002-1177-6109)}}
\author[1]{Jan Awrejcewicz\thanks{E-mail: \texttt{jan.awrejcewicz@p.lodz.pl} (ORCID: 0000-0003-0387-921X)}}

\affil[1]{Department of Automation, Biomechanics and Mechatronics, Lodz University of Technology, 1/15 Stefanowskiego Str., 90-924 Lodz, Poland}
\date{}
\begin{document}
\maketitle
\begin{abstract}
	The dynamics of a strongly nonlinear magnetic pendulum subjected to phase-controlled magnetic excitation is investigated experimentally and numerically. The main novelty of the study lies in introducing a state-dependent phase law, in which the phase of the coil current depends on the instantaneous angular position of the pendulum. In contrast to classical magnetic pendulum with external constant-phase excitation, this modification creates a feedback-type phase modulation and changes the timing of energy transfer between the electromagnetic actuator and the mechanical oscillator.	
	A mathematical model of the system is derived using an experimentally identified magnetic torque and validated against laboratory measurements. The validation performed for the classical case confirms that the model reproduces the principal dynamical features observed experimentally. For the phase-controlled system, the results reveal a rich organization of regular and chaotic regimes, broad families of coexisting attractors, and a pronounced reorganization of the bifurcation structure relative to the constant-phase case.	
	A particularly important result is the existence of a robust intrawell period-1 branch that persists over a wide range of excitation amplitudes despite substantial changes in the magnetic forcing strength. Energy analysis shows that this response is sustained by localized, phase-dependent bursts of magnetic power input that compensate dissipation on a cycle-by-cycle basis. Basin of attraction analysis demonstrates multistability and fractal basin boundaries, quantified using both box-counting dimension and boundary basin entropy. The obtained results provide new insight into phase-controlled energy transfer and attractor selection in nonlinear magneto-mechanical oscillators.
\end{abstract}



\vspace{0.5em}
\noindent\textbf{Keywords:} magnetic pendulum, non-autonomous oscillator, bifurcation, adaptation system, basin entropy.

\section{Introduction}\label{sec1}

Magnetic pendulums constitute a convenient class of nonlinear magneto-mechanical oscillators for investigating complex vibration phenomena in a relatively simple and experimentally accessible setting. Owing to the localized character of magnetic interaction and the strong nonlinearity of the restoring and excitation torques, such systems can exhibit a broad range of responses, including periodic, multi-periodic, and chaotic oscillations. For this reason, magnetic pendulum models have long been used to study fundamental issues of nonlinear dynamics, such as resonance, instability, multistability, and energy exchange. Early investigations focused on sustained oscillations generated by alternating magnetic fields and on their analytical description in terms of simplified or Mathieu-type models \cite{Bethenod1938,Rocard,Landa1996b,Minorsky}. These studies established the basic mechanisms of self-oscillation and parametric instability, while later works showed that magnetic pendulums may also exhibit chaotic dynamics, limit cycles, and multiple equilibrium states \cite{Khomeriki2016,Skubov2014}.

Subsequent research broadened this picture by demonstrating discrete or quantized oscillation amplitudes, one-sided oscillations, and complex nonlinear responses caused by spatially inhomogeneous magnetic forcing \cite{DamgovPopov,Wijata2021,Skurativskyi2022a}. Related configurations, such as Duboshinski's pendulum and its variants, revealed amplitude quantization, subharmonic resonance, and argumental oscillations \cite{Penner,Doubochinski2013,Tennenbaum2005,Luo2020a,Shumaev2017}. Other studies reported bistability, hysteresis, chaotic attractors, and strange attractors in magnetically forced pendulum systems, including configurations modeled by Duffing-type equations or extended by additional nonlinear effects \cite{Tran2013,Nana2017,Nana2018,Siahmakoun1997,Mann2009f,Kwuimy2012,Pedersen2020,Moon1979a,Moon1987,Donnagain2006}. More recently, vertically driven magnetic pendulums interacting with conducting plates and magnetic spring-pendulum systems have further confirmed the richness of these dynamics, including instability tongues, coexisting periodic solutions, internal resonances, and chaotic regimes \cite{Boeck2020a,Boeck2021,Becker2022,Meng2023}. A similarly rich structure of periodic responses and bifurcations has also been reported in other pendulum-type systems subjected to nonstandard forcing, such as the impact inverted pendulum under impulsive excitation studied by Lenci and Rega \cite{Lenci2000}.

Taken together, the available literature shows that magnetic pendulums can display a wide variety of nonlinear phenomena depending on the form of excitation and magnetic coupling. At the same time, most of these studies consider externally prescribed forcing with fixed phase, or more generally with excitation parameters that are independent of the instantaneous state of the oscillator.

Related studies on magnetically coupled and multi-degree-of-freedom oscillators have further confirmed the broad dynamical potential of magnetically interacting systems, including synchronization, nonlinear normal modes, controlled energy transfer, and enhanced energy harvesting \cite{Fradkov2005,Surganova2022,Kumar2019a,Darula2011}.

Despite the broad literature on magnetic pendulums, an important gap remains in the understanding of systems with state-dependent phase control. In most existing models, the magnetic excitation is prescribed externally, so that its phase is independent of the instantaneous motion of the oscillator. Under this assumption, the excitation acts as a time-dependent input superimposed on the pendulum dynamics, while the timing of energy transfer is not influenced by the current state of the system. In contrast, introducing a position-dependent phase shift creates an additional feedback path, because the excitation phase is updated according to the instantaneous angular position of the pendulum. This modification changes the structure of the forcing term from an exogenous periodic input into a state-dependent phase-modulated excitation, thereby altering both the timing and localization of energy injection. As a result, the controlled-phase formulation is not just a parametric variation of classical models, but a qualitatively different forcing mechanism that may generate dynamical responses not readily accessible under constant-phase excitation.

The aim of this work is to investigate the dynamics of a strongly nonlinear magnetic pendulum subjected to time-varying magnetic excitation with a phase that depends on the instantaneous angular position of the pendulum. The study is motivated in part by the work of Krylosova et al. \cite{Krylosova2020}, who showed that state-dependent control of forcing parameters in a non-autonomous oscillator can produce a rich hierarchy of regular and chaotic responses.

In contrast to classical magnetic pendulum models, where the excitation torque is typically written as
\[
T_{\text{mag}}(\theta,t)=f(\theta)\sin(\omega t+\phi_0),
\]
with a constant phase \(\phi_0\) independent of the pendulum motion \cite{Luo2020a,Boeck2020a,Tran2013}, the present study allows the phase of the alternating current in the coil to depend on the instantaneous angular position of the pendulum, i.e.
\[
\phi = p\,y(t).
\]
This transforms the excitation from a position-dependent magnetic forcing with externally prescribed time dependence into a state-dependent closed-loop interaction in the phase domain. From a broader perspective, the present study belongs to the class of approaches aimed at controlling complex dynamics in mechanical systems by suitably modifying the excitation and exploiting nonlinear response mechanisms, as discussed in the review by \cite{Rega2010}. In contrast to conventional harmonic forcing, where energy is supplied uniformly in time, phase-controlled excitation may concentrate the energy input within selected regions of motion, making the response strongly dependent on the current state of the system. Such a mechanism may be useful in tunable nonlinear oscillators, adaptive vibration control strategies, and systems in which one seeks to promote, suppress, or switch between coexisting attractors.
\textcolor{black}{
	The digital implementation of state-dependent excitation introduces sampled data effects that are absent from ideal continuous-time models. In hardware-in-the-loop systems, the measurement, numerical processing, and actuation stages introduce an effective delay, while the sample-and-hold operation may modify the stability properties when the sampling period is not sufficiently short \cite{Miklos2017,Miklos2018}. Consequently, the update frequency of a digitally implemented feedback law should be considered in relation to the characteristic frequencies of the physical system.}

While previous studies on magnetic pendulums have widely reported multistability, bifurcations, and chaotic behavior \cite{Nana2017,Nana2018,DamgovPopov}, the phase-controlled system investigated here exhibits a distinct combination of dynamical features, including fractal basin boundaries, impulsive energy transfer confined to a narrow angular range, and robust periodic responses that persist over a broad range of forcing amplitudes. These features arise specifically from the phase-feedback mechanism and therefore distinguish the proposed system from classical magnetic pendulum models driven by externally prescribed excitation.

The paper is structured as follows. Section~\ref{sec2} presents the experimental setup and the adopted experimental procedure. Section~\ref{sec3} derives the mathematical model and introduces the phase-controlled formulation. Section~\ref{sec4} provides preliminary studies and validation of the classical reference system. The nonlinear dynamics of the phase-controlled pendulum are analyzed in Section~\ref{sec5}, including the organization of dynamical regimes, bifurcation scenarios, coexisting attractors, basin structures, and energy-transfer properties. Section~\ref{sec6} discusses the physical interpretation and implications of the obtained results, while Section~\ref{sec7} concludes the paper.

\section{Experimental setup}\label{sec2}
\subsection{Description}
\begin{figure}[b!]
	\centering
	\begin{subfigure}{0.65\textwidth}
		\includegraphics[scale=0.4]{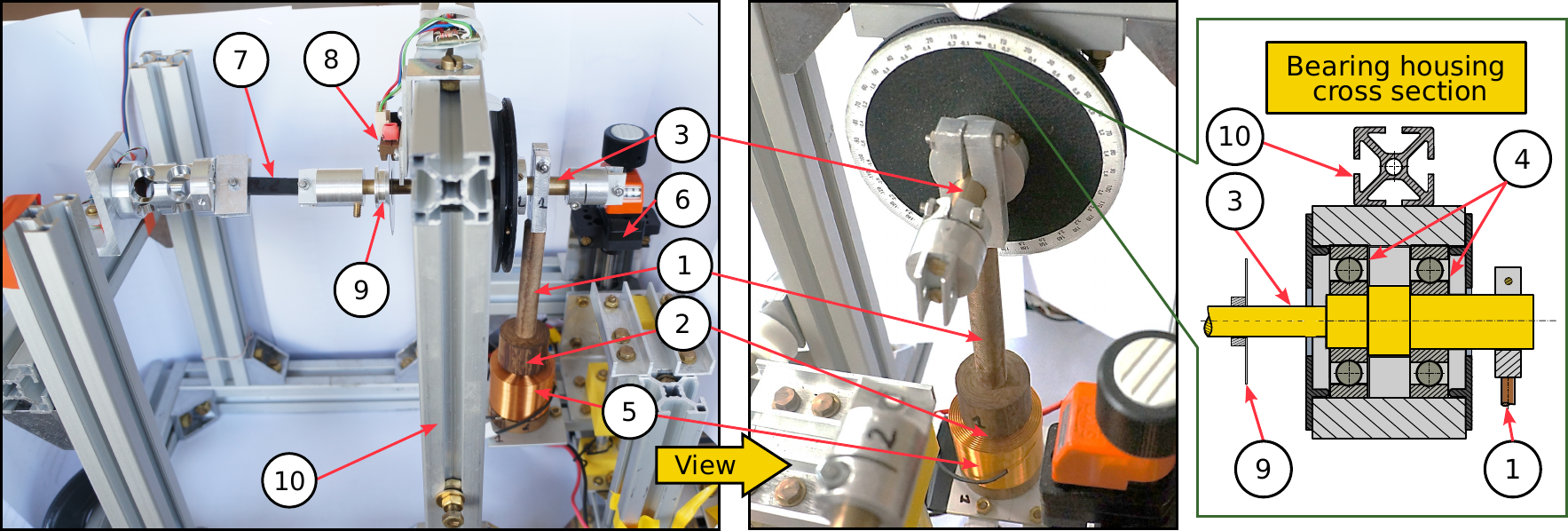}
		\caption{} \label{}
	\end{subfigure}\\
	\begin{subfigure}{0.3\textwidth}
		\includegraphics[scale=0.55]{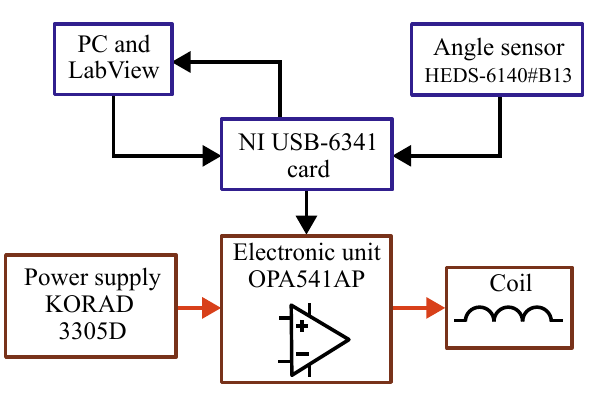}
		\caption{} \label{}
	\end{subfigure}
	\caption{Experimental setup (a): (1)~--~pendulum, (2)~--~neodymium magnet, (3)~--~brass shaft, (4)~--~rolling bearing (5)~--~electric coil, (6)~--~linear lift, (7)~--~elastic element, (8)~--~optical sensor, (9)~--~code wheel, and (10)~--~aluminium frame. Signal-flow diagram (b): red arrows denote current signals, whereas black arrows denote voltage signals.}
	\label{fig:experimental_rig_1}
\end{figure}

The experimental investigation was carried out using a specially designed laboratory setup shown in Fig.~\ref{fig:experimental_rig_1}a.
The setup comprises a magnetic pendulum (1) equipped with a neodymium magnet (2) mounted at one end of a brass shaft (3). The magnet has a diameter of 22~mm and a height of 10~mm. The shaft is supported by a pair of rolling bearings (4). An electric coil (5) is placed below the pendulum on a vertically adjustable platform. The platform position can be regulated by means of a linear lift (6), which makes it possible to set the clearance between the coil and the pendulum in the rest position. In the present study, this clearance was fixed at 1.6~mm. The coil has an inductance of 22~mH, a resistance of 10.6~$\Omega$, a copper-wire diameter of 0.5~mm, an outer diameter of 40~mm, an inner diameter of 17~mm, and a height of 31~mm.
The opposite end of the brass shaft is connected to the fixed base by means of an elastic rubber element (7). The role of this element is to increase the restoring torque in the system. The elastic element has a rectangular cross-section of $8.5\times6.5$~mm and a length of 40.3~mm. To minimize unwanted magnetic interactions, all structural components of the setup, including the frame (10), were manufactured from non-magnetic materials such as aluminium alloys, composite materials, and polymers.
The electric coil is supplied by a laboratory power supply (KORAD KA3305D) through a dedicated electronic unit based on the operational amplifier OPA541AP. The power supply operates in the current-source mode, which ensures a nearly constant current amplitude despite changes in the coil resistance caused by thermal effects or inductance effect. The waveform of the coil current follows the voltage signal delivered from the NI USB-6341 data acquisition card to the electronic unit. The voltage signal generated by the NI USB-6341 card is controlled in LabView. The current flowing through the coil generates a magnetic field that interacts with the pendulum magnet. According to the adopted polarity convention, a positive current repels the magnet from the coil, whereas a negative current attracts it toward the coil.
\textcolor{black}{
	Therefore, the coil intrinsic electrical time constant, calculated as $\tau_{RL}=L/R$, is approximately 2.08 ms. However, the coil is supplied by a current-controlled amplifier rather than directly by a prescribed voltage. Consequently, $\tau_{RL}$ characterises the intrinsic electrical time scale of the coil, but it should not be interpreted as the effective time constant or a pure delay of the complete actuation path. The effective relationship between the commanded and actual coil currents also depends on the closed-loop dynamics and bandwidth of the current amplifier. These dynamics are not explicitly represented in the numerical model, in which the coil current is prescribed directly.}
The angular position of the pendulum is measured by an optical incremental sensor (HEDS-6140\#B13), denoted by (8), operating together with a code wheel (9). The angular resolution of the sensor is 0.36\degree. A schematic diagram of the signal flow between the individual components of the experimental system is shown in Fig.~\ref{fig:experimental_rig_1}b.

Real-time adjustment of the excitation phase was implemented in LabView. During the experiment, the angular position measured by the optical encoder was continuously acquired and used to update the phase of the output signal applied to the coil. In this way, the excitation phase was made dependent on the instantaneous position of the pendulum. 

\textcolor{black}{
	Although this strategy enabled closed-loop phase control, the experimental implementation was affected by delays originating from several hardware and software components. The operational amplifier (OPA541AP) introduced a signal delay of approximately $10~\mu$s, whereas the optical sensor (HEDS-6140\#B13) had a response time of approximately $250$~ns. These hardware-related time scales were considerably shorter than the delays associated with the digital control implementation. The control loop was executed on a standard Windows-based PC rather than on a dedicated real-time platform. Consequently, the execution intervals of the LabVIEW control routine were not strictly deterministic and depended on the operating system, processor load, background tasks, signal acquisition, numerical processing, and output-signal updating. Additional latency and jitter were associated with the USB 2.0 communication interface and the internal buffering and driver operation of the NI USB-6341 data acquisition card. The resulting control-loop execution intervals were typically in the range of $1$--$10~\mathrm{ms}$. This range represents the combined timing variability of the digital control path and should not be interpreted as the latency of the USB interface alone. Consequently, the experimental system should be regarded as a near-real-time rather than a strictly real-time implementation. For the ideal phase-control law
\begin{equation} \phi_{\mathrm{ideal}}(t)=p\,y(t), \end{equation} 
the delayed experimental implementation can be approximated as 
\begin{equation} \phi_{\mathrm{exp}}(t)\approx p\,y(t-\tau), \end{equation} 
where $\tau$ denotes the effective delay of the measurement and control path. The instantaneous error in the commanded excitation phase is therefore 
\begin{equation} \delta\phi(t) =\phi_{\mathrm{exp}}(t)-\phi_{\mathrm{ideal}}(t) =p\left[y(t-\tau)-y(t)\right]. 
	\label{eq:real_phase shift error}
\end{equation} 
For a sufficiently small delay, this error can be approximated as 
 \begin{equation} \delta\phi(t)\approx-p\tau\dot{y}(t). \end{equation} 
Thus, the phase error depends not only on the effective delay, but also on the phase-control coefficient $p$, the instantaneous pendulum velocity, and the amplitudes and frequencies of the components present in the response. For the excitation frequencies investigated experimentally, ranging from approximately $1.00$ to $1.17~\mathrm{Hz}$, a delay of $1$--$10~\mathrm{ms}$ corresponds to a time-domain phase shift of approximately $0.36^\circ$--$4.09^\circ$. These values describe the phase shift of the respective components of the pendulum response and should not be interpreted directly as the complete error in the commanded excitation phase, which is governed by (\ref{eq:real_phase shift error}). In addition, the electromagnetic actuation path introduces its own dynamics because the coil current, and hence the resulting magnetic field, does not change instantaneously. Therefore, the total effective delay between the measurement of the pendulum displacement and the corresponding electromagnetic action may be greater than the control-loop execution interval and may vary between successive iterations. The numerical model used in this study assumes ideal, delay free phase feedback and does not reproduce the non-deterministic timing variations of the experimental implementation. These effects, together with measurement noise and environmental disturbances, may have contributed to some of the local discrepancies between the experimental and numerical results, including the spread of the Poincar\'e points shown later in Fig.~\ref{fig:1T_phaseplot_exp}. Although the controller operated considerably faster than the dominant mechanical response, the influence of delay and timing variability cannot be excluded entirely, particularly near bifurcation points, basin boundaries, and regions of coexisting or chaotic attractors.}

\subsection{Procedure}
Before each experiment, the mechanical configuration of the setup was adjusted and the clearance between the pendulum and the coil was set to 1.6~mm. The desired excitation signal was then generated in LabView and applied to the coil through the current driver. Depending on the test under consideration, the parameters of the excitation were selected so as to investigate the response of the system under phase-controlled magnetic forcing.
\textcolor{black}{
	The initial conditions were imposed manually. Before each experimental run, the pendulum was positioned at the prescribed initial angular displacement using the encoder signal as a reference. The pendulum was held stationary before release; therefore, the initial angular velocity was assumed to be zero, $\dot{\theta}(0)=0$. The initial angular velocity was not controlled independently.}
During the motion, the angular position of the pendulum was recorded continuously by the optical encoder, while the phase of the excitation signal was updated online according to the measured angular position. This enabled the implementation of the state-dependent phase control considered in the mathematical model.
The recorded time histories of the pendulum motion were used for further analysis and for comparison with numerical simulations. In particular, the experimental data served to validate the mathematical model and to construct phase portraits, Poincar\'e maps, and bifurcation diagrams for selected operating conditions. The experimental observations were also used to assess the influence of feedback delay, measurement uncertainty, and environmental disturbances on the system response.

\section{Mathematical model of the studied problem}\label{sec3}
\label{sec:mathematical_model}
\begin{figure}[t!]
	\centering
	\includegraphics[scale=1]{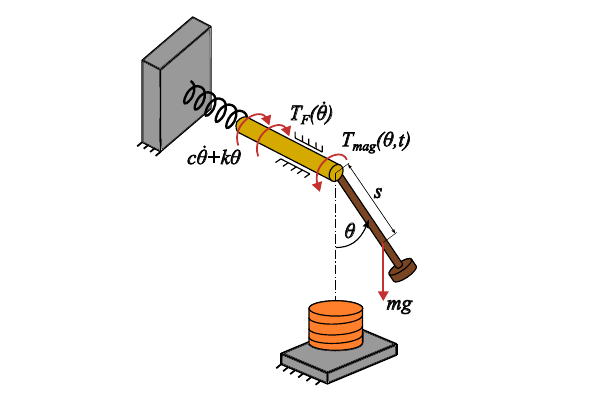}
	\caption{Physical model of the studied system: $mg$ -- gravitational force, $s$ -- distance between axis of rotation and pendulum center of mass, $T_F$ - motion resistance torque, $T_{mag}$ -- magnetic driven torque and $c\dot{\theta}+k\theta$ -- damping and stiffness torques of elastic element. }
	\label{fig:scheme_experimental_rig_2}
\end{figure}
In this section, a mathematical model of the investigated magnetic pendulum system is derived. The derivation is based on the physical model shown in Fig.~\ref{fig:scheme_experimental_rig_2}. The pendulum is subjected to torques associated with gravity, viscous dissipation, elastic restoring action, friction in the rolling bearings, and magnetic interaction. The following assumptions are adopted: the pendulum motion is planar, pendulum and shaft are treated as rigid, moreover the pendulum shaft is joined elastically with fixed support what is represented by a linear torsional stiffness and damping, and the magnetic interaction is described by an experimentally identified torque function.

\subsection{Equation of motion}
According to Newton's second law for rotational motion, the governing equation of motion takes the form
\begin{equation}
	J\ddot{\theta}+c\,\dot{\theta}+mgs\sin{\theta}+k\,\theta+T_F(\dot{\theta})=T_{mag}(\theta,i(t)),
	\label{eq:Newton_motion_1}
\end{equation}
where $\theta$ denotes the angular position of the pendulum, $J$ is the moment of inertia, $c$ is the viscous damping coefficient, and $k$ is the stiffness of the elastic element. The gravitational force $mg$ acts at a distance $s$ from the axis of rotation. The term $T_F(\dot{\theta})$ represents the friction torque generated in the rolling bearings, whereas $T_{mag}(\theta,i(t))$ denotes the magnetic excitation torque produced by the interaction between the coil and the permanent magnet.

\subsection{Friction model}
\begin{figure}[h!]
	\centering
	\includegraphics[scale=0.4]{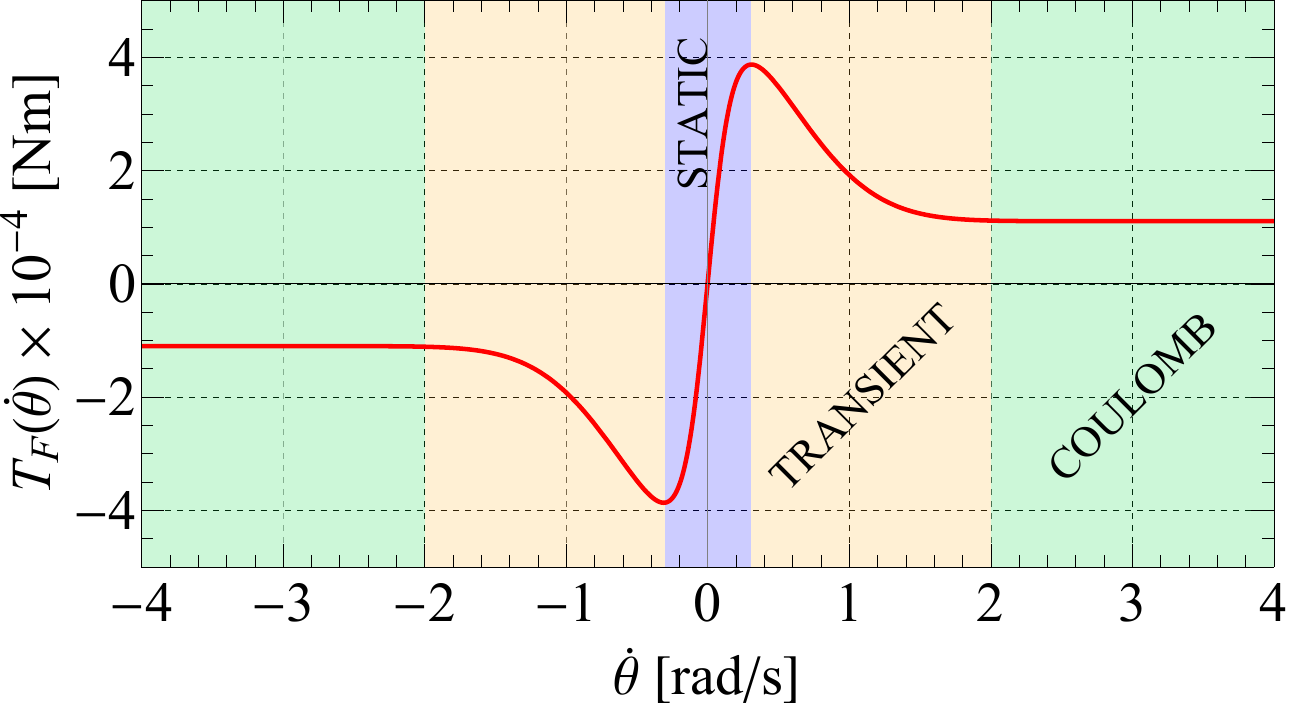}
	\caption{Torque of the resistance forces inside the rolling bearing. Static, transient, and Coulomb character of resistance are shown by purple, yellow and green areas, respectively.}
	\label{fig:resistant torque}
\end{figure}

The friction torque in the rolling bearings is described by the following Stribeck type expression \cite{Wijata2021}
\begin{equation}
	T_F(\dot{\theta})=
	\left[
	\tau_c+(\tau_s-\tau_c)\exp{\left(-\frac{\dot{\theta}^2}{v_s^2}\right)}
	\right]
	\tanh{(\epsilon\dot{\theta})},
	\label{eq:bearing_damping_2}
\end{equation}
where $\tau_c$ and $\tau_s$ are the Coulomb and static friction torques, respectively, $v_s$ is the Stribeck velocity, and $\epsilon$ is a regularization parameter. The corresponding torque vs. velocity characteristic is shown in Fig.~\ref{fig:resistant torque}. For low angular velocities, the friction torque exhibits a peak associated with static resistance, whereas for larger velocities it approaches the Coulomb friction level.

\begin{figure}[b!]
   	\begin{subfigure}{0.59\textwidth} 
 		 \includegraphics[scale=0.37]{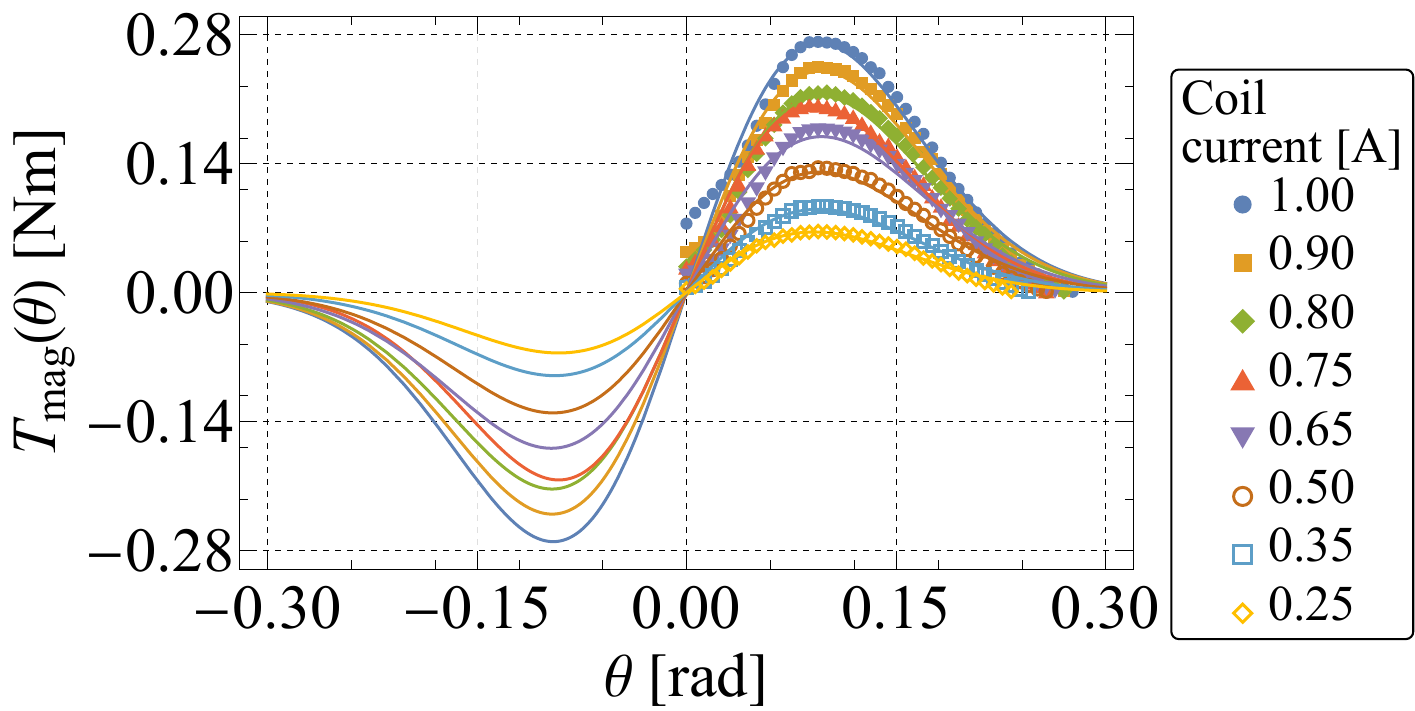} \caption{}
 	\end{subfigure}
 	\begin{subfigure}{0.4\textwidth} 
 		\includegraphics[scale=0.37]{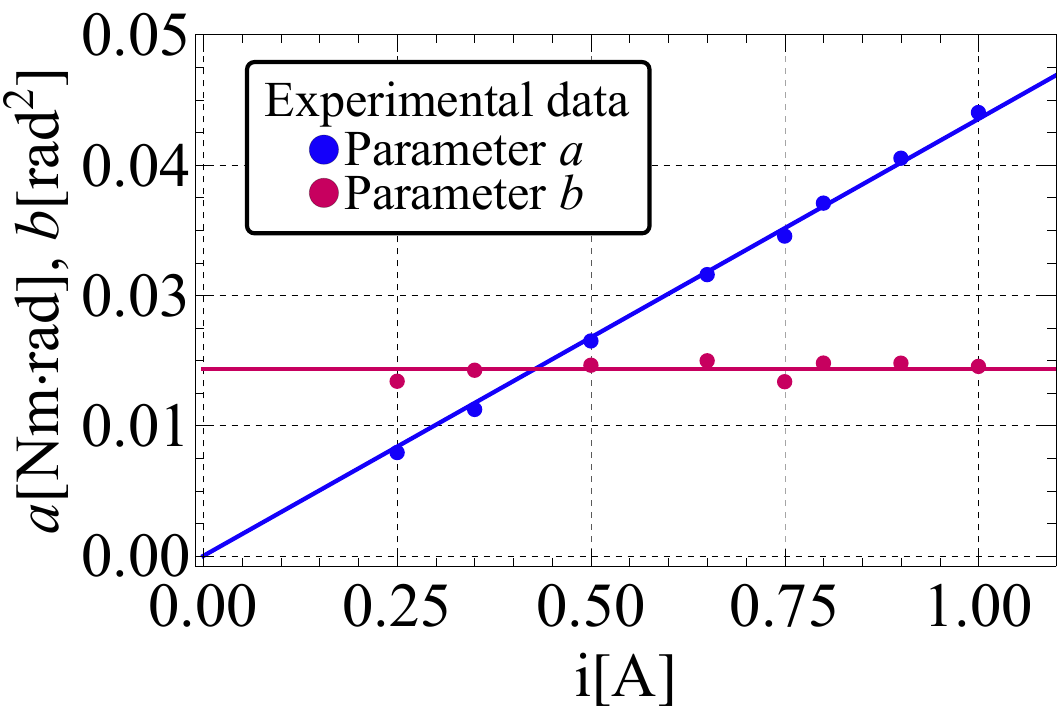} \caption{}
 	\end{subfigure}
 	\caption{Experimental identification of the current-dependent magnetic torque model. (a)~	Measured magnetic torque-angle characteristics for selected constant values of the coil current (dots) and the corresponding Gaussian-type fitted models given by Eq.~(\ref{eq:magnetic_func_3}) (solid curves). (b)~Values of the effective parameters $a$ and $b$ obtained by fitting Eq.~(\ref{eq:magnetic_func_3}) independently to the torque-angle characteristics measured at constant coil current levels. The identified values of $a$ are approximated by the linear relation given in Eq.~(\ref{eq:a_parameter_function_4}), whereas $b$ is approximated by its mean value.	}
 	 \label{fig:Current_parameters_3} 
 \end{figure}

\textcolor{black}{
	The parameters of the Stribeck friction model were identified using a two-stage procedure. First, free-decay experiments were conducted with the electromagnetic excitation disabled and the elastic element disconnected.}
 The resistive torque was reconstructed from the measured angular displacement, velocity, and acceleration, and the Stribeck model, supplemented by the viscous damping term, was fitted to the resulting torque-velocity characteristic. The estimated parameters were subsequently used as initial values in a time-domain optimization based on the complete nonlinear equation of motion. In the second stage, the parameters were refined by minimizing the discrepancy between the simulated and experimentally measured free-decay responses.
\textcolor{black}{
	\subsection{Magnetic torque model and its experimental identification} 
The magnetic torque depends on both the angular position of the pendulum and the instantaneous current $i(t)$~flowing through the coil. In \cite{Wijata2021,PilipchukPolczynski2022}, the localized torque-angle characteristic was represented using a Gaussian-type function with constant parameters $a$ and $b$, identified for a specified constant coil current. In the present study, this representation is extended to account explicitly for variations in the coil current. For this purpose, the constant parameters employed in the previous formulation are generalized as potentially current-dependent quantities $a(i)$ and $b(i)$. This leads to the following magnetic torque model}
\begin{equation}
	 T_{mag}(\theta,i(t))= \frac{2a(i(t))}{b(i(t))} \exp{\left(-\frac{\theta^2}{b(i(t))}\right)}\theta. \label{eq:magnetic_func_3}
 \end{equation} 
 \textcolor{black}{
 	Thus, the Gaussian-type dependence on the angular displacement is adopted from the earlier representation, whereas the experimental identification of the current dependence of its parameters constitutes the extension introduced in the present study. This extension is required because the phase-dependent excitation produces a time-varying coil current, and the magnetic torque must therefore be evaluated using the instantaneous value of $i(t)$. The functional form in Eq.~(\ref{eq:magnetic_func_3}) reflects the localized nature of the magnetic interaction presented in Fig. \ref{fig:Current_parameters_3}a.}
 \textcolor{black}{
 	In the experimental setup, the magnetic torque acts significantly only within a relatively narrow angular range around the lower equilibrium position. The Gaussian-type factor captures this localization, while the multiplicative factor $\theta$ ensures odd symmetry with respect to $\theta=0$. Consequently, the torque tends to zero for large angular deflections and changes sign with the pendulum displacement, which is consistent with the observed physical behaviour of the magnetic interaction. The parameterization through $a(i)$ and $b(i)$ allows both the magnitude and the effective angular width of the torque profile to be examined as possible functions of the coil current. The parameter $a(i)$ represents the overall strength of the magnetic interaction and incorporates the combined influence of the coil current, magnetic field intensity, permanent magnet properties, and coil-magnet geometry. The parameter $b(i)$ determines the effective angular width of the interaction region and describes how rapidly the magnetic torque decreases as the pendulum moves away from the lower equilibrium position. In this sense, $b(i)$ is primarily associated with the spatial distribution of the magnetic field and the geometry of the coil-magnet arrangement, including the clearance between the interacting components. Both parameters should be interpreted as effective lumped quantities rather than isolated fundamental magnetic constants. Their possible dependence on the coil current was identified experimentally in the present study, as shown in Fig.~\ref{fig:Current_parameters_3}b. For each prescribed constant value of the coil current, the magnetic torque was measured as a function of the pendulum angle. Each experimentally obtained torque-angle characteristic was then fitted independently using Eq.~(\ref{eq:magnetic_func_3}), yielding a separate parameter pair $(a,b)$. The identified values were subsequently analyzed as functions of the coil current. The identified parameter $a$ was found to vary approximately linearly with the coil current within the investigated operating range. Accordingly, its current dependence was represented as}
 \begin{equation}
 	 a(i(t))=K_p\,i(t),
 	  \label{eq:a_parameter_function_4} 
 	  \end{equation}
\textcolor{black}{
	where $K_p$ is the experimentally identified proportionality coefficient. The linear fit yields a coefficient of determination equal to $R^2=0.9997$, confirming the suitability of this approximation within the investigated current range. Although $b$~was initially treated as a potentially current-dependent parameter, its identified values showed no systematic variation with the coil current. The mean value was found to be
 \begin{equation}
	b(i(t))\equiv{b}=0.0179\:\mathrm{rad^2},
	\label{eq:b_parameter_mean}
\end{equation}
with a standard deviation of $\sigma_{SD}=0.0008$ $\mathrm{rad^2}$.}
 \textcolor{black}{
 	This result indicates that, within the investigated current range, the coil current primarily modifies the strength of the magnetic interaction, whereas the effective angular width of the interaction region remains approximately unchanged. Consequently, the experimentally identified magnetic torque model used in numerical analysis takes the form}
 \begin{equation}
 	T_{mag}(\theta,i(t))= \frac{2K_p i(t)}{{b}} \exp{\left(-\frac{\theta^2}{{b}}\right)}\theta. 
 	\label{eq:magnetic_func_final}
 \end{equation} 
\textcolor{black}{
	Equation~(\ref{eq:magnetic_func_final}) constitutes a current-dependent extension of the constant-parameter Gaussian torque profile employed in \cite{Wijata2021,PilipchukPolczynski2022}. It enables the magnetic torque to be evaluated continuously for the instantaneous coil current generated by the phase-dependent excitation law. For all considered current levels, the fitted curves shown in Fig.~\ref{fig:Current_parameters_3}a reproduce the experimental torque-angle characteristics with coefficients of determination exceeding $R^2=0.98$. This confirms the suitability of the adopted representation for the investigated operating range. The remaining discrepancies between the fitted curves and the experimental data may be associated with measurement uncertainty and residual mechanical resistance present during the torque measurements. In particular, the contribution of bearing friction could not be eliminated completely and was most noticeable at small angular deflections.}

\subsection{Classical harmonic excitation}

In the classical formulation, the current supplied to the coil is assumed to be harmonic
\begin{equation}
	i(t)=I_0\sin{(\omega_0 t+\phi_0)},
	\label{eq:coil_current_signal_5}
\end{equation}
where $I_0$ is the current amplitude, $\omega_0$ is the angular frequency, and $\phi_0$ is a constant phase shift. Substituting Eqs.~(\ref{eq:a_parameter_function_4}) and (\ref{eq:coil_current_signal_5}) into Eq.~(\ref{eq:magnetic_func_3}) yields
\begin{equation}
	T_{mag}(\theta,t)=
	\frac{2K_p I_0}{b}
	\exp{\left(-\frac{\theta^2}{b}\right)}
	\theta
	\sin{(\omega_0 t+\phi_0)}.
	\label{eq:magnetic_torque_sine_6}
\end{equation}

From a theoretical point of view, the magnetic excitation described by Eq.~(\ref{eq:magnetic_torque_sine_6}) may be interpreted as an additional nonlinear stiffness term that appears periodically in time. Depending on the instantaneous phase of the current signal, this contribution may act in a stabilizing or destabilizing manner. In the classical case, however, the phase $\phi_0$ remains constant and is independent of the pendulum motion.

\subsection{Dimensionless formulation}

To reduce the number of independent parameters and to obtain a more general representation of the system dynamics, Eq.~(\ref{eq:Newton_motion_1}) is transformed into dimensionless form using the substitutions
\begin{equation}
	y=\frac{\theta}{\theta_s},
	\qquad
	x=\frac{t}{t_s},
	\label{eq:dimensionless_substitutions_7}
\end{equation}
where
\[
\theta_s=\sqrt{b},
\qquad
t_s=\sqrt{\frac{J}{mgs}}.
\]
The angular scaling $\theta_s=\sqrt{b}$ is chosen so that the Gaussian factor in the magnetic torque reduces to the normalized form $\exp(-y^2)$. After substitution, the governing equation takes the form
\begin{equation}
	y''+\beta y'+\alpha y+\gamma\sin{\left(\frac{1}{\gamma}y\right)}
	+\left[\delta+\zeta\exp\left(\nu {y'}^2\right)\right]\tanh{(\sigma y')}
	=
	A_0\exp{(-y^2)}y\sin{(\Omega x+\phi_0)},
	\label{eq:motion_dimensionless_8}
\end{equation}
where the dimensionless parameters are defined as
\begin{equation}
	\begin{split}
		\alpha&=\frac{k}{mgs},
		\qquad
		\beta=\frac{c\sqrt{J}}{J\sqrt{mgs}},
		\qquad
		\gamma=\frac{1}{\sqrt{b}},
		\qquad
		\delta=\frac{\tau_c}{mgs\sqrt{b}},
		\qquad
		\zeta=\frac{\tau_s-\tau_c}{mgs\sqrt{b}},
		\\[2mm]
		\nu&=-\frac{bmgs}{v_s^2J},
		\qquad
		\sigma=\frac{\epsilon\sqrt{bmgs}}{\sqrt{J}},
		\qquad
		A_0=\frac{2K_pI_0}{bmgs},
		\qquad
		\Omega=\frac{\omega_0}{\sqrt{mgs/J}}.
	\end{split}
	\label{eq:parameters_dimensionless_9}
\end{equation}
Equation~(\ref{eq:motion_dimensionless_8}) corresponds to the classical case of externally prescribed magnetic forcing with constant phase shift $\phi_0$.

\subsection{Extension to phase-controlled excitation}
\label{sec:phase_control_model}

The main extension considered in the present study consists in allowing the phase of the excitation signal to depend on the instantaneous angular position of the pendulum. In dimensionless form, the phase shift is assumed to vary linearly with the generalized coordinate according to
\begin{equation}
	\phi_0(y)=py,
	\label{eq:phase_shift_control}
\end{equation}
where $p$ is a constant phase-control coefficient.

This modification introduces a feedback-type coupling between the system state and the excitation signal, because the timing of the magnetic forcing is no longer prescribed externally but is adjusted according to the instantaneous pendulum position. Substituting Eq.~(\ref{eq:phase_shift_control}) into Eq.~(\ref{eq:motion_dimensionless_8}) leads to the final dimensionless equation of motion considered in the present work
\begin{equation}
	y'' + \beta y' + \alpha y + \gamma \sin{\left(\frac{1}{\gamma} y\right)}
	+ \left[\delta + \zeta \exp\left(\nu {y'}^2\right)\right] \tanh{(\sigma y')}
	=
	A_0 \exp{(-y^2)} y \sin{(\Omega x + py)}.
	\label{eq:motion_control_phase_10}
\end{equation}

The linear dependence $\phi_{0}(y)=py$ was adopted deliberately as the simplest non-trivial form of state-dependent phase control. On the one hand, it can be interpreted as a first-order approximation of a more general smooth function $\phi_{0}(y)$ in the vicinity of the operating point, since
$\phi_{0}(y)=\phi_{0}(0)+\phi'_{0}(0)y+\mathcal{O}(y^{2}).$
Furthermore, it represents a realistic proportional feedback law that can be implemented directly in real time using the measured angular position of the pendulum. This choice therefore serves both as a minimal theoretical model, allowing the isolated effect of phase-state coupling to be investigated, and as a practically realizable control strategy.

It should be emphasized that Eq.~(\ref{eq:motion_control_phase_10}) is not obtained simply by changing the numerical value of a parameter in the classical forcing term. Rather, the excitation is transformed from a position-dependent magnetic forcing with externally prescribed harmonic time dependence into a state-dependent phase-modulated forcing. As a consequence, the timing and localization of energy transfer between the electromagnetic actuator and the pendulum become dependent on the current state of motion. In this sense, the phase-controlled formulation introduces an additional feedback path into the system and therefore distinguishes the present model from classical magnetic pendulum systems driven by constant-phase excitation.

From a physical viewpoint, the relation $\phi_{0}=py$ may also be interpreted as an effective state-dependent modulation of the excitation timing, since the instantaneous phase evolution yields a local frequency shift proportional to the velocity, i.e. $\Omega + py'$. Such a mechanism is relevant in systems where the instant of energy injection must adapt to the current state of motion, for example in synchronization and phase-locking problems, resonance control, adaptive vibration suppression, and energy transfer regulation in mechatronic oscillators. By modifying the excitation phase according to the instantaneous pendulum position, the system can supply or withdraw energy at dynamically favorable moments, which is essential for controlling multistability, maintaining desired periodic responses, or avoiding unwanted transitions to chaotic motion.
\begin{table}[ht]
	\caption{Values of dimensionless parameters.}
	\label{tab:parameters_1}
	\centering
	\begin{tabular}{ccccccccc}
		\hline \hline
		Parameter & $\alpha$ & $\beta$ & $\gamma$  & $\delta$ & $\zeta$ & $\nu$  & $\sigma$ \\
		Value     & 0.33220  & 0.03431 & 7.46857 & 0.01443  & 0.04492 & -2.33268  & 7.34449  &\\
		\hline \hline
	\end{tabular}
\end{table}
\subsection{Parameter identification}
The numerical computations presented in this paper were performed in \textit{Wolfram Mathematica}; additional details concerning the numerical integration procedure are provided in the Appendix \ref{app:numerical_methods}. The dimensionless parameters defined by Eq.~(\ref{eq:parameters_dimensionless_9}) are listed in Table~\ref{tab:parameters_1}, except for the excitation amplitude $A_0$, whose value is specified separately for each result discussed later.

The dimensionless parameters were derived from the corresponding dimensional quantities. The dimensional parameters were identified using a combination of experimental measurements and numerical fitting. In particular, the quantities $mgs$, $k$, and $J$ were first estimated experimentally and then used as reference values in the numerical identification procedure. Next, the equation of motion (\ref{eq:Newton_motion_1}) was fitted to the experimental time histories describing the free oscillations of the pendulum. This procedure made it possible to refine the reference estimates and to determine the damping-related parameters, namely $c$, $\tau_c$, $\tau_s$, and $v_s$. The magnetic parameters $K_p$ and $b$ were identified independently from dedicated measurements of the magnetic torque as described above.

\section{Preliminary studies}\label{sec4}

In this section, preliminary investigations are carried out in order to provide a basic understanding of the system behavior before the detailed nonlinear analysis presented later. First, the equilibrium structure of the associated frozen system is examined and the local stability of selected equilibrium states is discussed in the context of the phase-controlled formulation. Next, the mathematical model is validated against experimental observations using phase portraits, Poincar\'e maps, and bifurcation diagrams obtained for the classical reference case with constant excitation phase (\(p=0\)). 
\textcolor{black}{
	In this way, the section provides both a without phase-controlled model and a baseline dynamical characterization of the experimentally validated system.}

\begin{figure}[ht]
	\centering
	\begin{subfigure}{0.4\textwidth}
		\includegraphics[scale=0.3]{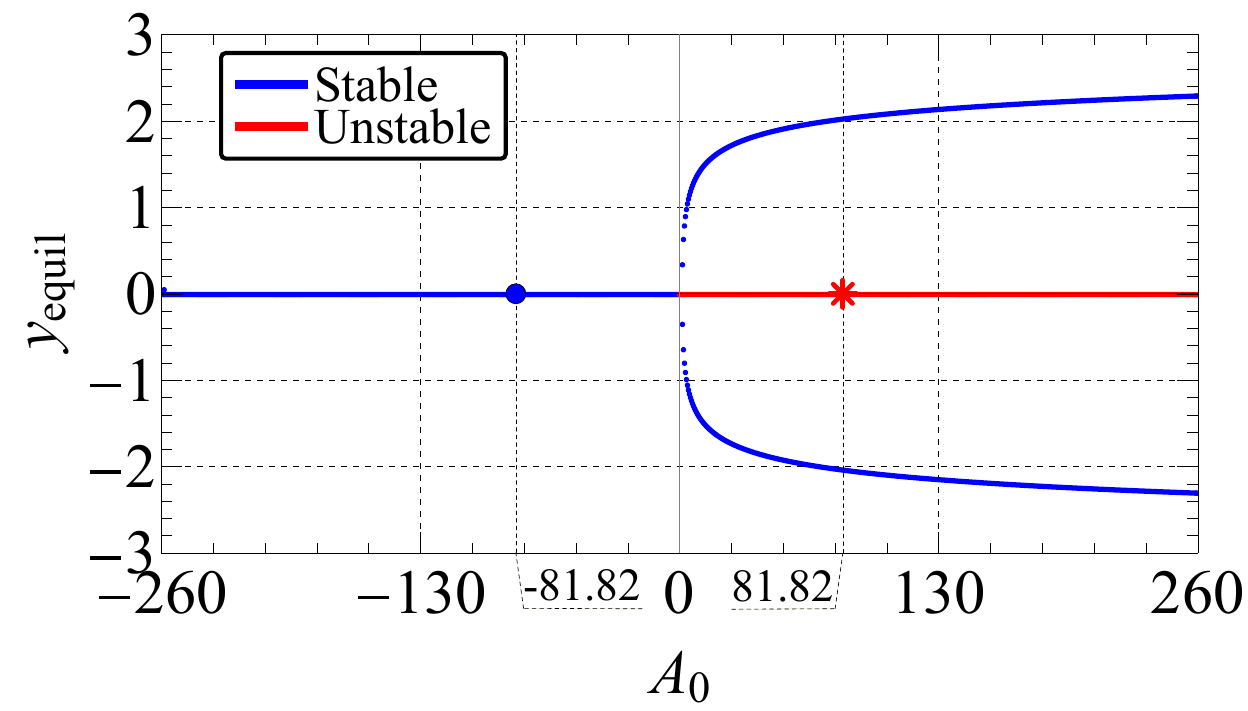}
		\caption{} \label{}
	\end{subfigure}
	\hfil
	\begin{subfigure}{0.3\textwidth}
		\includegraphics[scale=0.3]{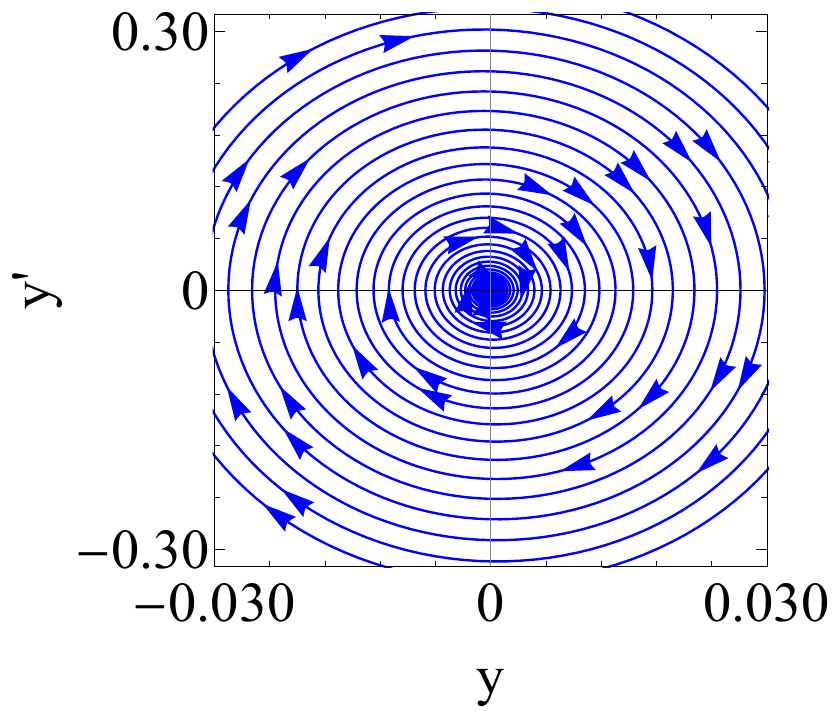}
		\caption{} \label{}
	\end{subfigure}
	\hfil
	\begin{subfigure}{0.25\textwidth}
		\includegraphics[scale=0.3]{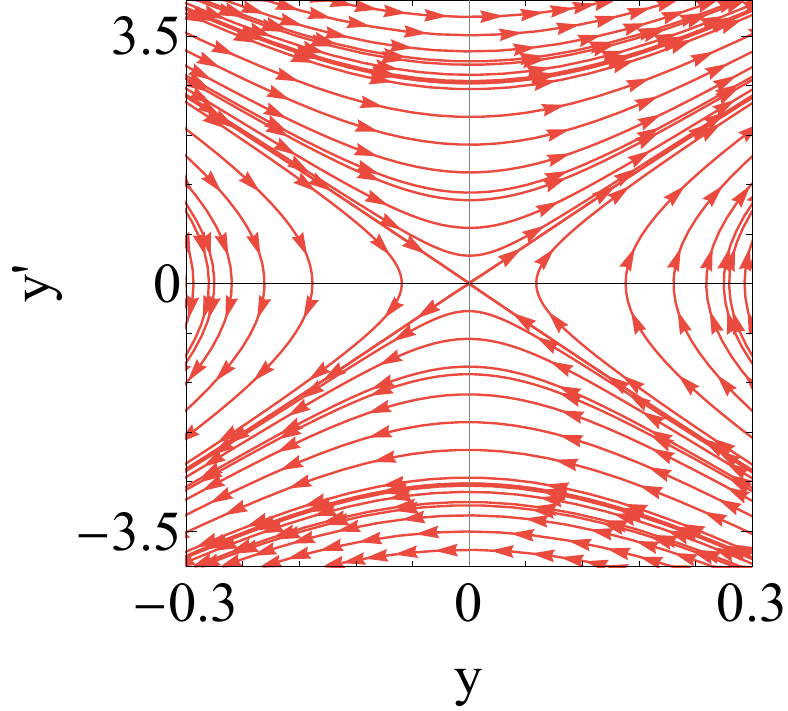}
		\caption{} \label{}
	\end{subfigure}
	\caption{Equilibrium structure and local stability of the frozen system. (a)~Equilibrium positions $y_{\mathrm{equil}}$ as functions of the control parameter $A_0$. The central equilibrium is stable for negative $A_0$, whereas for positive $A_0$ it loses stability and two symmetric stable non-zero branches emerge. (b)~Local phase portrait of the lower equilibrium for the attractive direction of the magnetic interaction, showing a stable focus [blue dot in (a)]. (c)~Local phase portrait of the central equilibrium for the repulsive direction of the magnetic interaction, showing an unstable saddle [red asterisk in (a)].}
	\label{fig:Stability_saddle_focus_4}
\end{figure}

Because the governing equation (\ref{eq:motion_control_phase_10}) is non-autonomous and state-dependent, the notion of equilibrium is considered here for the associated frozen system obtained by fixing the excitation at its extreme values. In this quasi-static formulation, the equilibrium positions are determined from Eq.~(\ref{eq:motion_control_phase_10}) by setting $y'=0$, $y''=0$, and replacing the harmonic excitation term by its instantaneous extreme value. This leads to the equilibrium condition
\begin{equation}
	\alpha y + \gamma \sin\left(\frac{1}{\gamma}y\right)
	-
	A_0 \exp(-y^2)\,y=0,
	\label{eq:equilibrium_condition_11}
\end{equation}
where the sign of $A_0$ reflects the effective direction of the magnetic interaction.
The equilibrium branches computed from Eq.~(\ref{eq:equilibrium_condition_11}) for $A_0 \in [-260,260]$ are shown in Fig.~\ref{fig:Stability_saddle_focus_4}a. For negative values of $A_0$, only the central equilibrium $y_{\mathrm{equil}}=0$ exists and remains stable. In contrast, for positive values of $A_0$, the trivial equilibrium loses stability and two symmetric non-zero stable equilibria emerge. This transition has the form of a supercritical pitchfork bifurcation, which reflects the symmetry of the governing equation with respect to the transformation $y \mapsto -y$.

\textcolor{black}{
	From a physical viewpoint, this means that the attractive magnetic interaction preserves the lower central equilibrium, whereas the repulsive interaction destabilizes it and promotes the formation of two symmetric shifted equilibrium configurations. To examine the local behaviour near the equilibrium states in more detail, a linear stability analysis was performed for selected cases corresponding to $A_0=\pm81.82$, which represents a coil current of $\pm1~\mathrm{A}$. Let
	$\mathbf{y}=[y,y']^{\mathrm{T}}$
	where $y$ and $y'$ denote the dimensionless angular displacement and angular velocity, respectively. For the considered equilibrium state $\mathbf{y}^{*}=[0,0]^{\mathrm{T}}$, the governing equations are linearized in its vicinity, yielding
	\begin{equation}
		 \mathbf{y}' = \mathbf{A}_J\mathbf{y}, \label{eq:linear_stability}
	\end{equation}
	where $\mathbf{A}_J$ is the Jacobian matrix of the autonomous system evaluated at the considered equilibrium state. The detailed derivation of the Jacobian matrices and the corresponding eigenvalue calculations are provided in Appendix~\ref{app:Jacobian_matrix}.}

\begin{figure}[ht]
	\centering
	\includegraphics[scale=0.15]{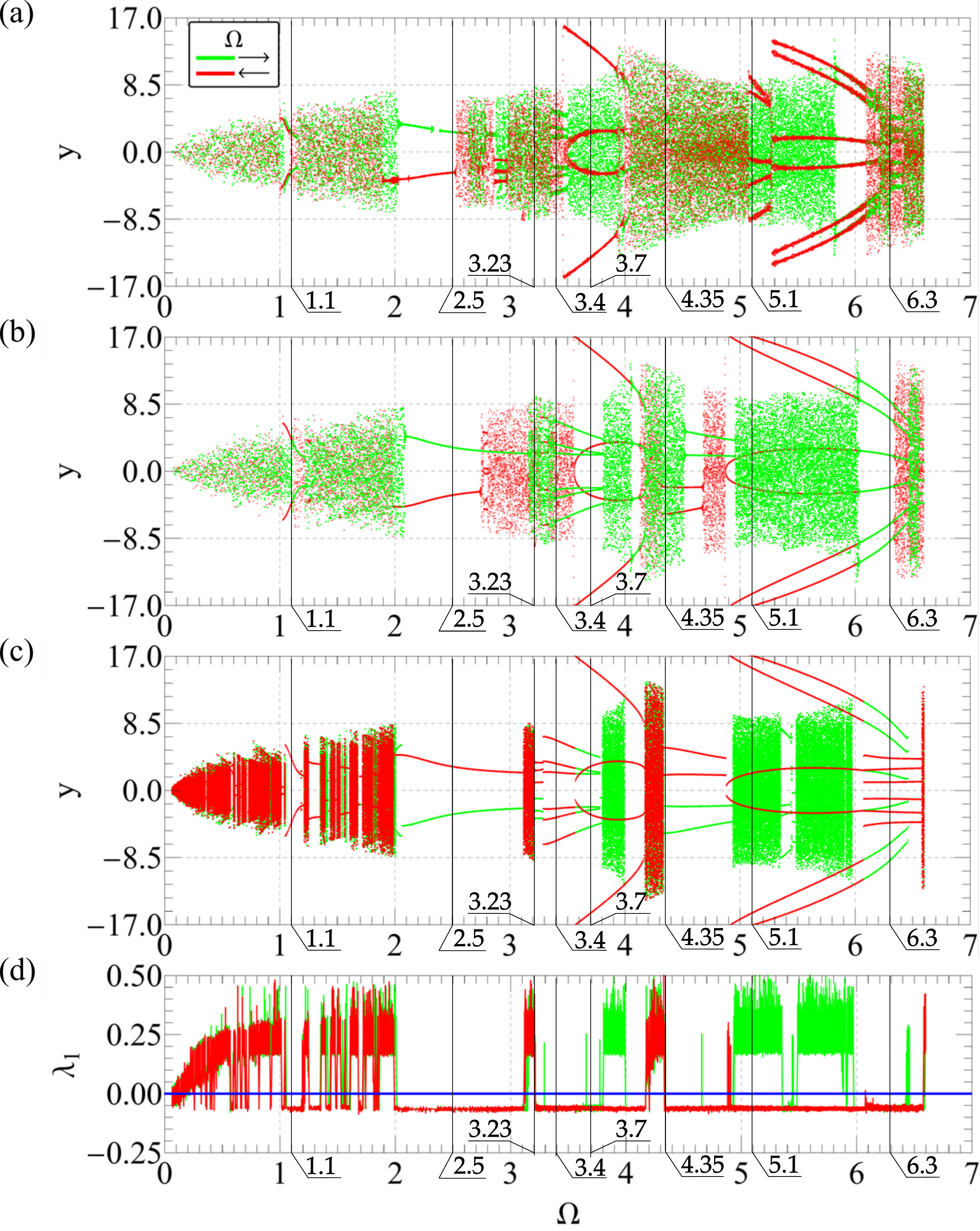}
	\caption{Bifurcation diagrams obtained for the system expressed by Eq.~(\ref{eq:motion_dimensionless_8}) with fixed $A_0=81.82$ and for increasing/decreasing control parameter $\Omega$: (a)~experiment, (b)~simulation imitating experiment, (c)~classic simulation and (d) the largest Lyapunov exponent $\lambda_1$ corresponding to the classic simulation.}
	\label{fig:biff_simulation_LLCE}
\end{figure}

For the attractive direction of the magnetic interaction, the lower central equilibrium behaves as a stable focus, as shown in Fig.~\ref{fig:Stability_saddle_focus_4}b. In contrast, for the repulsive direction, the central equilibrium becomes an unstable saddle, as illustrated in Fig.~\ref{fig:Stability_saddle_focus_4}c. These results are consistent with the equilibrium diagram in Fig.~\ref{fig:Stability_saddle_focus_4}a and confirm that the direction and magnitude of the magnetic interaction play a decisive role in determining the local structure of the phase space.
As a reference case, a bifurcation analysis was first performed for the classical magnetic pendulum model with constant excitation phase, i.e. for \(p=0\). The aim of this analysis was twofold: first, to validate the mathematical model against experimental observations, and second, to identify the basic dynamical capabilities of the system before introducing the phase-controlled excitation considered later in the paper. The dimensionless excitation frequency \(\Omega\) was selected as the control parameter, and bifurcation diagrams were constructed for both increasing and decreasing values of \(\Omega\), as shown in Fig.~\ref{fig:biff_simulation_LLCE}. 
The diagrams were based on Poincar\'e sections sampled at successive upward zero-crossings of the coil current signal.
The experimental bifurcation diagram is shown in Fig.~\ref{fig:biff_simulation_LLCE}a. In this case, the control parameter \(\Omega\) was varied continuously in time in both directions. To reproduce the experimental conditions numerically, the same procedure was implemented in the simulations by replacing the constant-frequency excitation with a linear frequency sweep, yielding the bifurcation diagram shown in Fig.~\ref{fig:biff_simulation_LLCE}b. The slope of the sweep was fixed at \(0.1~\mathrm{Hz/min}\). This sweep-type procedure makes it possible to track the evolution of a single attractor as the control parameter changes, and therefore reproduces the path-dependent behavior observed experimentally. At the same time, it may not reveal all coexisting attractors, because the system follows only one branch selected by its previous dynamical state.
For this reason, a classical bifurcation diagram was also computed for fixed values of the control parameter \(\Omega\), as shown in Fig.~\ref{fig:biff_simulation_LLCE}c. In this case, after each simulation the last Poincar\'e point was used as the initial condition for the next, slightly higher or lower frequency value, allowing the attractor to be continuously tracked in parameter space. This standard continuation-like procedure reveals hysteresis and multistability more clearly, since different branches may be followed depending on the direction of frequency change.
The experimental and numerical bifurcation diagrams show good qualitative agreement. In particular, both the experimental data and the numerical results reveal parameter intervals with regular oscillations, regions of irregular broadband response, and pronounced hysteresis between forward and backward frequency sweeps. 
\textcolor{black}{
	The experimental diagram exhibits a larger spread of points, which can be attributed to measurement noise and other imperfections of the laboratory setup, whereas the numerically obtained diagrams are sharper and reveal the underlying bifurcation structure more clearly.}

To further support the interpretation of the bifurcation structure, the largest Lyapunov exponent corresponding to the classical bifurcation diagram is shown in Fig.~\ref{fig:biff_simulation_LLCE}d. Positive values of \(\lambda_1\) correlate with parameter regions where irregular responses are observed in the bifurcation diagrams, indicating chaotic dynamics, whereas non-positive values correspond to regular oscillatory regimes. The Lyapunov exponents were computed numerically after discarding an initial transient of \(n_t=1000\) forcing cycles. The tangent-space vectors were re-orthonormalized every \(n_n=0.1\) forcing cycle, and the exponents were accumulated over \(n_c=5000\) forcing cycles starting from the initial condition \(\{y(0),y'(0)\}=\{0.001,0\}\).

Overall, the bifurcation analysis for \(p=0\) confirms that the identified model is capable of reproducing the principal dynamical transitions observed experimentally. It also provides a baseline description of the classical magnetic pendulum dynamics, which serves as a reference for assessing the additional effects introduced later by the state-dependent phase control.

\begin{figure}[ht]
	\begin{subfigure}{0.25\textwidth}
		\includegraphics[scale=0.95]{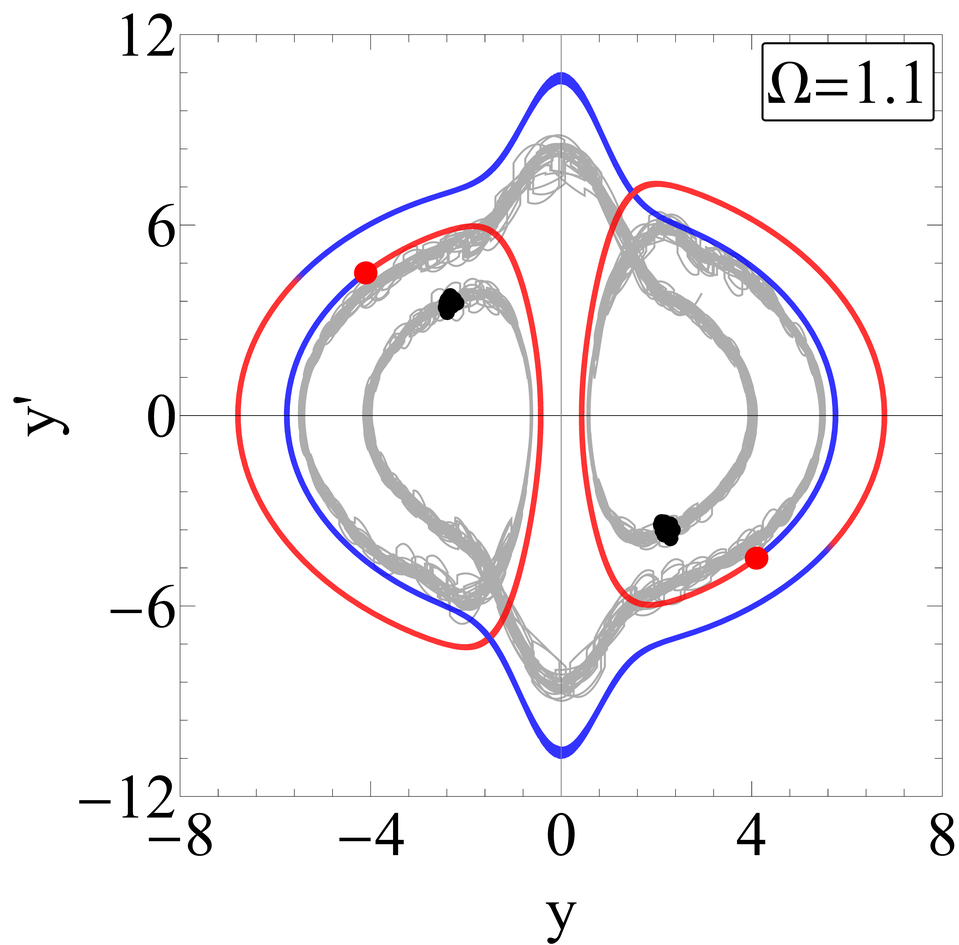}
		\caption{} \label{}
	\end{subfigure}
	\begin{subfigure}{0.25\textwidth}
		\includegraphics[scale=0.95]{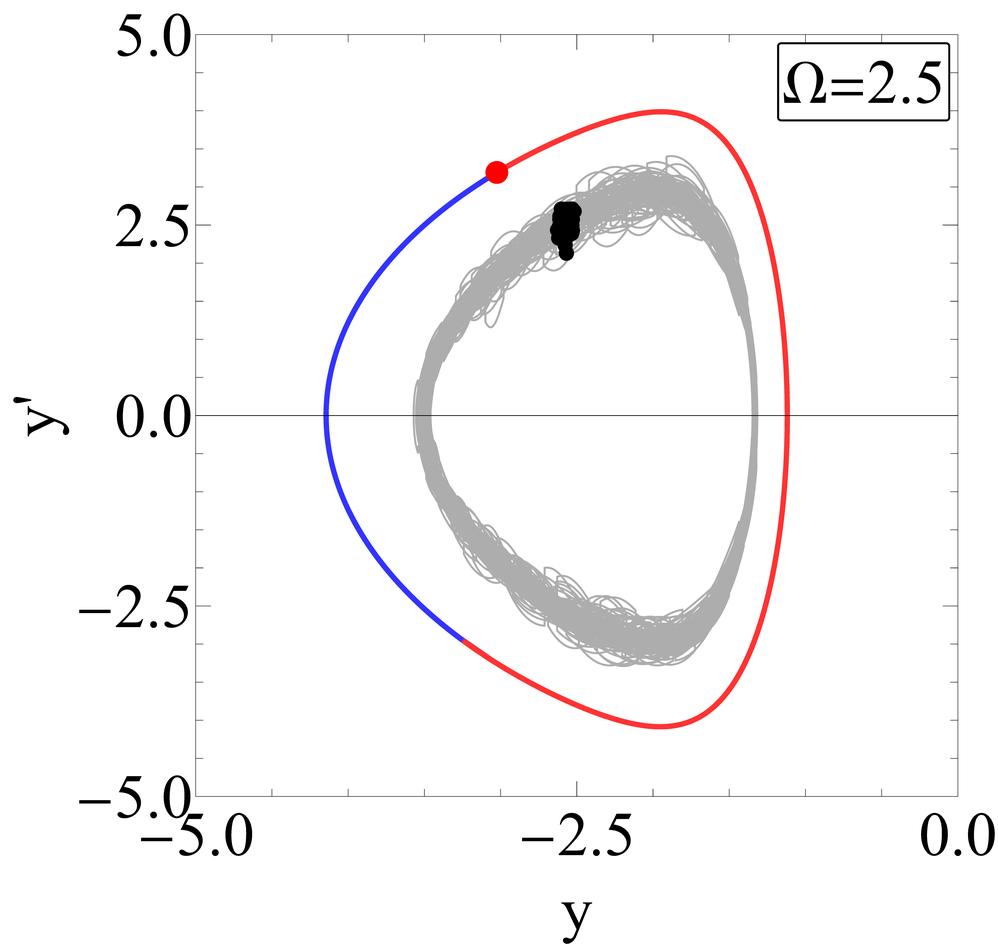}	
		\caption{} \label{}
	\end{subfigure}
	\begin{subfigure}{0.24\textwidth}
		\includegraphics[scale=0.95]{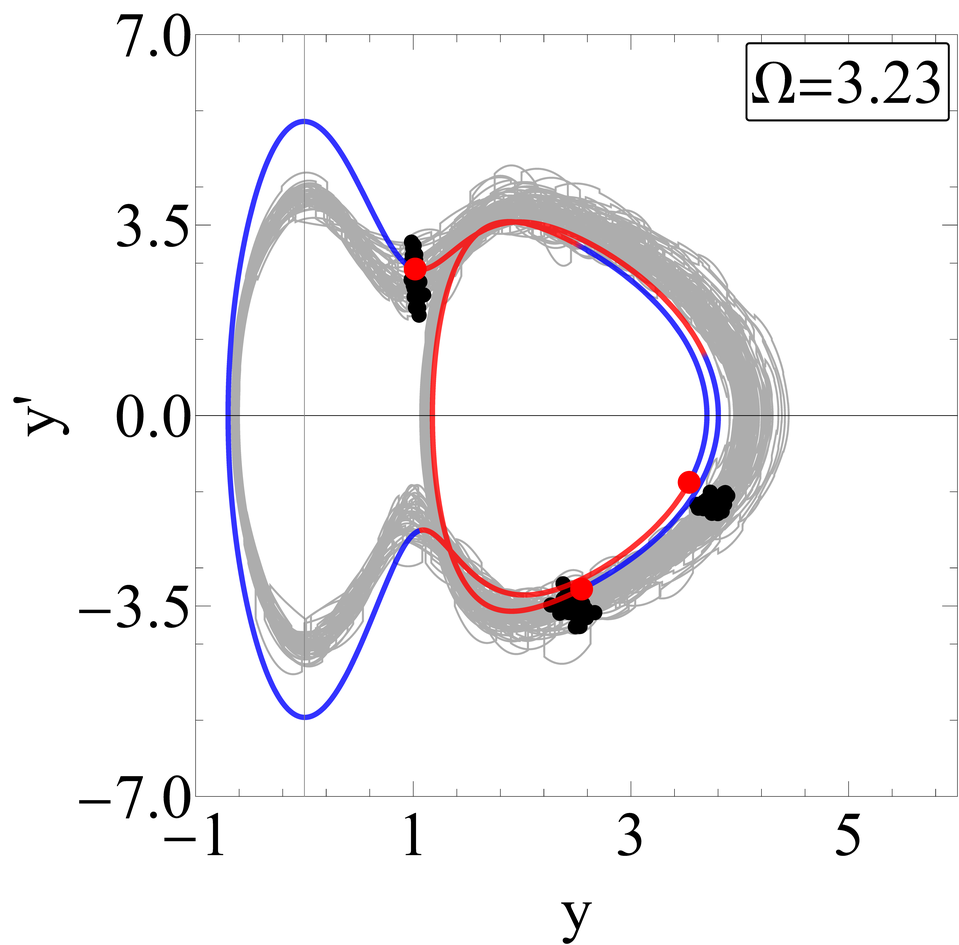}	
		\caption{} \label{}
	\end{subfigure}
	\begin{subfigure}{0.23\textwidth}
		\includegraphics[scale=0.95]{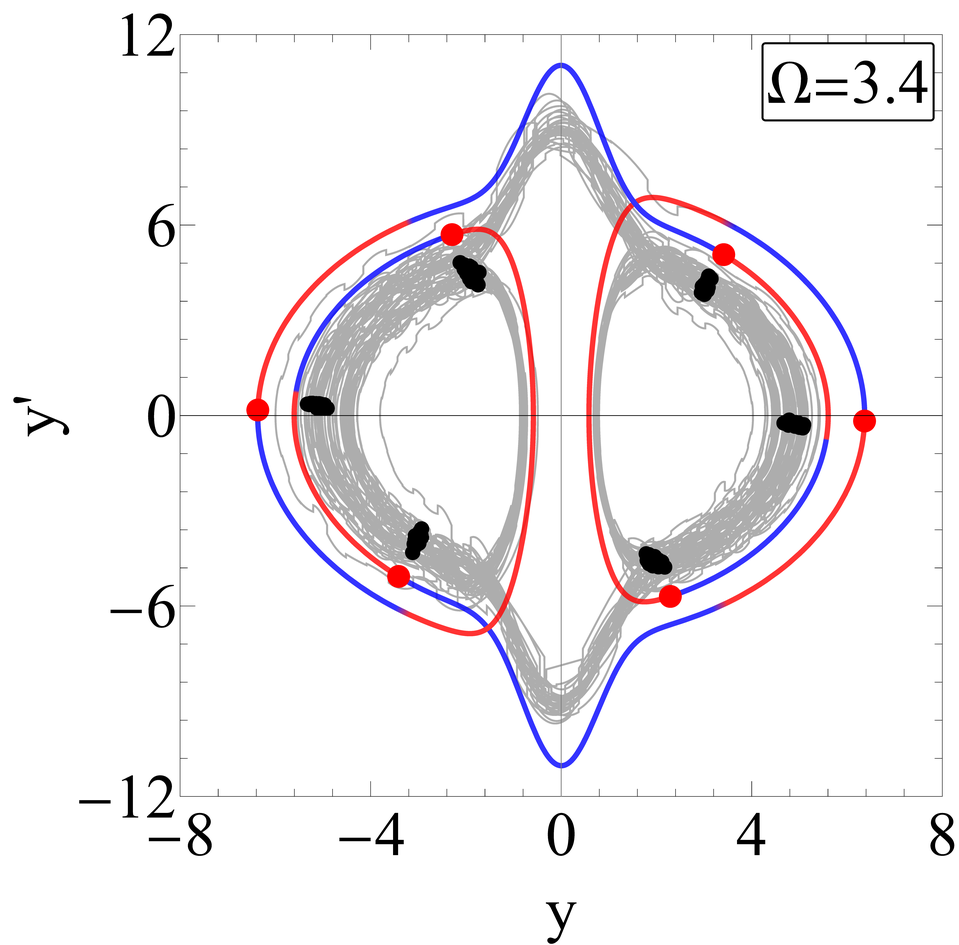}	
		\caption{} \label{}
	\end{subfigure}
	\begin{subfigure}{0.25\textwidth}
		\includegraphics[scale=0.95]{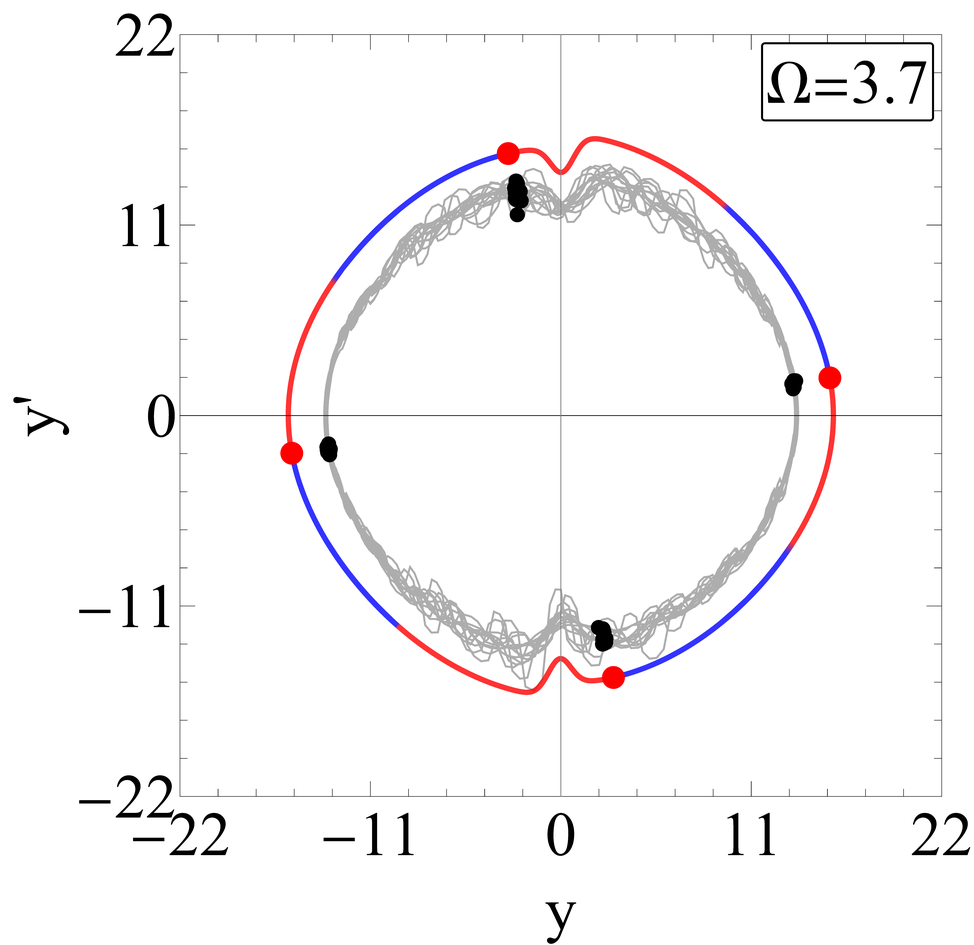}	
		\caption{} \label{}
	\end{subfigure}
	\begin{subfigure}{0.255\textwidth}
		\includegraphics[scale=0.95]{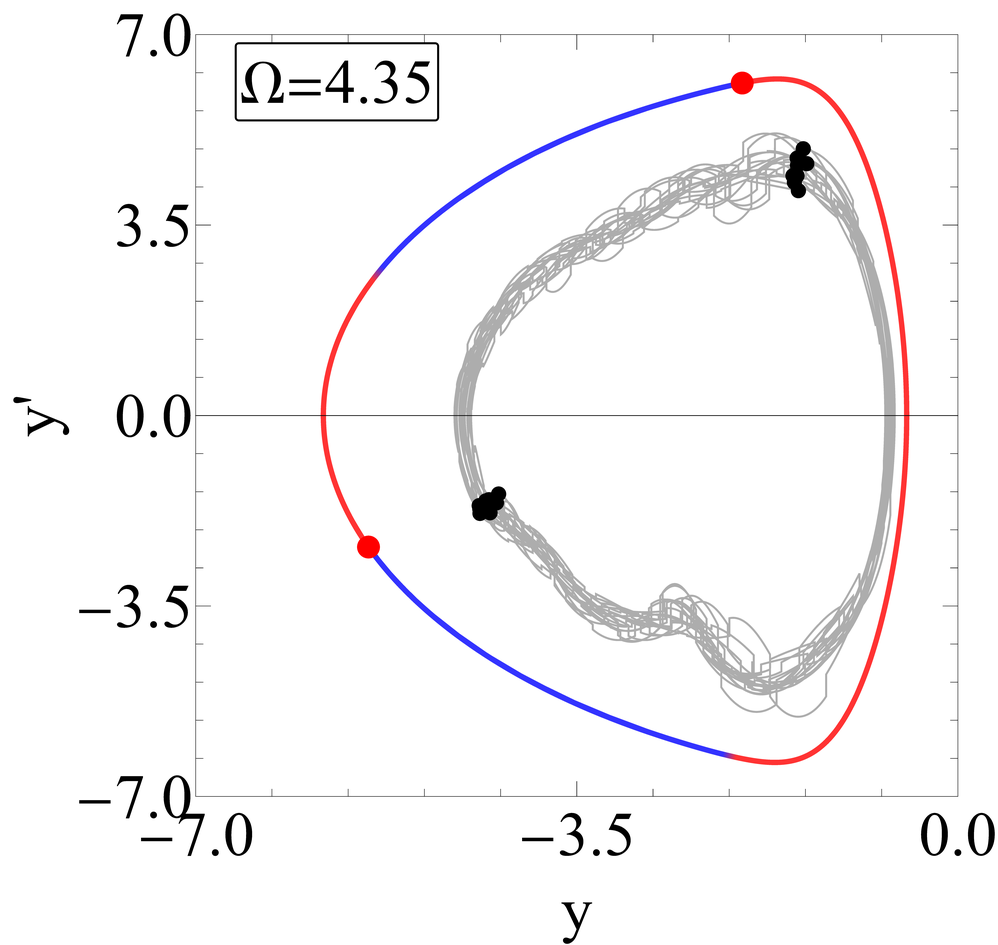}	
		\caption{} \label{}
	\end{subfigure}
	\begin{subfigure}{0.24\textwidth}
		\includegraphics[scale=0.95]{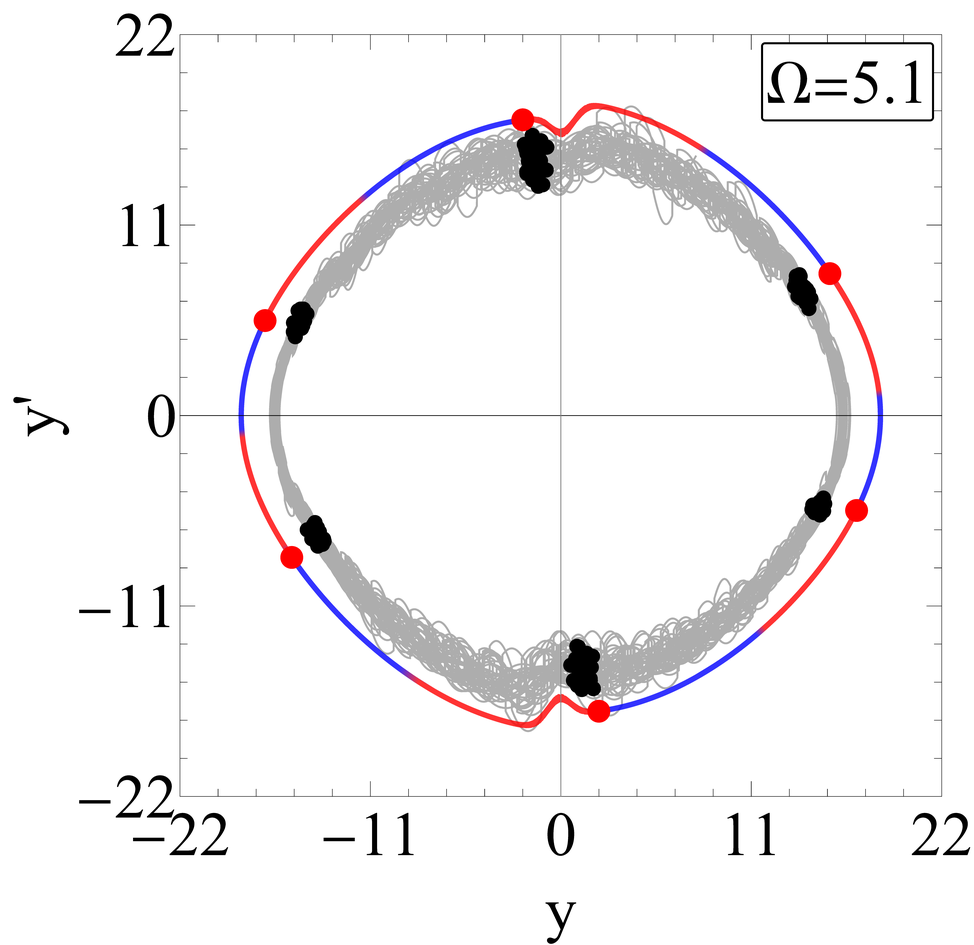}	
		\caption{} \label{}
	\end{subfigure}
	\begin{subfigure}{0.23\textwidth}
		\includegraphics[scale=0.95]{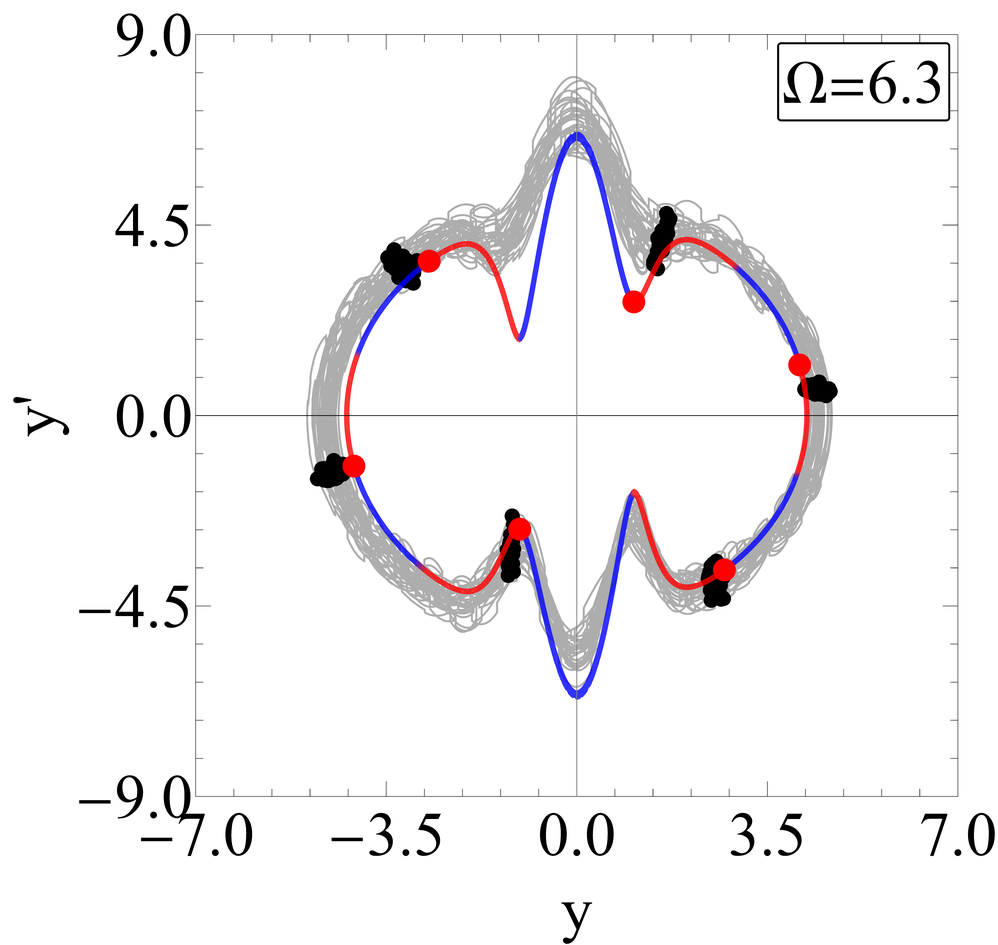}	
		\caption{} \label{}
	\end{subfigure}
	\caption{Comparison of numerical and experimental phase trajectories with the corresponding Poincar\'e sections for selected values of the control parameter \(\Omega\) in the classical system governed by Eq.~(\ref{eq:motion_dimensionless_8}). The numerical phase trajectories are shown by blue-red curves: the red segments correspond to positive coil current (repulsive magnetic interaction), whereas the blue segments correspond to negative coil current (attractive magnetic interaction). The experimental trajectories are shown by gray curves. Dark gray and dark red points denote the experimental and numerical Poincar\'e sections, respectively.}
	\label{fig:preliminary_phase trajectories}
\end{figure}

Selected phase trajectories together with the corresponding Poincar\'e sections are shown in Fig.~\ref{fig:preliminary_phase trajectories}. These examples complement the bifurcation diagrams by illustrating the geometric structure of representative attractors observed for selected values of the control parameter \(\Omega\). A comparison between the numerically computed and experimentally recorded phase trajectories indicates good overall agreement in both the shape of the trajectories and the location of the associated Poincar\'e points.

\textcolor{black}{
The experimental responses are visibly more diffuse, which can be attributed to measurement noise and other imperfections of the laboratory setup; nevertheless, the principal topology of the attractors is reproduced correctly by the mathematical model.}

The presented examples also help interpret the bifurcation structure shown previously in Fig.~\ref{fig:biff_simulation_LLCE}. In particular, some regular oscillatory states visible in Fig.~\ref{fig:preliminary_phase trajectories}c,f are not clearly exposed in the experimental sweep bifurcation diagram in Fig.~\ref{fig:biff_simulation_LLCE}a, but they are recovered in the classical numerical bifurcation diagram in Fig.~\ref{fig:biff_simulation_LLCE}c. This observation confirms that the experimental sweep follows only one path-dependent attractor branch, whereas the classical numerical continuation is able to reveal additional coexisting regular solutions.

Recall that the sign of the coil current determines whether the magnetic interaction acts attractively or repulsively, which corresponds to a temporary change of the effective potential from single-well to double-well form. In this context, the trajectories shown in Fig.~\ref{fig:preliminary_phase trajectories}b,f correspond to intrawell oscillations, i.e. oscillations confined to one potential well. The responses presented in Fig.~\ref{fig:preliminary_phase trajectories}e,g,h correspond to inter-well oscillations, in which the pendulum moves between two potential wells. By contrast, the trajectories in Fig.~\ref{fig:preliminary_phase trajectories}a,c,d exhibit mixed behavior, combining intrawell and interwell motion within one oscillatory cycle.

Taken together, these examples demonstrate that the identified model captures not only the global bifurcation structure of the classical system but also the detailed geometry of representative periodic responses. This agreement provides additional support for using the model as a reference baseline before analyzing the more complex dynamics induced by the state-dependent phase control.

Overall, the preliminary investigations confirm that the experimentally identified model provides a reliable baseline description of the classical system and therefore offers a sound starting point for analyzing the additional dynamical effects introduced by the state-dependent phase control.

\section{Magnetic pendulum under external periodic magnetic field with a~controlled phase} \label{sec5}

In this section, the dynamical response of the magnetic pendulum described by Eq.~(\ref{eq:motion_control_phase_10}) is presented for the case of state-dependent phase control. The results are organized so as to illustrate how the introduction of the phase-control parameter \(p\) modifies the system response in comparison with the classical reference case \(p=0\). Particular attention is paid to the distribution of dynamical regimes in the parameter space, representative attractors, and the structure of the attraction basins for selected values of the excitation parameters. 

\subsection{Dynamical regimes in the \((p,A_0)\) parameter plane}

\begin{figure}[h!]
	\begin{subfigure}{0.32\textwidth}
		\includegraphics[scale=0.8]{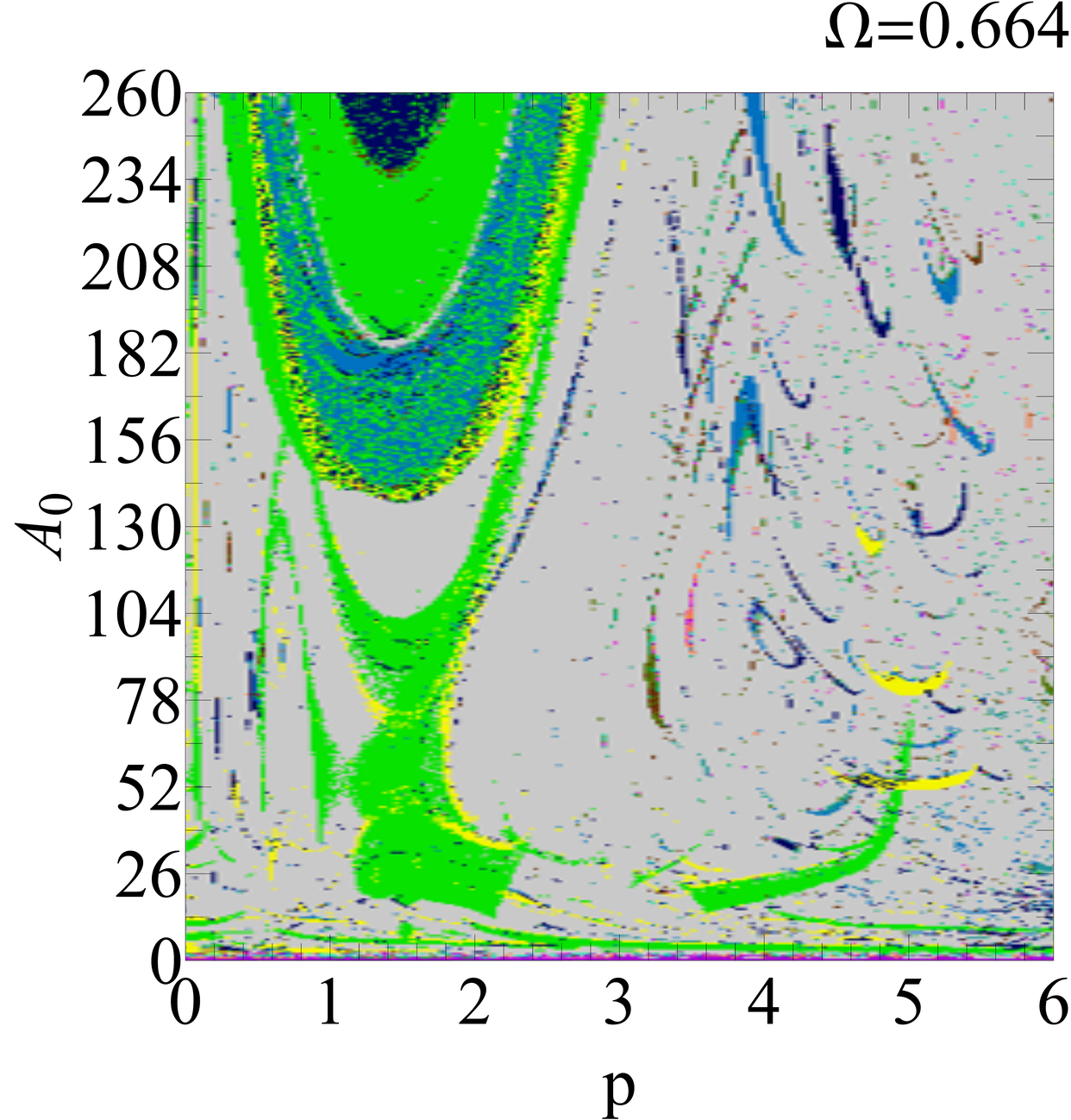}
		\caption{} \label{}
	\end{subfigure}
	\begin{subfigure}{0.32\textwidth}
		\includegraphics[scale=0.8]{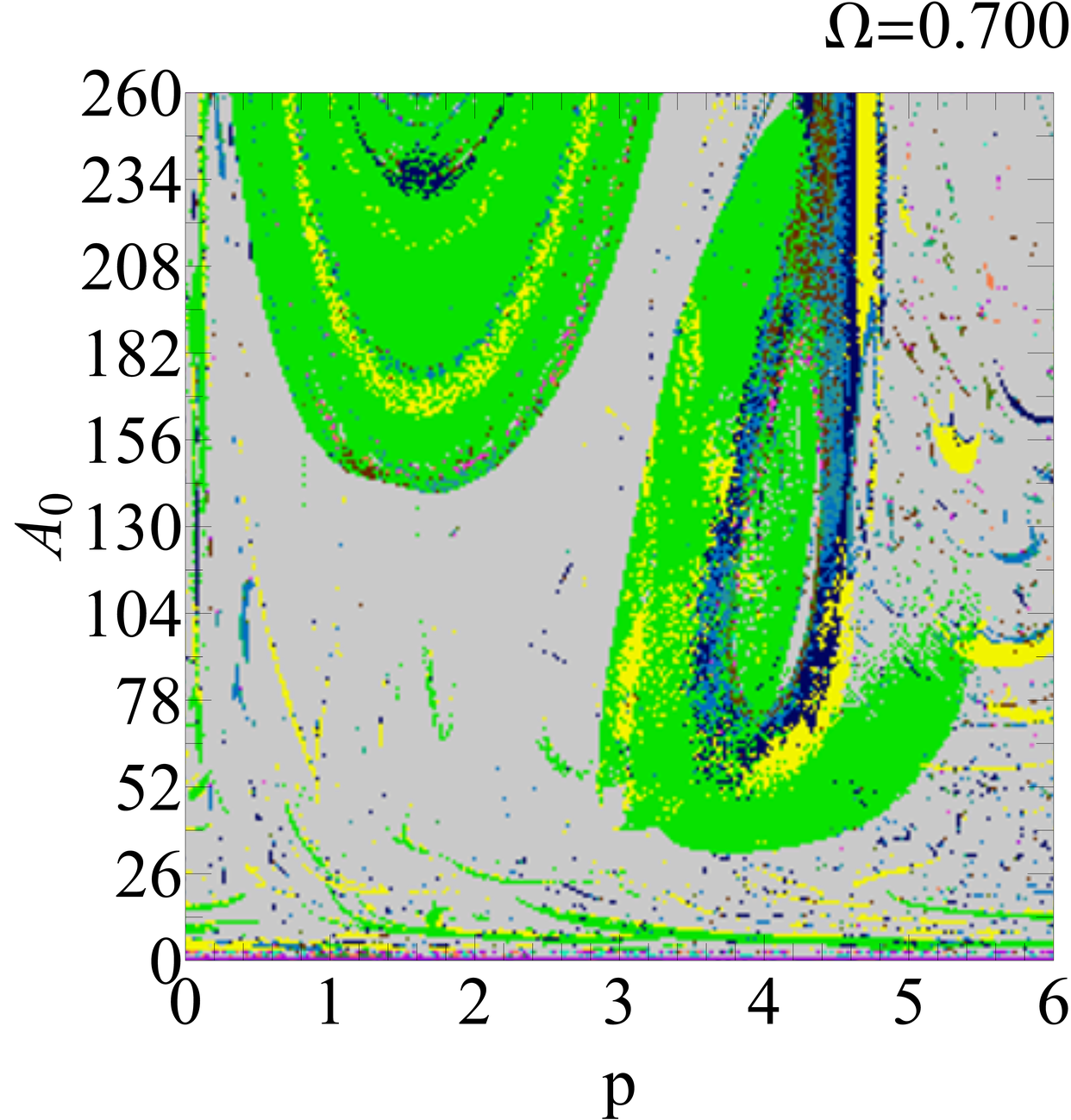}
		\caption{} \label{}
	\end{subfigure}
	\begin{subfigure}{0.32\textwidth}
		\includegraphics[scale=0.8]{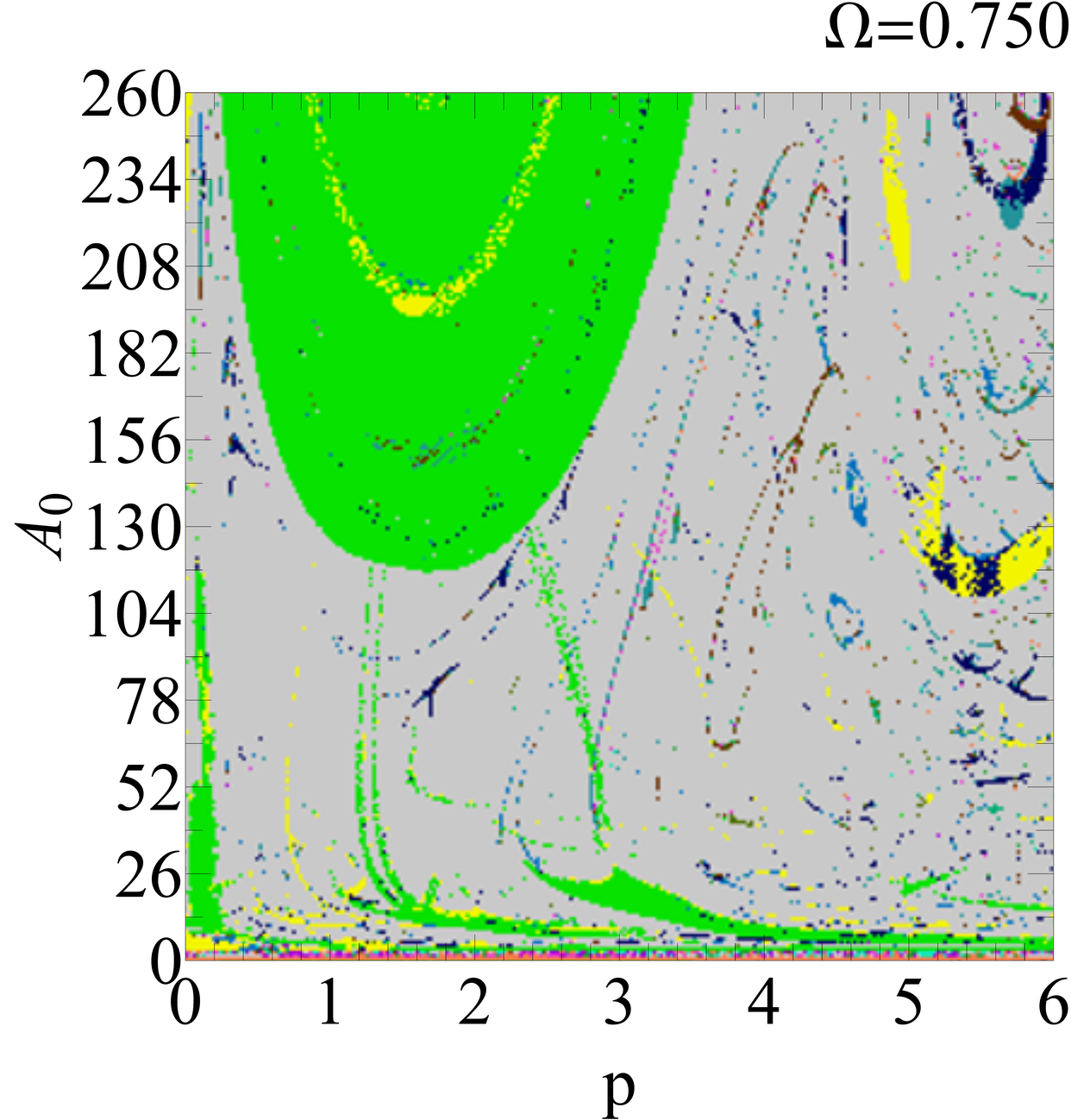} 
		\caption{} \label{}
	\end{subfigure}
	\includegraphics[scale=0.3]{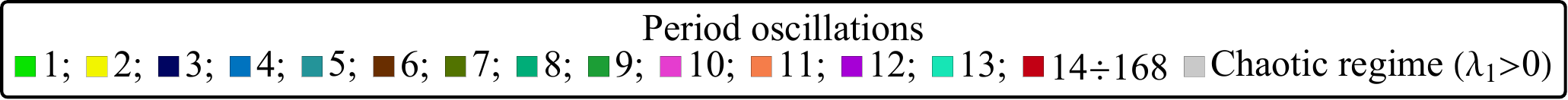} 
	\caption{Charts of dynamical regimes in the \((p,A_0)\) parameter plane for the phase-controlled system governed by Eq.~(\ref{eq:motion_control_phase_10}) and for fixed initial conditions \((y,y')=(0.001,0)\): (a)~\(\Omega=0.664\), (b)~\(\Omega=0.700\), and (c)~\(\Omega=0.750\). Colored regions correspond to periodic solutions of different periods, while gray regions indicate chaotic oscillations identified by positive values of the largest Lyapunov exponent \(\lambda_1\). The color legend is shown below the charts.}
	\label{fig:phase_control_charts}
\end{figure}

Figure~\ref{fig:phase_control_charts} presents charts of dynamical regimes governed by Eq.~(\ref{eq:motion_control_phase_10}) on the parameter plane \((p,A_0)\) for three selected values of the excitation frequency, namely \(\Omega=\{0.664,\,0.700,\,0.750\}\). The purpose of these charts is to illustrate how the system response changes with the phase-control coefficient \(p\) and the excitation amplitude \(A_0\) for representative values of the excitation frequency.
All numerical integrations were performed for the fixed initial conditions \((y,y')=(0.001,0)\) and over a time interval corresponding to 5000 excitation periods. Figure~\ref{fig:phase_control_charts}a shows the regime distribution in the \((p,A_0)\) plane for \(\Omega=0.664\), whereas Fig.~\ref{fig:phase_control_charts}b,c correspond to \(\Omega=0.700\) and \(\Omega=0.750\), respectively. Different colors on the charts correspond to different periodic solutions of the system, while the gray color denotes chaotic oscillations. 
The chaotic regions were identified on the basis of positive values of the largest Lyapunov exponent \(\lambda_1\). The color legend associated with the periodic responses is shown below the charts.

The obtained parameter-plane charts reveal a rich distribution of periodic and chaotic regimes over the investigated \((p,A_0)\) domain. Regions associated with low-period oscillations coexist with parameter intervals where period-doubling sequences and chaotic responses are observed. In addition, the charts indicate strong sensitivity of the long-term response to parameter variation, as different periodic attractors may occur for nearly the same initial conditions but slightly different values of \(p\) and \(A_0\). A comparison of Fig.~\ref{fig:phase_control_charts}a--c shows that moderate changes in the excitation frequency \(\Omega\) do not lead to major qualitative changes in the global structure of the parameter planes. The main effect of varying \(\Omega\) is a shift of the boundaries separating the different dynamical regimes, including the bifurcation values of \(p\) and \(A_0\), rather than a complete reorganization of the regime distribution.
\begin{figure}[b!] 
	\centering 
	\begin{subfigure}{0.49\textwidth} 
		\centering 
		\includegraphics[scale=1] {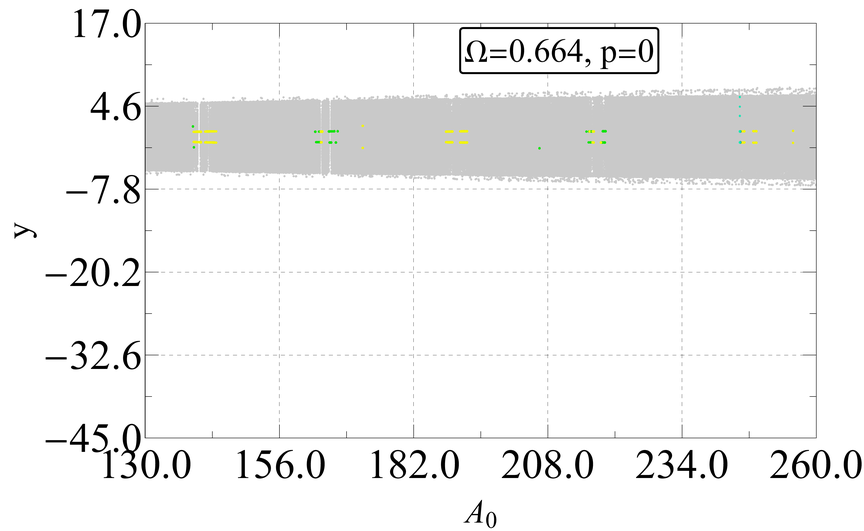} 
		\caption{} 
		\label{fig:bif_Omega0664_p0}
	\end{subfigure} 
	\hfill 
	\begin{subfigure}{0.49\textwidth} 
		\centering 
		\includegraphics[scale=1] {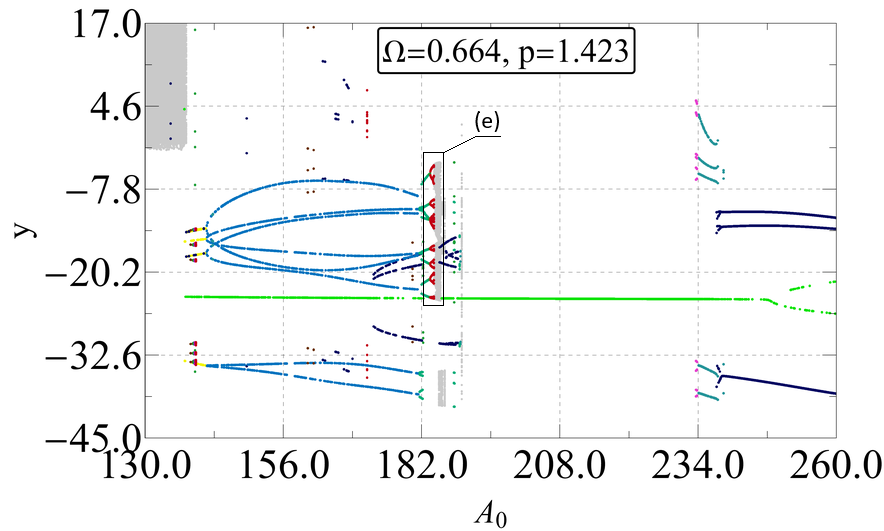} 
		\caption{}
		\label{fig:bif_Omega0664_p1423} 
	\end{subfigure} 
	\begin{subfigure}{0.49\textwidth} 
		\centering 
		\includegraphics[scale=1] {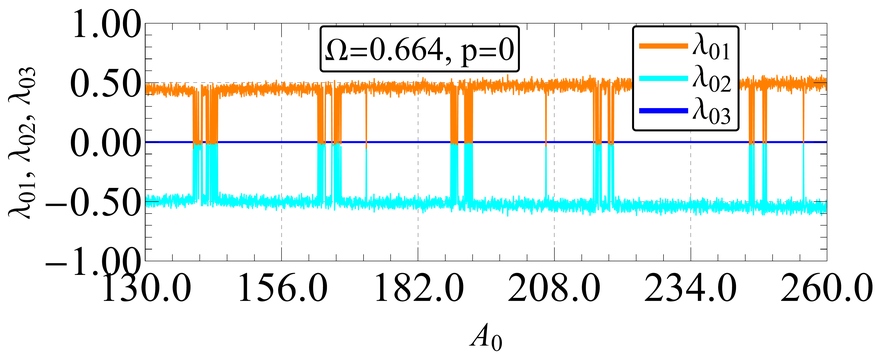}
		\caption{} 
		\label{fig:lyap_Omega0664_p0} 
	\end{subfigure} 
	\hfill 
	\begin{subfigure}{0.49\textwidth}
		\centering
		\includegraphics[scale=1] {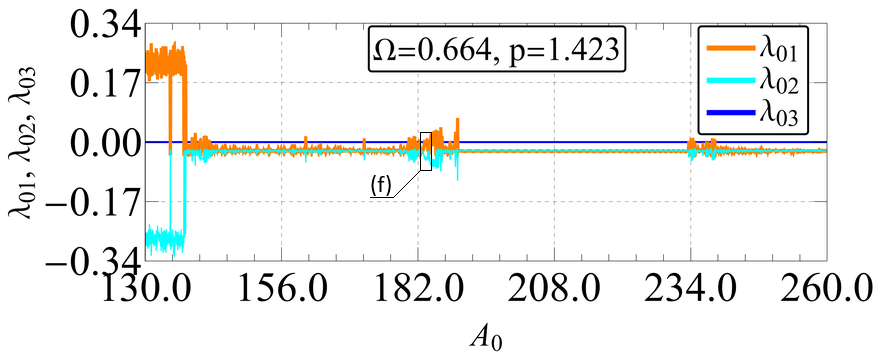}
		\caption{}
		\label{fig:lyap_Omega0664_p1423}
	\end{subfigure}
	\begin{subfigure}{0.49\textwidth}
		\includegraphics[scale=1]{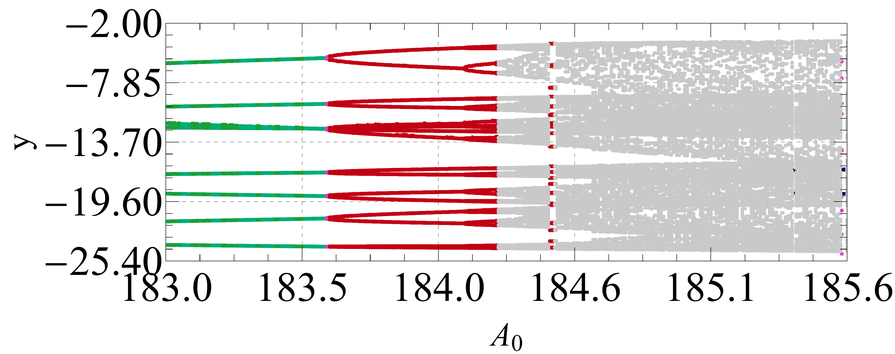}
		\caption{}
	\end{subfigure}
	\begin{subfigure}{0.49\textwidth}
		\includegraphics[scale=1]{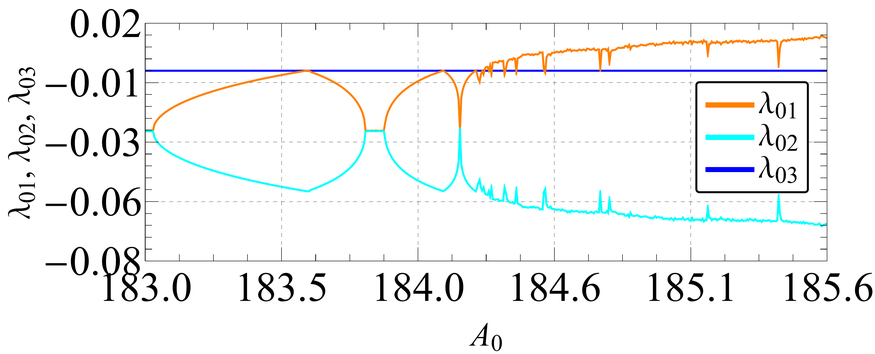}
		\caption{}
	\end{subfigure}
	\caption{Bifurcation diagrams and corresponding Lyapunov spectra for $\Omega=0.664$ and $A_0\in(130,260)$. Panels (a) and (b) show the bifurcation diagrams $y(A_0)$, whereas panels (c) and (d) show the corresponding Lyapunov spectra $(\lambda_{01},\lambda_{02},\lambda_{03})$. The left column corresponds to the classical case $p=0$, while the right column corresponds to the phase-controlled case $p=1.423$. Panels (e) and (f) present detailed view of the transition to chaos for  the phase-controlled system from (b) and (d). Colors in the bifurcation diagrams denote the periodicity of the response, while gray regions indicate chaotic dynamics. Positive values of the largest Lyapunov exponent $\lambda_{01}$ correspond to chaotic motion, whereas non-positive values indicate regular motion.} \label{fig:phase_control_bifurcation_Omega0664} 
\end{figure}

\textcolor{black}{
	To examine the influence of the phase-control coefficient $p$ in more detail, one-dimensional bifurcation diagrams with respect to the excitation amplitude $A_0$ were constructed for the fixed value $p=1.423$ and  $\Omega=0.664$ (see Fig.~\ref{fig:phase_control_bifurcation_Omega0664}b), selected as a representative vertical section of the regime chart shown in Fig.~\ref{fig:phase_control_charts}a. For comparison, the corresponding bifurcation diagrams were also computed for the classical case $p=0$ (see Fig.~\ref{fig:phase_control_bifurcation_Omega0664}a). The corresponding Lyapunov spectra are shown in Fig.~\ref{fig:phase_control_bifurcation_Omega0664}c,d.}

\textcolor{black}{
	For $\Omega=0.664$, the bifurcation diagram obtained for $p=0$ is dominated by a broad chaotic band over nearly the entire investigated interval $A_0\in(130,260)$, with only a few isolated windows of regular motion. By contrast, the diagram presented in Fig.~\ref{fig:phase_control_bifurcation_Omega0664}b and computed for $p=1.423$ contains several extended periodic branches occupying a substantial part of the parameter range, followed by a sequence of bifurcations and transitions to irregular dynamics at larger values of $A_0$. The corresponding Lyapunov spectra confirm this difference.	For $p=1.423$, $92.43\%$ of the investigated parameter values are characterised by a non-positive largest Lyapunov exponent, compared with only $6.27\%$ for $p=0$.}
\textcolor{black}{
	 The enlarged bifurcation structure presented in Fig.~\ref{fig:phase_control_bifurcation_Omega0664}e suggests that, for this parameter section, the onset of chaos is preceded by a cascade of period-doubling bifurcations. The Lyapunov exponent plot in Fig.~\ref{fig:phase_control_bifurcation_Omega0664}f provides quantitative support for this observation by identifying the parameter interval in which the irregular response becomes chaotic.}

\textcolor{black}{
 Supplementary results obtained for $\Omega=0.700$ and $\Omega=0.750$ are provided in Appendix~\ref{app:additional_frequencies}. These additional cases confirm that the principal effect of the phase-dependent excitation is not restricted to $\Omega=0.664$. The additional calculations show a qualitatively similar stabilising influence of the phase-dependent excitation, although the detailed organisation of the periodic and chaotic intervals depends on the excitation frequency.}

\begin{figure}[h!]
	\begin{subfigure}{0.32\textwidth}
		\includegraphics[scale=0.8]{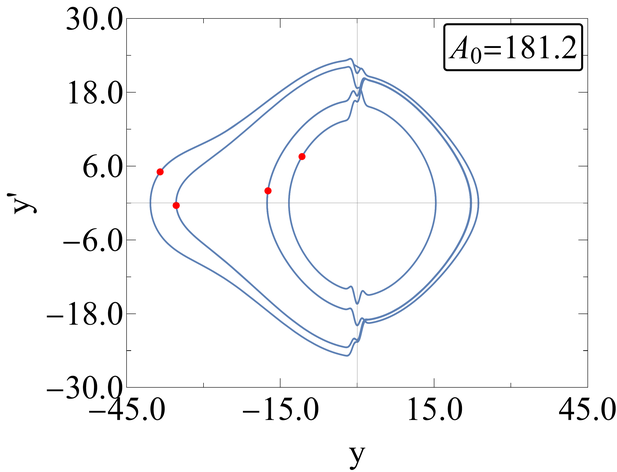} 
		\caption{} \label{}
	\end{subfigure}
	\begin{subfigure}{0.32\textwidth}
		\includegraphics[scale=0.8]{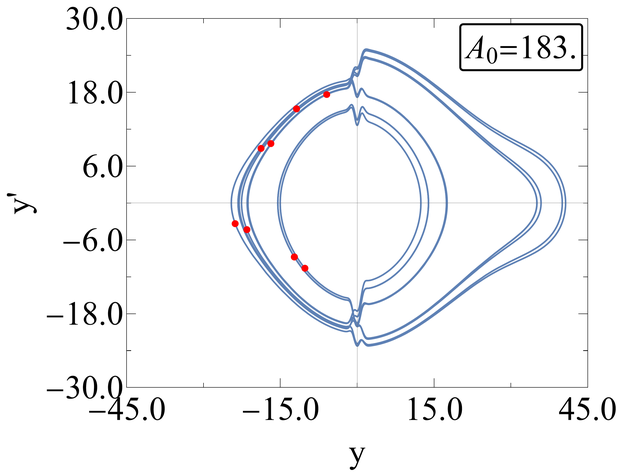} 
		\caption{} \label{}
	\end{subfigure}
	\begin{subfigure}{0.32\textwidth}
		\includegraphics[scale=0.8]{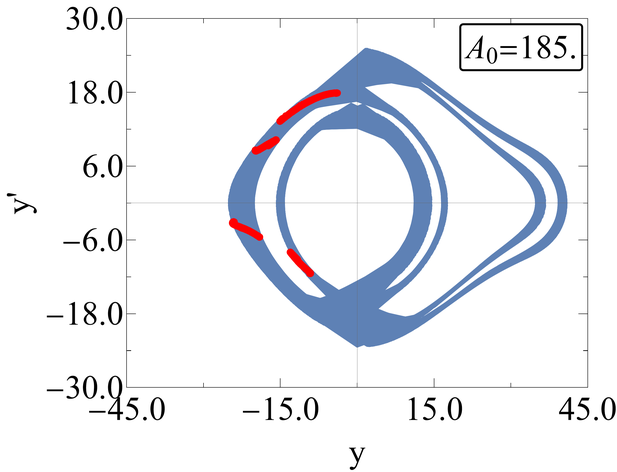} 
		\caption{} \label{}
	\end{subfigure}
	\begin{subfigure}{0.32\textwidth}
		\includegraphics[scale=0.75]{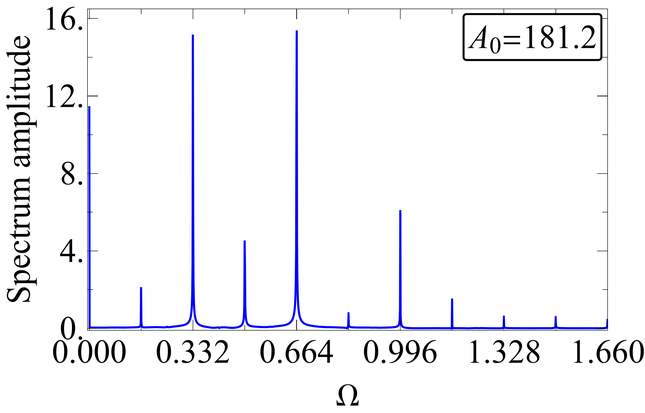} 
		\caption{} \label{}
	\end{subfigure}
	\begin{subfigure}{0.33\textwidth}
		\includegraphics[scale=0.75]{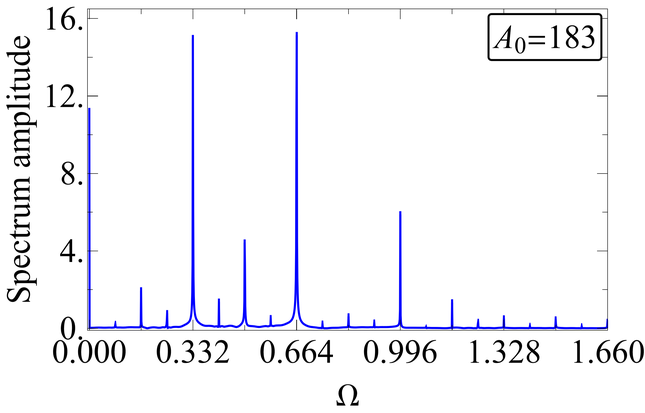} 
		\caption{} \label{}
	\end{subfigure}
	\begin{subfigure}{0.33\textwidth}
		\includegraphics[scale=0.75]{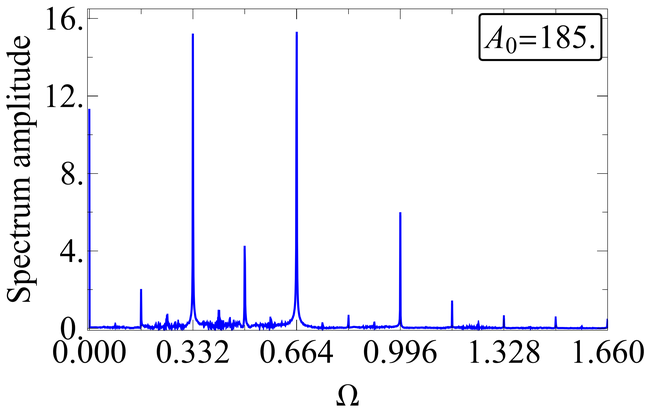} 
		\caption{} \label{}
	\end{subfigure}
	\caption{Representative phase trajectories with the corresponding Poincar\'e sections (a--c) and Fourier spectra (d--f) for the phase-controlled system with \(\Omega=0.664\) and \(p=1.423\), computed for three selected values of the excitation amplitude: \(A_0=181.2\), \(183\), and \(185\). Panels (a,d) correspond to a regular low-period response, panels (b,e) illustrate a higher-period oscillation, and panels (c,f) show a chaotic regime. In the phase portraits, red points denote the Poincar\'e sections. In the chaotic case, the corresponding Lyapunov spectrum is \(\{\lambda_{01},\lambda_{02},\lambda_{03}\}=\{0.0117,\,0.0000,\,-0.0623\}\).}
	\label{fig:phasePlot_PS_control_phase}
\end{figure}

To illustrate the changes in attractor geometry associated with the bifurcation sequence shown in Fig.~\ref{fig:phase_control_bifurcation_Omega0664}e, representative phase trajectories together with the corresponding Poincar\'e sections were computed for \(\Omega=0.664\), \(p=1.423\), and three selected values of the excitation amplitude, namely \(A_0=181.2\), \(183\), and \(185\). The results are shown in Fig.~\ref{fig:phasePlot_PS_control_phase}a--c, while the corresponding Fourier spectra are presented in Fig.~\ref{fig:phasePlot_PS_control_phase}d--f.
For \(A_0=181.2\), the phase trajectory forms a regular closed structure with a finite set of Poincar\'e points, corresponding to a low-period oscillatory response. For \(A_0=183\), the attractor becomes more complex and the number of Poincar\'e points increases, indicating a higher-period oscillation. For \(A_0=185\), the Poincar\'e points form a dense set distributed along the attractor, which is consistent with chaotic motion.
The Fourier spectra are in agreement with these observations. For \(A_0=181.2\) and \(A_0=183\), the spectra consist of discrete peaks associated with the excitation frequency \(\Omega=0.664\), its harmonics, and additional subharmonic components. In the chaotic case \(A_0=185\), the dominant peaks remain visible, but they are superimposed on a broadband spectral background, which is characteristic of irregular motion. Thus, the phase portraits, Poincar\'e sections, and Fourier spectra consistently illustrate the progression from regular oscillations to chaos in the analyzed parameter interval.
In the online version of the paper, this transition is additionally illustrated by GIF~1, which shows the evolution of the phase trajectory and the corresponding Poincar\'e section as the parameter \(A_0\) varies within the bifurcation interval.

\subsection{Intrawell period-1 oscillations}
An interesting feature of the phase-controlled system is the presence of a pronounced period-1 (green color) branch, visible in the bifurcation diagrams for \(p=1.423\)  (see Fig.~\ref{fig:phase_control_bifurcation_Omega0664}b) and all three considered excitation frequencies (see Appendix~\ref{app:additional_frequencies}). For \(\Omega=0.664\), this branch extends over a wide interval of the control parameter \(A_0\) and remains nearly horizontal in the bifurcation diagram, indicating only weak variation of the corresponding oscillatory response with increasing excitation amplitude.

\begin{figure}[ht]
	\centering
	\begin{subfigure}{0.55\textwidth}
		\includegraphics[scale=0.33]{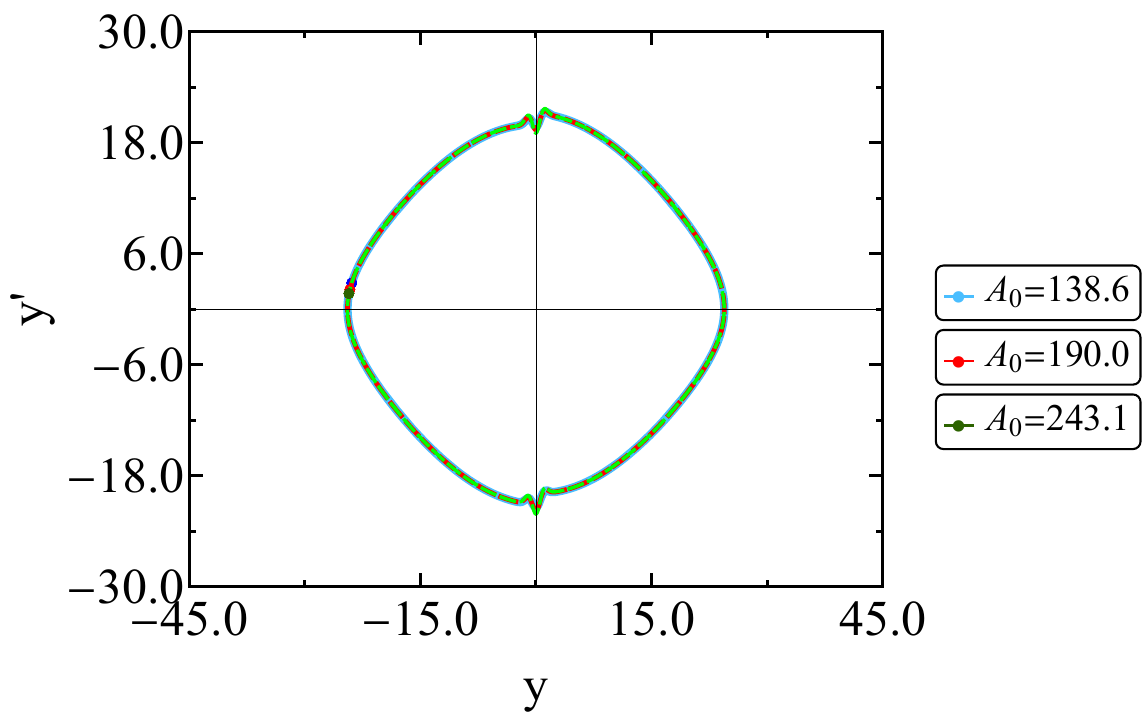}
		\includegraphics[scale=0.33]{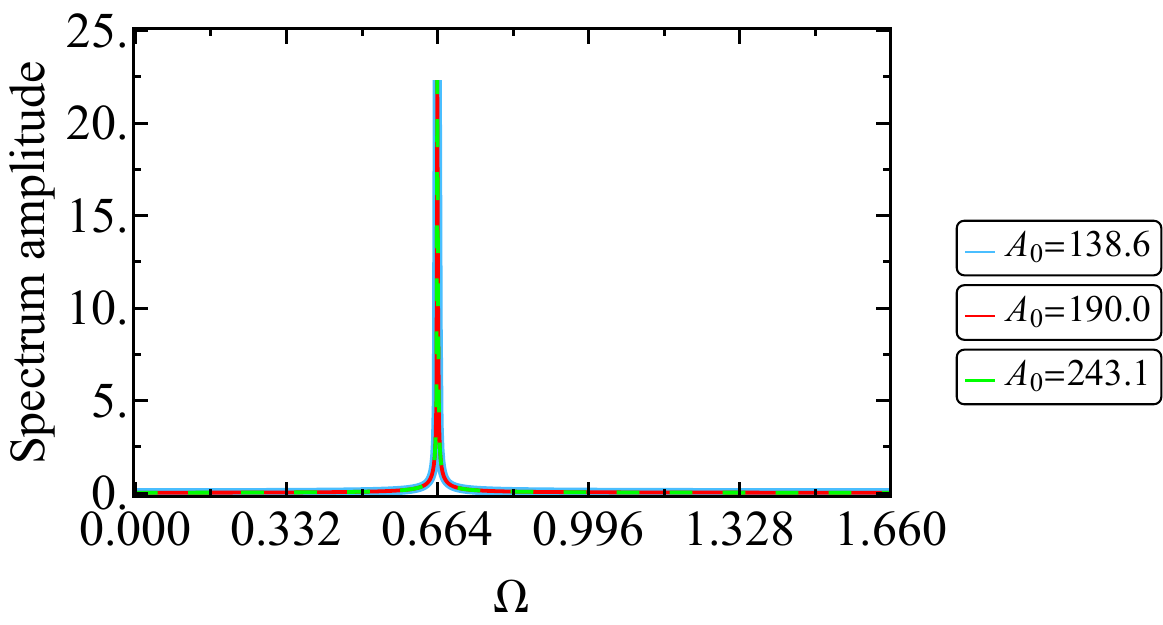}
		\caption{} \label{}
	\end{subfigure}
	\begin{subfigure}{0.4\textwidth}	
		\includegraphics[scale=0.33]{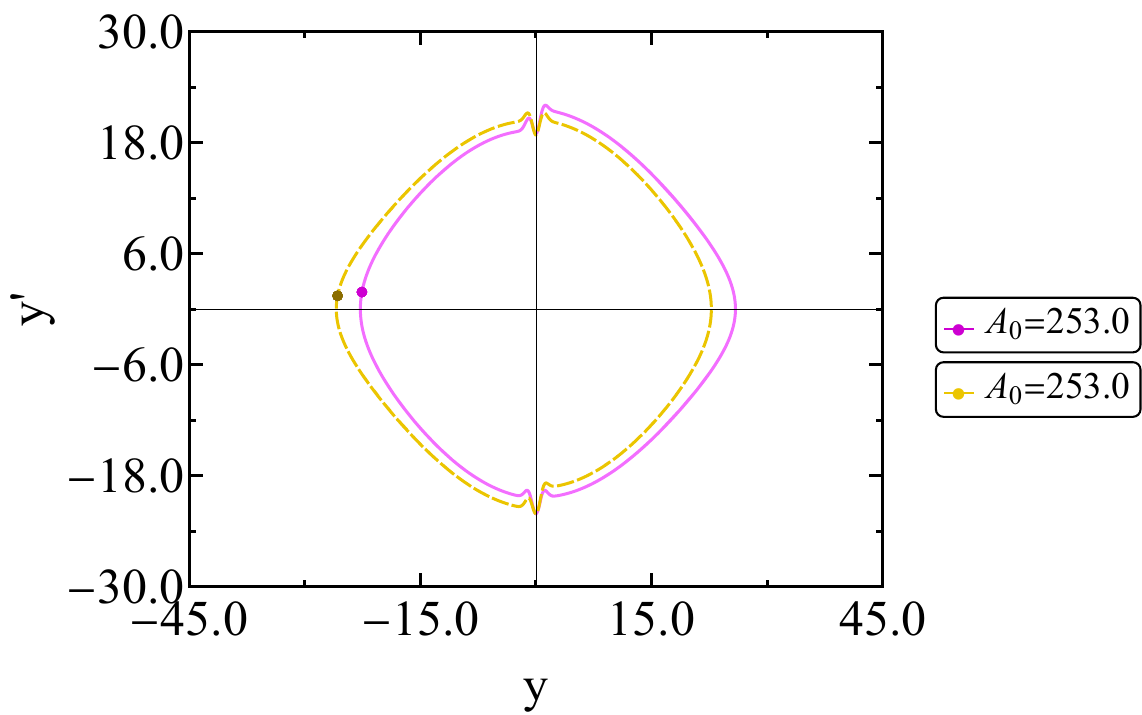} 
		\includegraphics[scale=0.33]{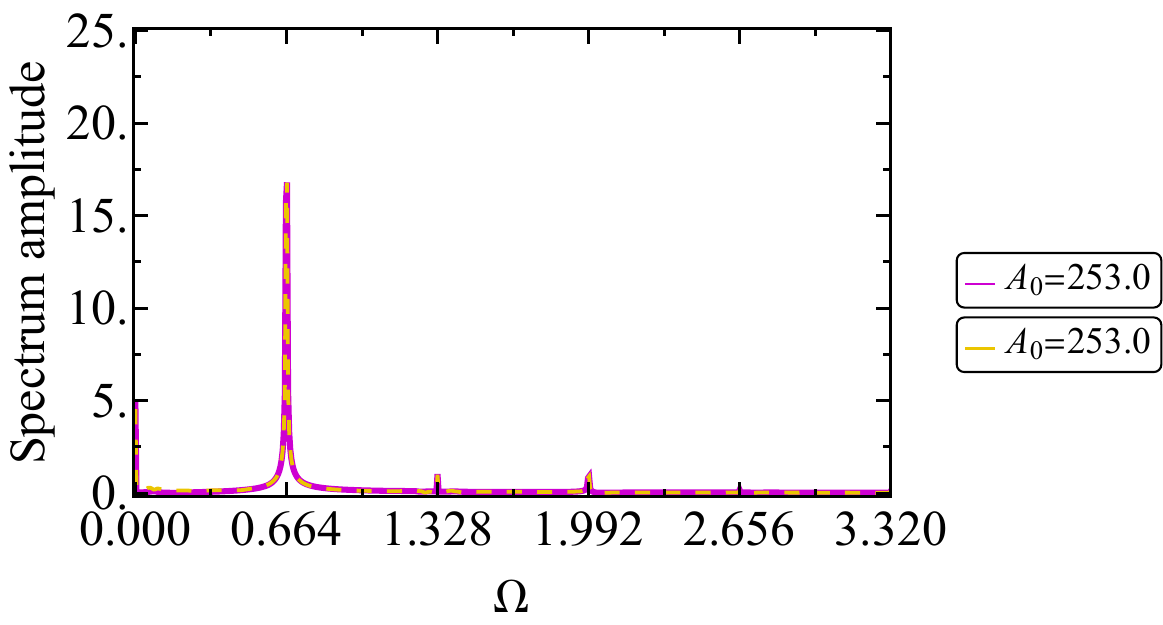} 
		\caption{} \label{}
	\end{subfigure}
	\caption{Phase plot and Fourier spectra of period-1 oscillations obtained for (a) set of three different $A_0=\left\lbrace 138.6, 190.0, 243.1\right\rbrace $ and (b) $A_0=253.0$ but different initial conditions (magenta -- \(\left\lbrace y,y'\right\rbrace =\left\lbrace -24.520,1.887\right\rbrace \), yellow -- \(\left\lbrace y,y'\right\rbrace =\left\lbrace -22.520,1.000\right\rbrace \)).}  
	\label{fig:1T_phaseplot_FFTplot}
\end{figure}

To illustrate this behavior, representative phase trajectories together with their Fourier spectra are shown in Fig.~\ref{fig:1T_phaseplot_FFTplot}a for three substantially different values of the excitation amplitude, namely \(A_0=138.6\), \(190.0\), and \(243.1\). In all three cases, the phase trajectories are qualitatively almost identical and differ only slightly in the location of the Poincar\'e point, which moves closer to the line \(y'=0\) as \(A_0\) increases. The corresponding Fourier spectra are also nearly indistinguishable and are dominated by a single main peak at the excitation frequency \(\Omega=0.664\).

A qualitative change is observed when \(A_0\) exceeds approximately 247. In this range, the original period-1 branch splits into two distinct period-1 attractors, which depend on the initial conditions. This is illustrated in Fig.~\ref{fig:1T_phaseplot_FFTplot}b for \(A_0=253.0\), where two different phase trajectories are obtained for the same parameter values but slightly different initial conditions. The corresponding Fourier spectra show the same fundamental frequency content, together with additional harmonics, and remain very similar for both solutions.
\textcolor{black}{Those two responses represent two distinct coexisting period-1 attractors. Although their phase-space trajectories exhibit very similar and approximately centrally reflected global geometries, they differ locally in the central magnetic-interaction region. Moreover, their stroboscopic Poincar\'e points, evaluated at the same excitation phase, are not related by the transformation $(y,y')\mapsto(-y,-y')$. Therefore, these responses are interpreted as distinct coexisting periodic attractors rather than as an exact symmetry-related pair or direct evidence of a symmetry-breaking bifurcation.}

\begin{figure}[ht]
	\centering
	\includegraphics[scale=0.4]{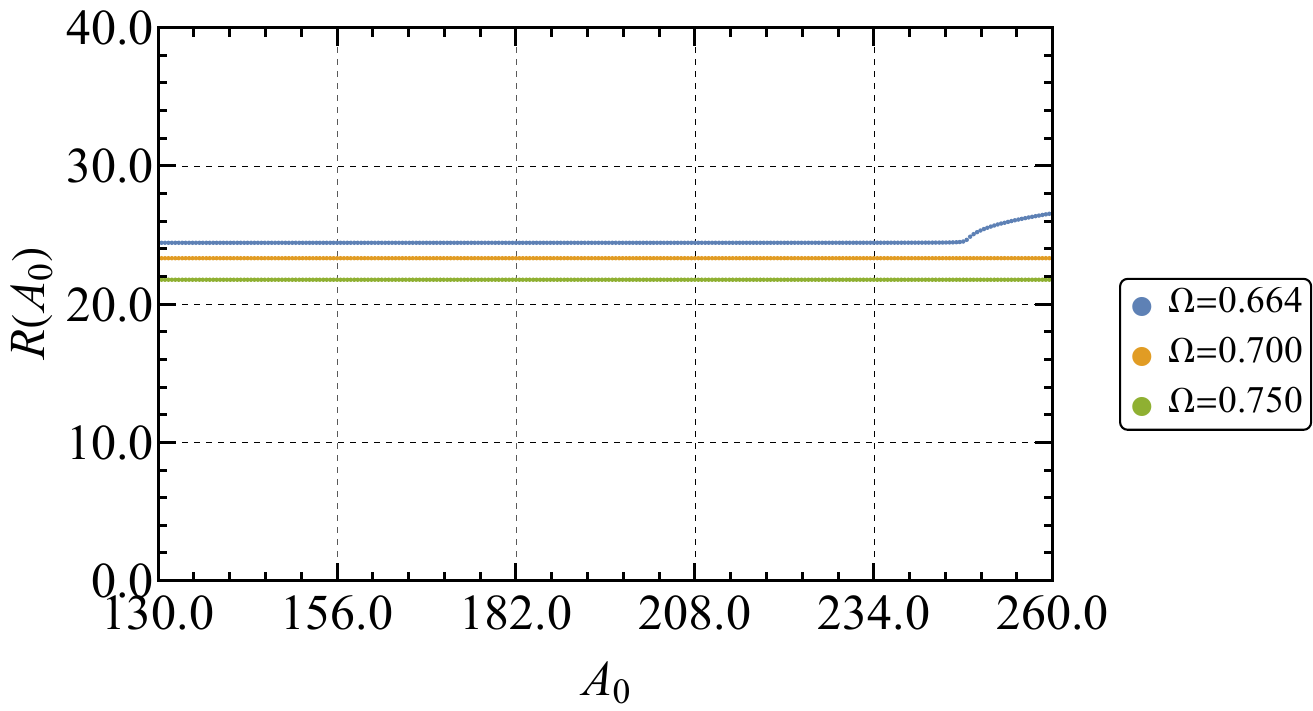}
	\caption{Maximum phase-space radius \(R(A_0)\) of the intrawell period-1 orbit as a function of the excitation amplitude \(A_0\) for the phase-controlled system with \(p=1.423\) and for three selected excitation frequencies: \(\Omega=0.664\), \(0.700\), and \(0.750\). The nearly flat character of the curves indicates that the size of the period-1 orbit depends only weakly on the forcing amplitude over the investigated parameter interval.}
	\label{fig:R_A0_period1}
\end{figure}

Although the detailed phase portraits and Fourier spectra are shown here only for \(\Omega=0.664\), the existence of qualitatively similar intrawell period-1 responses for \(\Omega=0.700\) and \(\Omega=0.750\) was confirmed using a more compact quantitative measure. To this end, the weak variation of the intrawell period-1 branch was characterized by the maximum phase-space radius
\begin{equation}
	R(A_0)=\max_{x\in[x_0,x_0+T]} \sqrt{y(x)^2+\left(y'(x)\right)^2},
	\label{eq:R_A0}
\end{equation}
evaluated for the asymptotic periodic response, where \(T\) denotes one period of the steady-state oscillation after the transient has decayed.
Figure~\ref{fig:R_A0_period1} shows the dependence of \(R(A_0)\) on the excitation amplitude for \(p=1.423\) and for the three selected values of \(\Omega\). In all cases, the curves remain nearly horizontal over the investigated range of \(A_0\), confirming that the size of the corresponding period-1 orbit is only weakly affected by changes in the forcing amplitude. A clear frequency dependence is nevertheless visible, with smaller excitation frequencies corresponding to slightly larger phase-space radii.

\begin{figure}[ht]
	\centering
	\begin{subfigure}{0.49\textwidth}
		\includegraphics[scale=0.31]{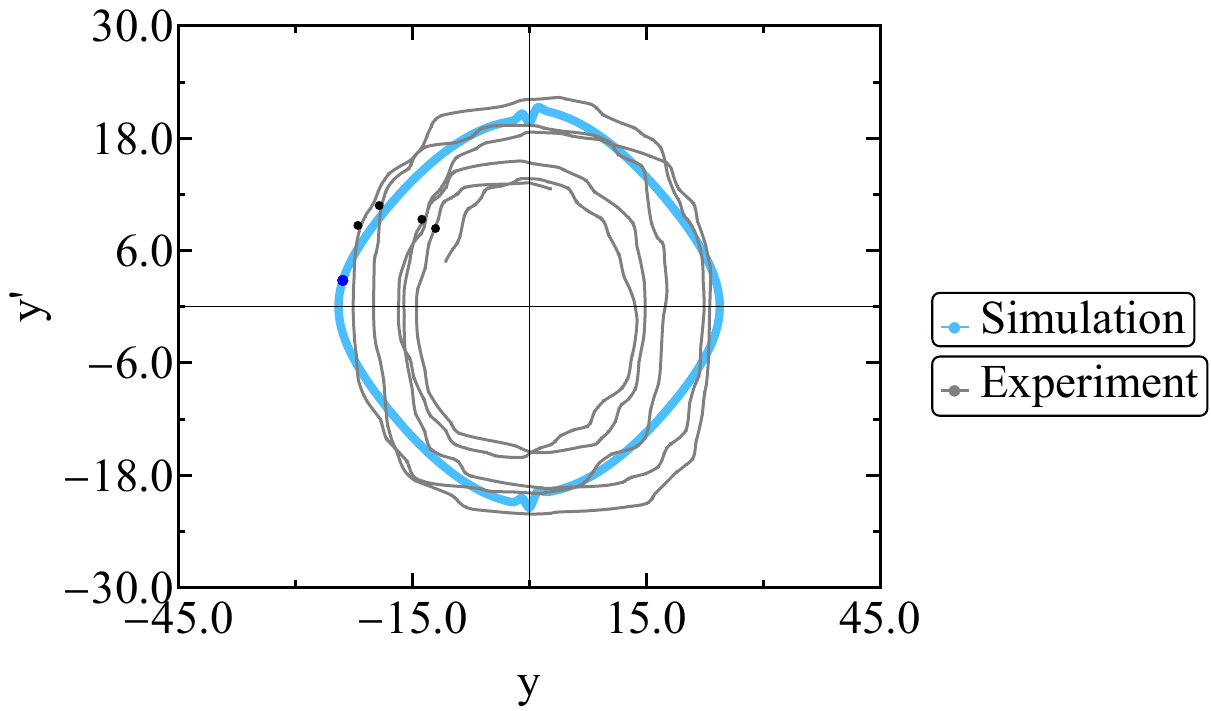}
		\caption{} \label{}
	\end{subfigure}
	\begin{subfigure}{0.35\textwidth}	
		\includegraphics[scale=0.31]{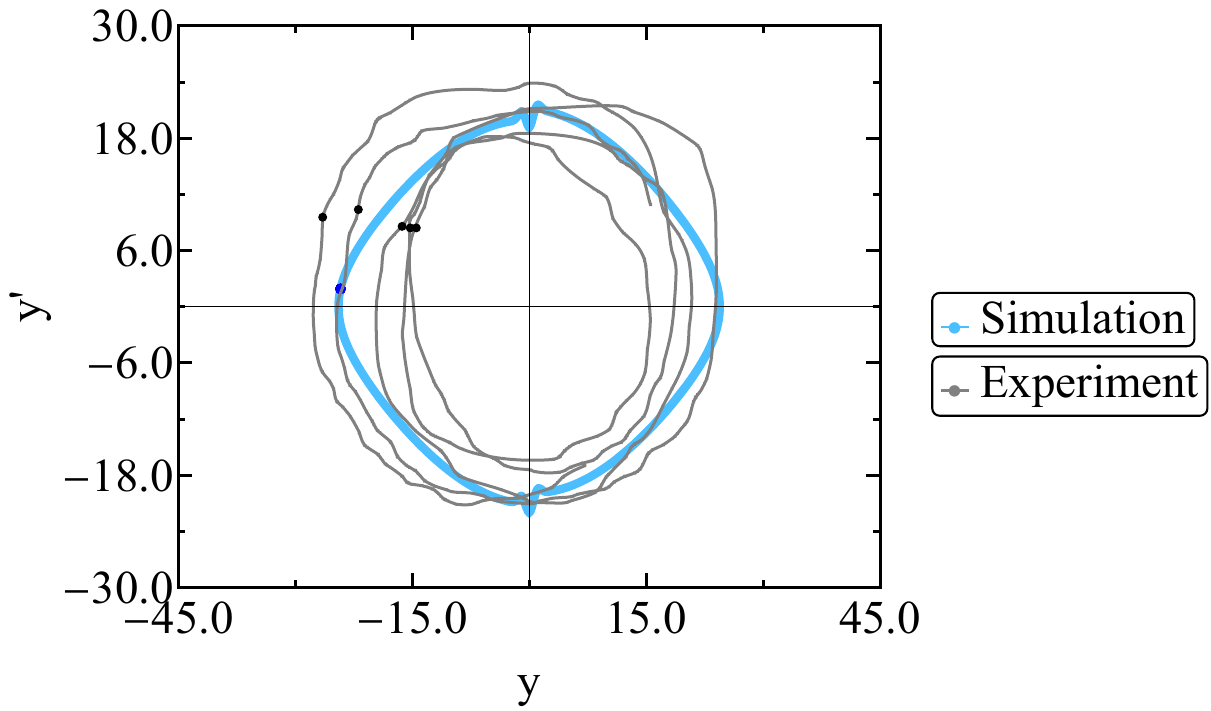} 
		\caption{} \label{}
	\end{subfigure}
	\caption{Comparison of numerical and experimental phase trajectories for the intrawell period-1 response of the phase-controlled system at \(\Omega=0.664\) and \(p=1.423\): (a)~\(A_0=138.6\), (b)~\(A_0=208.65\). The blue curves denote numerical trajectories, the gray curves denote experimental trajectories, and the corresponding Poincar\'e points are marked by dots.}  
	\label{fig:1T_phaseplot_exp}
\end{figure}

To verify that the numerically identified intrawell period-1 response is also observable experimentally, selected simulation and experimental phase trajectories were compared for \(\Omega=0.664\), \(p=1.423\), and two representative values of the excitation amplitude. The results are shown in Fig.~\ref{fig:1T_phaseplot_exp}. In both cases, the experimental phase trajectories occupy approximately the same region of phase space as their numerical counterparts and exhibit a similar closed-loop structure characteristic of period-1 motion.
The corresponding experimental Poincar\'e points form compact clusters rather than a single point, indicating a small spread of the measured response around the numerically predicted orbit. Although the quantitative overlap between simulation and experiment is not exact, the overall size, shape, and location of the trajectories remain in good qualitative agreement. This confirms that the period-1 branch identified numerically corresponds to a physically realizable oscillatory regime of the phase-controlled system.

\begin{figure}[ht]
	\centering
	\begin{subfigure}{0.45\textwidth}
		\includegraphics[scale=0.8]{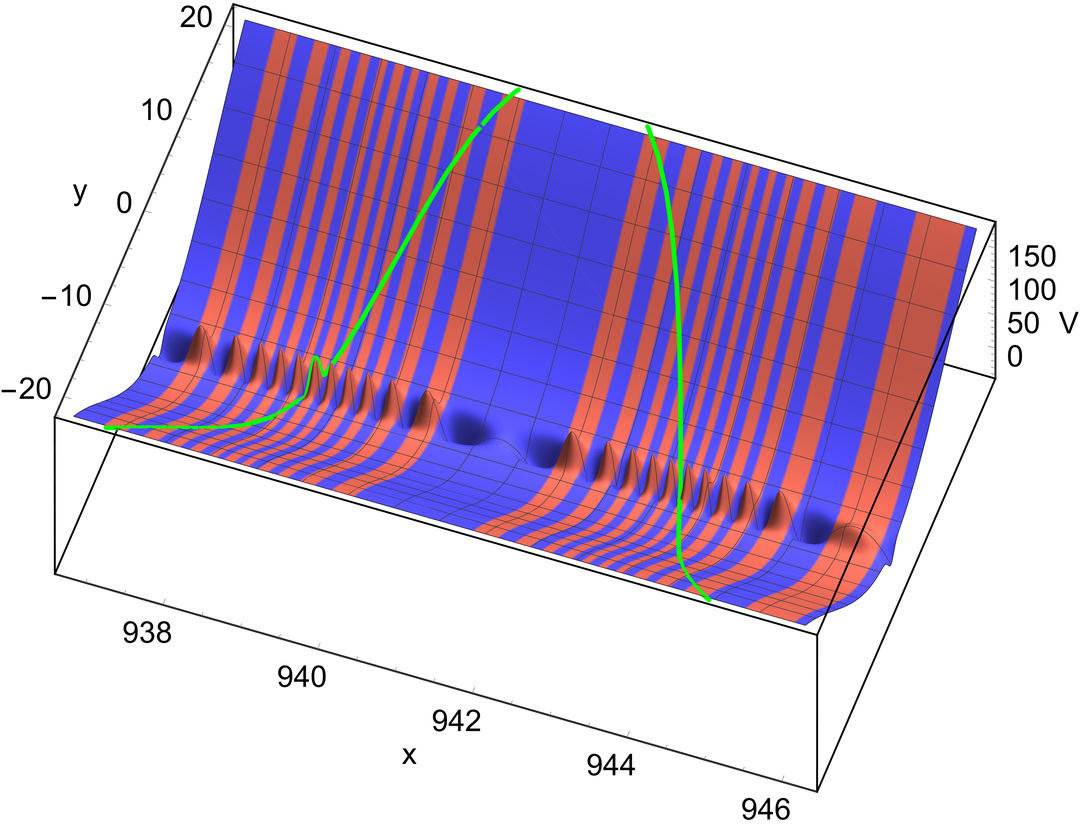} 
		\caption{} \label{}
	\end{subfigure}
	\begin{subfigure}{0.5\textwidth}
		\includegraphics[scale=0.8]{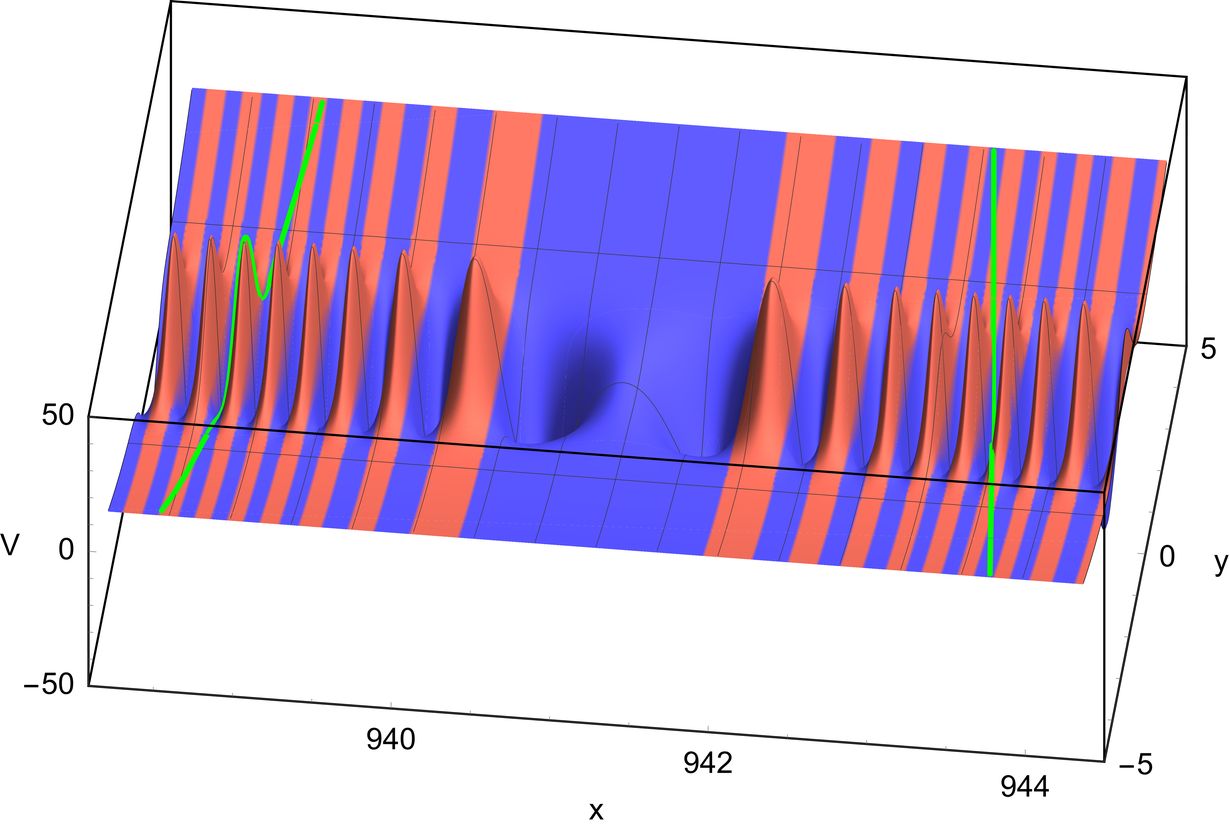}
		\caption{} \label{}
	\end{subfigure}
	\caption{Three-dimensional representation of the time-dependent effective potential \(V(x,y)\) and the corresponding pendulum trajectory for the period-1 response shown previously in Fig.~\ref{fig:1T_phaseplot_FFTplot}a: (a)~global view and (b)~zoomed view. The pendulum trajectory is marked by the green curve. Red and blue regions correspond to positive and negative coil current, respectively, i.e. to repelling and attracting magnetic interaction.}
	\label{fig:potential_FFT}
\end{figure}

To provide an intuitive interpretation of the observed motion, a trajectory-conditioned effective potential was constructed for visualization purposes. Instead of introducing an exact scalar potential for the full state-dependent forcing term, the phase shift was evaluated along the numerically computed solution \(y_s(x)\), yielding the following effective time-dependent potential landscape:
\begin{equation}
	V_{\mathrm{eff}}(x,y)=
	\frac{\alpha}{2}y^2
	+\gamma^2\left(1-\cos\left(\frac{y}{\gamma}\right)\right)
	+\frac{A_0}{2}\exp(-y^2)\sin\bigl(\Omega x + p\,y_s(x)\bigr),
	\label{eq:effective_potential_visual}
\end{equation}
where \(y\) is treated as an independent coordinate for visualization, while \(y_s(x)\) denotes the numerically computed trajectory of the pendulum. In this representation, the first two terms describe the gravitational and elastic restoring contributions, whereas the last term represents the magnetic modification of the potential profile along the actual phase-controlled motion.
The three-dimensional plots shown in Fig.~\ref{fig:potential_FFT} were generated from Eq.~(\ref{eq:effective_potential_visual}) and illustrate how the pendulum trajectory evolves in a time-varying effective potential landscape. In particular, the plots show that the magnetic interaction periodically reshapes the local wells and barriers of the potential, so that the pendulum alternates between motion toward an attracting well and motion away from a repelling ridge.

\subsection{Basins of attraction of the intrawell period-1 branch}

\begin{figure}[ht]
	\begin{subfigure}{0.32\textwidth}
		\includegraphics[scale=0.27]{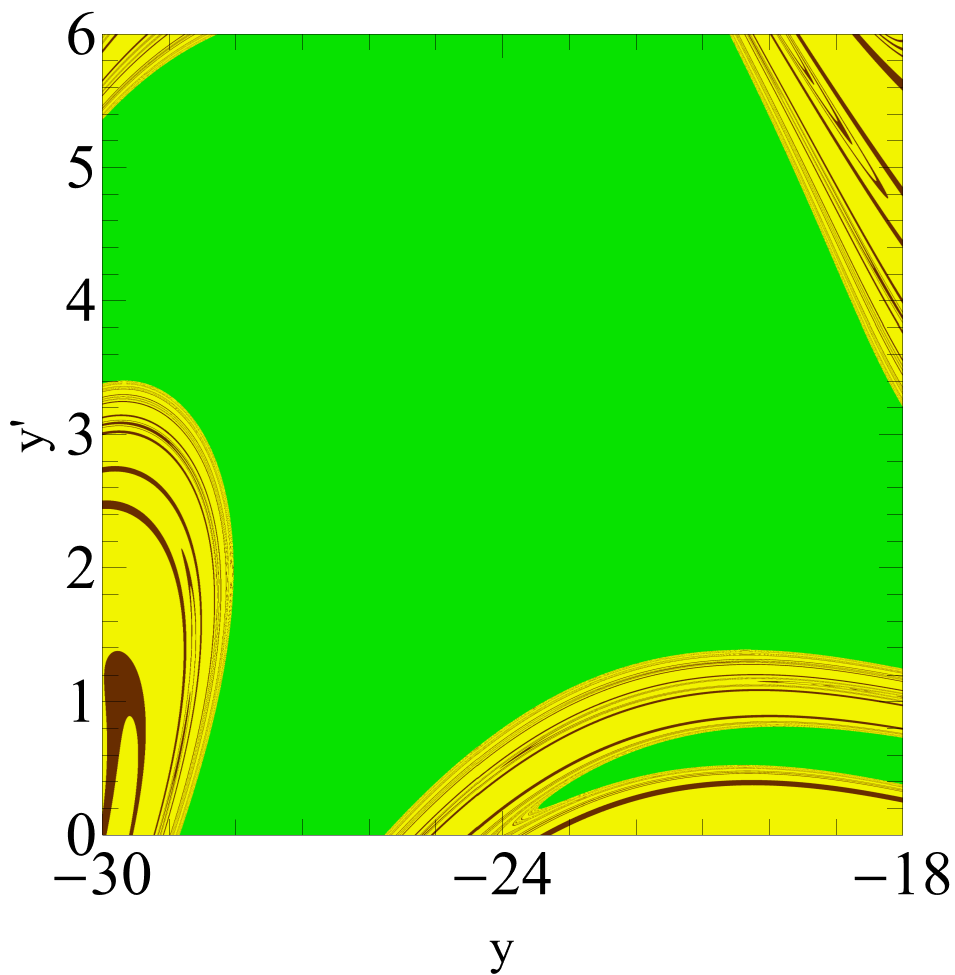}
		\caption{} \label{}
	\end{subfigure}
	\begin{subfigure}{0.32\textwidth}
		\includegraphics[scale=0.27]{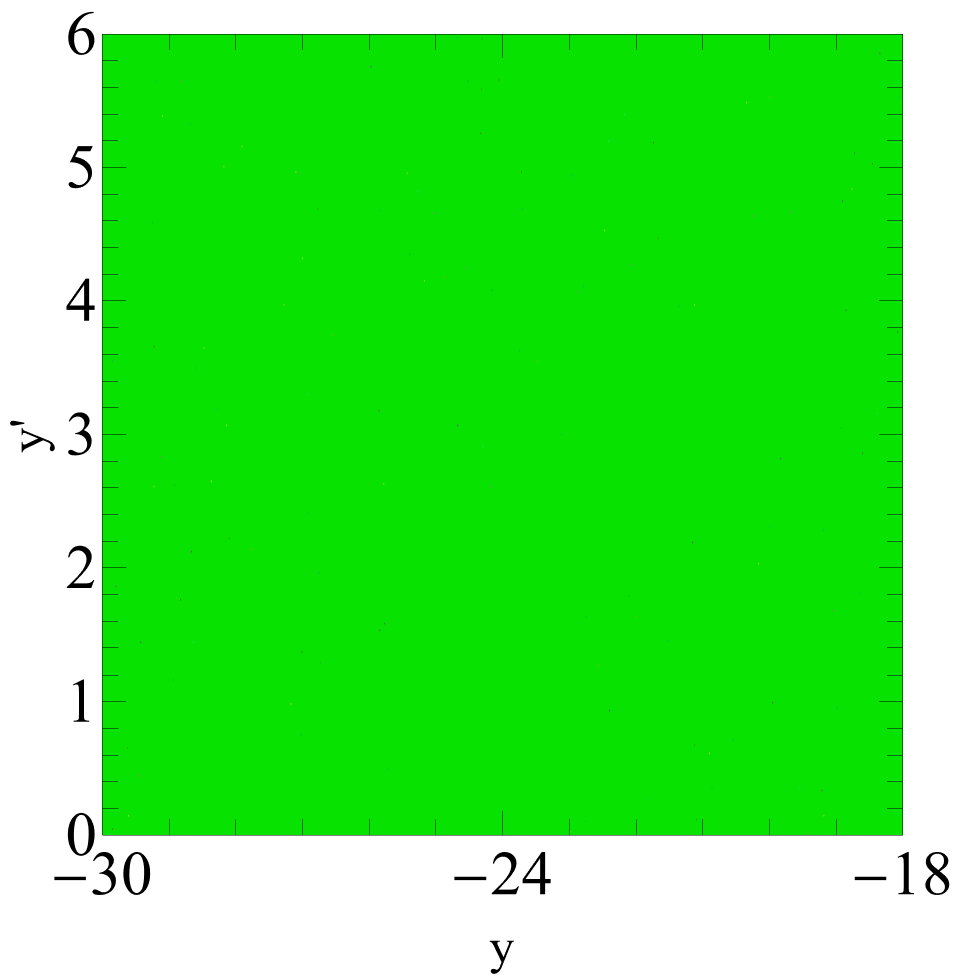}
		\caption{} \label{}
	\end{subfigure}
	\begin{subfigure}{0.32\textwidth}	
		\includegraphics[scale=0.27]{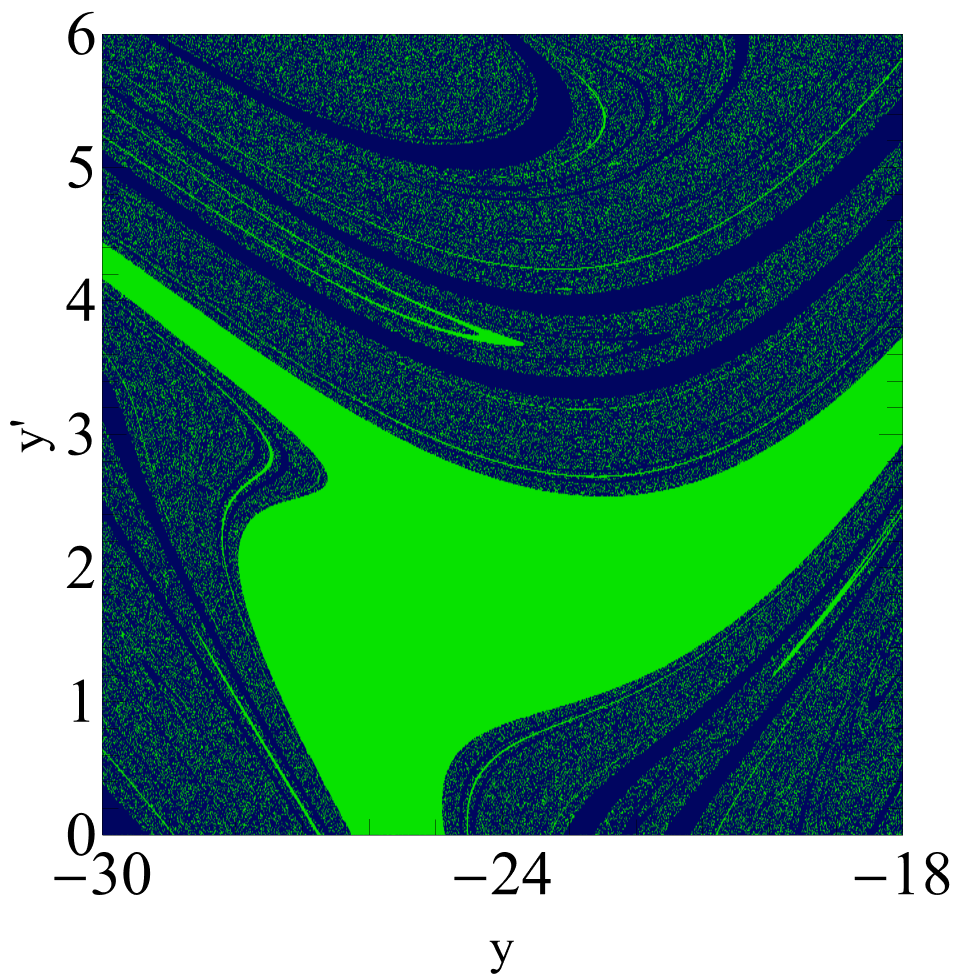} 
		\caption{} \label{}
	\end{subfigure}
	\includegraphics[scale=0.3]{legend.pdf} 
	\caption{Basins of attraction computed in the vicinity of the Poincar\'e point associated with the intrawell period-1 branch for \(\Omega=0.664\) and \(p=1.423\): (a)~\(A_0=138.6\), (b)~\(A_0=210.0\), and (c)~\(A_0=243.1\). The color of each region indicates the periodicity of the attractor reached from the corresponding initial condition, according to the color convention used in Fig.~\ref{fig:phase_control_charts}.}
	\label{fig:1T_phaseplot_basins}
\end{figure}

To examine the local robustness of the intrawell period-1 response, basins of attraction were computed in the vicinity of the corresponding Poincar\'e point for \(\Omega=0.664\), \(p=1.423\), and three selected values of the excitation amplitude \(A_0\). The results are shown in Fig.~\ref{fig:1T_phaseplot_basins}.

In all cases, the initial conditions were chosen from a neighborhood of the Poincar\'e point associated with the period-1 solution.
For \(A_0=138.6\) (Fig.~\ref{fig:1T_phaseplot_basins}a), the basin of the period-1 attractor (green) occupies the largest part of the investigated domain. At the same time, additional regions corresponding to period-2 (yellow) and period-6 (brown) responses are also present, indicating the coexistence of several stable periodic solutions in the considered neighborhood of initial conditions.
For \(A_0=210.0\) (Fig.~\ref{fig:1T_phaseplot_basins}b), the basin of the period-1 attractor becomes dominant over almost the entire investigated region. In this case, nearly all tested initial conditions converge to the same period-1 response, and competing attractors are reduced to a negligible extent.

A different structure is observed for \(A_0=243.1\) (Fig.~\ref{fig:1T_phaseplot_basins}c). Although the period-1 solution remains present, a substantial part of the investigated domain is now occupied by the basin of a period-3 attractor (dark blue). The basin boundaries are considerably more intricate than in the previous cases, and the distribution of colors indicates a stronger dependence of the long-term response on the initial conditions.
Taken together, these basin plots show that the intrawell period-1 branch remains the dominant local attractor for moderate values of \(A_0\), whereas for larger amplitudes its basin shrinks and competes with other periodic solutions, in particular with the period-3 response.

\begin{figure}[ht]
	\centering
	\includegraphics[scale=0.26]{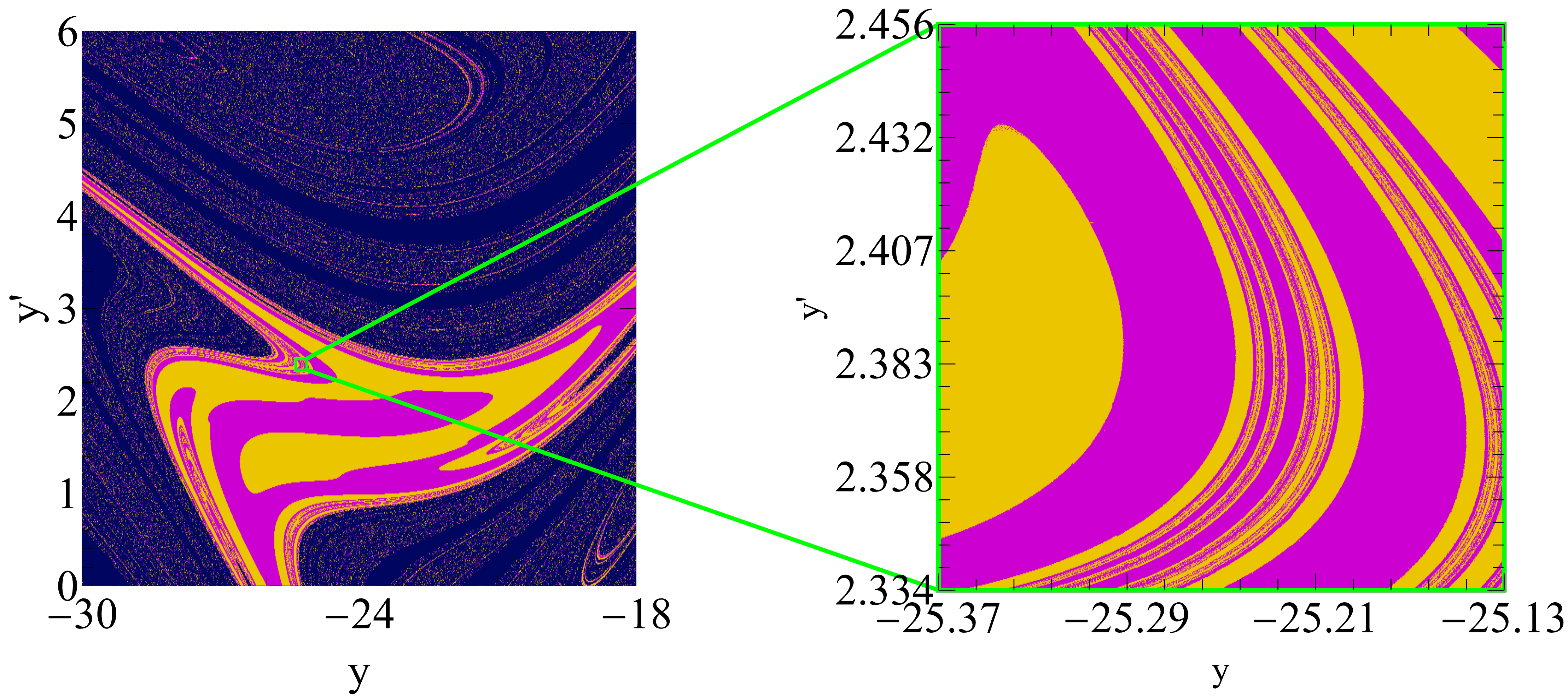}
\hspace{0.5cm}
	\includegraphics[scale=0.285]{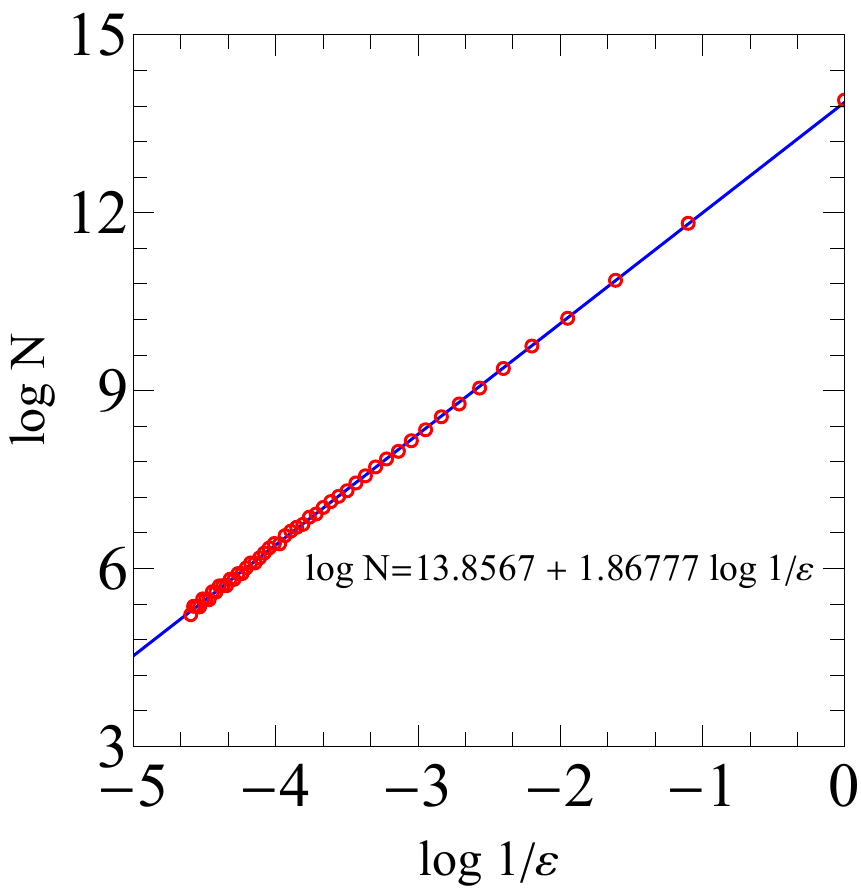} 
	\\
	(a) \hspace{4.5cm} (b) \hspace{5cm} (c) 
\caption{Basins of attraction for \(\Omega=0.664\), \(p=1.423\), and \(A_0=253.0\): (a)~full basin structure in the neighborhood of the Poincar\'e points corresponding to the two coexisting period-1 solutions shown in Fig.~\ref{fig:1T_phaseplot_FFTplot}b, (b)~magnified view of the region marked in panel (a), and (c)~log-log plot used to estimate the box-counting dimension of the basin boundary. Dark blue corresponds to the period-3 attractor, while magenta and yellow correspond to the two distinct period-1 attractors.}
	\label{fig:1T_phaseplot_basins_fractalNATURE}
\end{figure}

To examine the coexistence structure in more detail, basins of attraction were computed for the parameter set \(\Omega=0.664\), \(p=1.423\), and \(A_0=253.0\), for which two distinct period-1 solutions were previously identified in Fig.~\ref{fig:1T_phaseplot_FFTplot}b. The resulting basin plot is shown in Fig.~\ref{fig:1T_phaseplot_basins_fractalNATURE}a. In addition to the two coexisting period-1 attractors (magenta and yellow), the basin structure also contains regions leading to a period-3 attractor (dark blue), indicating the coexistence of at least three different long-term responses for the same parameter values.

A magnified view of the selected region marked in Fig.~\ref{fig:1T_phaseplot_basins_fractalNATURE}a is shown in Fig.~\ref{fig:1T_phaseplot_basins_fractalNATURE}b. The zoom reveals a strongly interwoven arrangement of the magenta and yellow basins, with thin alternating bands and highly irregular interfaces between neighboring attractor regions.
To quantify this boundary complexity, the box-counting dimension was evaluated for the extracted basin-boundary set shown in Fig.~\ref{fig:1T_phaseplot_basins_fractalNATURE}b. The box-counting dimension \(FD\) is defined as
\begin{equation}
	FD=\lim_{\varepsilon\to 0}\frac{\log N(\varepsilon)}{\log(1/\varepsilon)},
\end{equation}
where \(N(\varepsilon)\) denotes the number of boxes of size \(\varepsilon\) required to cover the analyzed set. In practice, \(FD\) was estimated from the slope of the linear fit in the \(\log N\) versus \(\log(1/\varepsilon)\) diagram shown in Fig.~\ref{fig:1T_phaseplot_basins_fractalNATURE}c. The obtained value, \[FD \approx 1.87,\]
is consistent with a fractal basin boundary in the considered region of the phase space.

\begin{figure}[ht]
	\centering
	\includegraphics[scale=0.3]{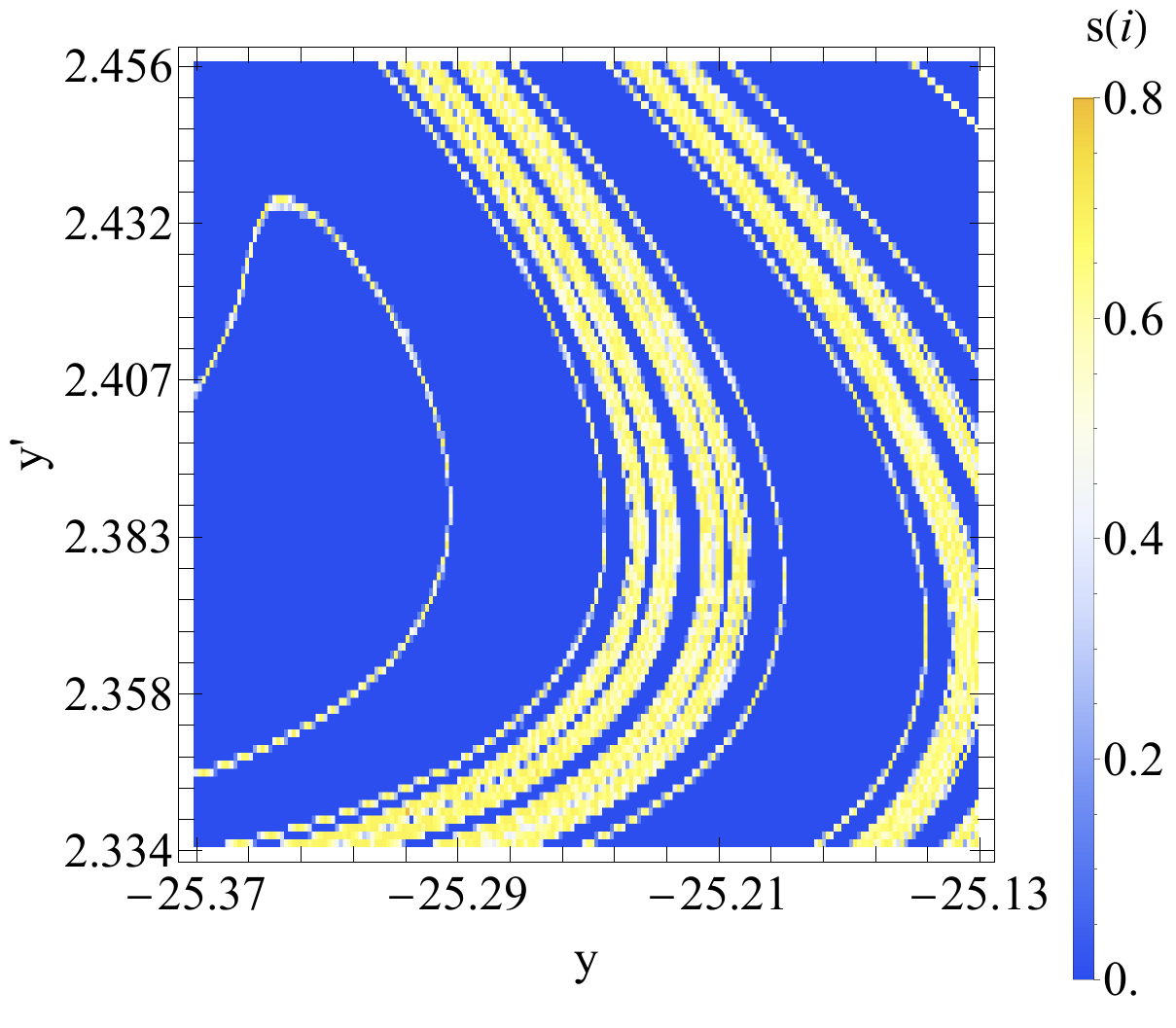}
	\caption{Distribution of the local box entropy \(s(i)\) for the zoomed two-basin region shown in Fig.~\ref{fig:1T_phaseplot_basins_fractalNATURE}b. Blue regions correspond to boxes containing only one basin, whereas higher values of \(s(i)\) (light and yellow tones) indicate strong local mixing of the two competing period-1 basins along the irregular boundary.}
	\label{fig:local_entropy_distribution}
\end{figure}

As an additional verification, the fractality of the two-basin boundary was analyzed using the \textit{boundary basin entropy test} proposed by A. Puy \cite{Puy2021a}. The boundary basin entropy is defined as
\begin{equation}
	S_{bb}=N_b^{-1}\sum_{i=1}^{N_b}s(i),
	\label{eq:Sbb formula}
\end{equation}
where $N_b$ is the number of disk boxes uniformly and randomly distributed on the basin boundaries, and \(s(i)\) is the local entropy associated with the \(i\)-th disk box,
\begin{equation}
	s(i)=\sum_{k=1}^{M}-p(k)\ln p(k),
	\label{eq:box_entropy}
\end{equation}
with \(M=2\) for the two competing period-1 basins in the zoomed region, term $k$ denotes a certain basin of attraction $(k=1,2)$, and finally $p(k)$ is a discrete probability distribution of having the basin $k$ for each disk box. The numerically obtained value for the considered boundary is
\[
S_{\mathrm{bb}} = 0.4818 \pm 0.0221.
\]
The method assumes that for a theoretically smooth two-basin boundary, the boundary basin entropy takes the value $S_{bb}=0.4395093$, whereas significant deviation from this value indicates a non-smooth, that is, fractal structure. Since this value exceeds the theoretical threshold corresponding to a smooth two-basin boundary, it supports the interpretation of the analyzed interface as fractal. The corresponding distribution of local entropy values \(s(i)\) is shown in Fig.~\ref{fig:local_entropy_distribution}, where the largest values are concentrated along the interwoven basin boundary. The local entropy map highlights the regions where the two basins are most strongly mixed; high values of \(s(i)\) are concentrated along the irregular boundary, whereas regions occupied by only one attractor are characterized by \(s(i)=0\).

\subsection{Energy transfer mechanism of the intrawell period-1 branch}
To clarify why the intrawell period-1 branch remains only weakly dependent on the excitation amplitude \(A_0\), the energy exchange in this regime was analyzed in detail. The magnetic pendulum can be viewed as a mechanical subsystem coupled to an electromagnetic actuation loop through the coil-magnet interaction. Because the excitation phase depends on the instantaneous pendulum position, the coil current signal is state-dependent and the corresponding magnetic torque becomes strongly localized in time.

\begin{figure}[t]
	\centering
	\begin{subfigure}{0.75\textwidth}
		\includegraphics[scale=0.65]{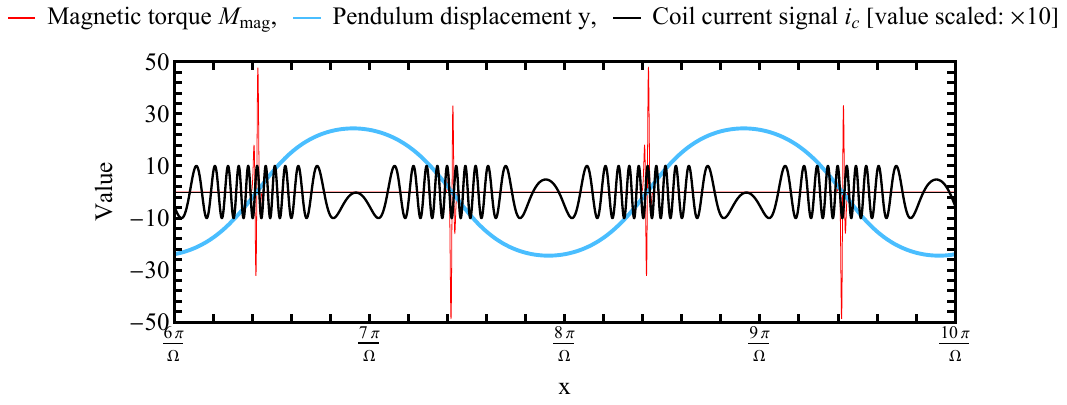}
		\caption{} \label{}
	\end{subfigure}\\
	\begin{subfigure}{0.45\textwidth}
		\includegraphics[scale=0.65]{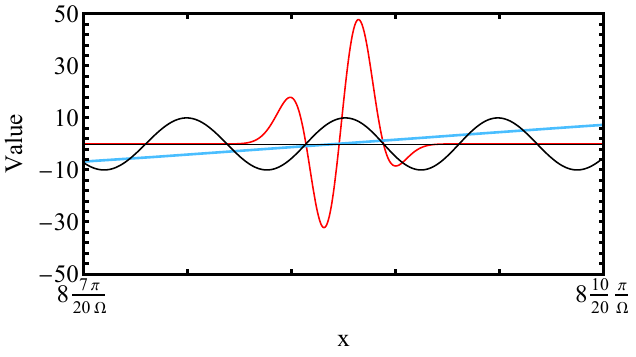}
		\caption{} \label{}
	\end{subfigure}
	\caption{Time histories illustrating the phase-controlled period-1 response: (a)~coil current signal \(i_c\), pendulum displacement \(y\), and magnetic torque \(M_{mag}\); (b)~zoomed view showing the localized torque peaks in relation to the current signal and pendulum motion.}
	\label{fig:FMplot}
\end{figure}

Representative time histories of the period-1 response are shown in Fig.~\ref{fig:FMplot}. The pendulum displacement is close to harmonic, whereas the coil current exhibits phase modulation induced by the feedback term \(py\). As a consequence, the magnetic torque does not act continuously over the whole oscillation cycle, but appears in short pulse-like intervals concentrated near passages through the vicinity of the lower equilibrium position. This localization is shown more clearly in the zoomed view in Fig.~\ref{fig:FMplot}b.
The temporal localization of the magnetic interaction and the corresponding evolution of the pendulum motion are additionally illustrated in Supplementary Material (GIF~2 and GIF~3).

\begin{figure}[t]
	\centering
	\begin{subfigure}{0.29\textwidth}
		\includegraphics[scale=0.63]{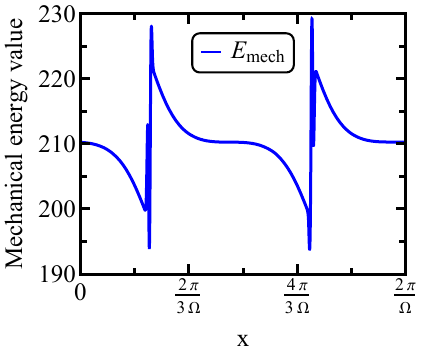}
		\caption{} \label{}
	\end{subfigure}
	\begin{subfigure}{0.29\textwidth}
		\includegraphics[scale=0.63]{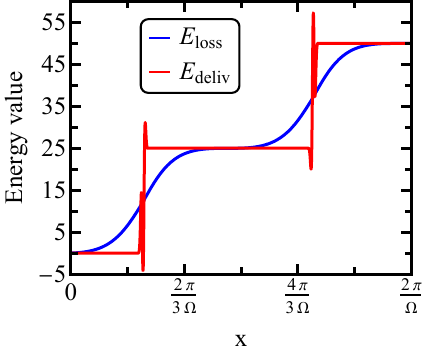}
		\caption{} \label{}
	\end{subfigure}
	\begin{subfigure}{0.4\textwidth}	
		\includegraphics[scale=0.63]{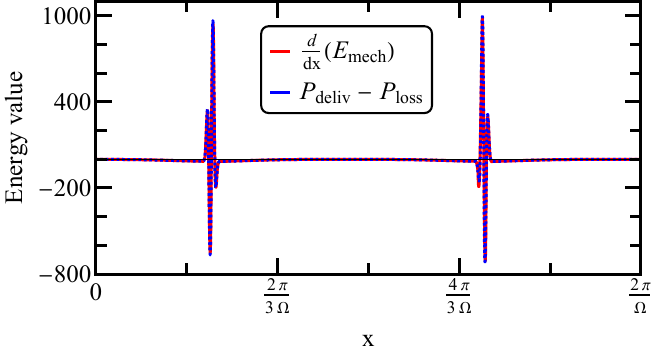} 
		\caption{} \label{}
	\end{subfigure}
	\caption{Energy balance for the representative intrawell period-1 response: (a)~dimensionless mechanical energy \(E_{mech}(x)\); (b)~cumulative energy loss \(E_{loss}(x)\) and cumulative energy delivered \(E_{deliv}(x)\); (c)~comparison between the time derivative of the mechanical energy and the instantaneous power balance \(P_{deliv}(x)-P_{loss}(x)\).}
	\label{fig:Energies}
\end{figure}

The dimensionless mechanical energy of the system is defined as
\begin{equation}
	E_{mech}(x)=T(x)+V(x),
	\label{eq:mechanical_energy_13}
\end{equation}
where
$T(x)=\frac{1}{2}{y'}^2$ is the kinetic energy and $V(x)=\frac{\alpha}{2}y^2+\gamma^2\left(1-\cos\frac{y}{\gamma}\right)$ is the potential energy associated with the conservative restoring part of the system. Figure~\ref{fig:Energies}a shows the time evolution of \(E_{mech}(x)\) for the representative period-1 oscillation. The energy varies almost periodically, but its evolution is characterized by sharp jumps superimposed on slower intervals of decay.

The instantaneous power dissipated in the system is given by
\begin{equation}
	P_{loss}(x)=
	\beta {y'}^2+
	\left[\delta+\zeta\exp\left(\nu {y'}^2\right)\right]\tanh(\sigma y')\,y',
	\label{eq:power_loss_14}
\end{equation}
whereas the power delivered by the magnetic excitation is
\begin{equation}
	P_{deliv}(x)=
	A_0\exp(-y^2)\,y\,\sin(\Omega x + py)\,y'.
	\label{eq:power_deliv_15}
\end{equation}
The corresponding cumulative energy loss and cumulative delivered energy are defined by
\begin{equation}
	E_{loss}(x)=\int_0^\chi P_{loss}(x)\,dx,
	\label{eq:energy_loss_deliv_16}
\end{equation}
\begin{equation}
	E_{deliv}(x)=\int_0^\chi P_{deliv}(x)\,dx.
	\label{eq:energy_loss_deliv_17}
\end{equation}

The numerical results shown in Fig.~\ref{fig:Energies}b indicate that the energy supplied to the system is concentrated in short increments, while the dissipated energy increases in a smoother manner between these events. Over one oscillation cycle, however, the net energy delivered to the system remains close to the energy dissipated, which is consistent with the persistence of a bounded period-1 response.

Finally, Fig.~\ref{fig:Energies}c compares the time derivative of the mechanical energy with the instantaneous power balance
\begin{equation}
	\frac{dE_{mech}(x)}{dx}=P_{deliv}(x)-P_{loss}(x).
	\label{eq:energy_power_balance_18}
\end{equation}
The close overlap of the two curves confirms the numerical consistency of the model and shows that the observed period-1 oscillation is sustained by a balance between localized energy injection and continuous dissipation.

\textcolor{black}{The localized bursts of magnetic power are associated with the passage
	of the pendulum through the central magnetic interaction region rather
	than with one precisely defined angular position. Since the magnetic
	torque vanishes at $y=0$, the extrema of the power bursts generally
	occur at small non-zero angular displacements. Their exact timing and
	magnitude depend on the instantaneous angular velocity, coil current,
	and excitation phase.
	For a stable periodic attractor, the bursts repeat at approximately the
	same phases of successive excitation cycles. Small perturbations of the
	initial conditions that remain within the same basin of attraction
	primarily affect the transient response. However, near basin boundaries
	or bifurcation points, small changes in the initial conditions or model
	parameters may lead to a different coexisting attractor or dynamical
	regime, thereby changing the timing and magnitude of the energy-transfer
	bursts. In chaotic regimes, the energy transfer remains spatially
	localized, but the individual bursts vary between successive cycles.}

\section{Discussion}\label{sec6}

\subsection{Preliminary studies}

The performed stability analysis is based on the extreme values of the time-varying excitation, \(\sin(\Omega x+\phi_0)=\pm 1\), and therefore corresponds to the associated frozen system with maximum attracting or repelling magnetic action. The resulting equilibrium states should be interpreted as instantaneous fixed points of this quasi-static formulation rather than as equilibria of the full non-autonomous oscillator. Consequently, the analysis provides local information about the phase-space structure during the strongest magnetic interaction, but it does not determine the stability of periodic orbits of the full system.

The equilibrium branches obtained as functions of \(A_0\) reveal a qualitative change in the local structure of the frozen system. For negative values of \(A_0\), the central equilibrium \(y=0\) remains stable, whereas for positive values it loses stability and two symmetric stable non-zero branches appear. This corresponds to a supercritical pitchfork bifurcation in the equilibrium structure of the frozen system. Physically, attractive magnetic interaction stabilizes the centered pendulum configuration, while repulsive interaction destabilizes it and promotes shifted equilibrium positions. This interpretation is confirmed by the stream plots presented in Fig.~\ref{fig:Stability_saddle_focus_4}. In the attractive case, the lower equilibrium behaves as a stable focus, whereas in the repulsive case it becomes an unstable saddle. These results indicate that the sign of the magnetic interaction strongly influences the local geometry of the phase space, which helps explain the sensitivity of the phase-controlled system to the timing and sign of energy injection. However, this quasi-static analysis should be regarded only as a preliminary diagnostic tool; the full dynamics of the non-autonomous system must be assessed using the periodic orbit, bifurcation, and basin analyses discussed later.
\textcolor{black}{It should be emphasised that the frozen-system analysis provides only a quasi-static interpretation of the instantaneous equilibrium configurations and does not constitute a stability analysis of the periodic solutions of the full non-autonomous system. Replacing the sinusoidal excitation by its extreme values reveals how the limiting magnetic-interaction configurations modify the number, positions, and local stability of the instantaneous equilibria. This approach is therefore useful for interpreting changes in the effective restoring torque. However, the actual response depends on the continuous time variation of the excitation, its phase relationship with the pendulum motion, and the energy transferred over the complete excitation cycle. Consequently, a stable equilibrium of the frozen system does not necessarily imply a stable periodic orbit of the full system. A rigorous local stability assessment of such an orbit would require linearization along the complete periodic solution and calculation of the corresponding Floquet multipliers. This analysis was not performed in the present study. Instead, the dynamical regimes of the full non-autonomous system were evaluated using direct numerical integration, stroboscopic Poincaré maps, bifurcation diagrams, and Lyapunov exponents, supported by experimental measurements.}

The bifurcation diagrams shown in Fig.~\ref{fig:biff_simulation_LLCE} provide a baseline characterization of the classical system with constant excitation phase (\(p=0\)). Both the experimental sweep diagram and the numerical bifurcation plots reveal alternating windows of regular and irregular response as the control parameter \(\Omega\) varies. The positive and non-positive values of the largest Lyapunov exponent in Fig.~\ref{fig:biff_simulation_LLCE}d support the interpretation of these regions as chaotic and regular regimes, respectively.
The comparison between Fig.~\ref{fig:biff_simulation_LLCE}a,b and Fig.~\ref{fig:biff_simulation_LLCE}c also highlights the difference between sweep-type and classical bifurcation analysis. The sweep procedure follows only one attractor selected by the system history and therefore reproduces the experimental path-dependent behavior, whereas the classical continuation-like diagram reveals hysteresis and multistability more clearly. The visible differences between increasing and decreasing frequency sweeps confirm that the system response depends strongly on the direction of parameter variation.

The experimental bifurcation diagram is more diffuse than the numerical ones, which can be attributed to measurement noise, feedback delay, and simplifications of the mathematical model, as well as to the fact that the experiment was performed with a continuously varying control parameter. Nevertheless, the agreement between experiment and simulation is sufficiently good to confirm that the model captures the principal dynamical transitions of the system. The multistability observed for \(\Omega=\{2.3,\,4.6\}\) and \(\Omega=\{3.7,\,6.4\}\) indicates that even the classical system may exhibit coexisting attractors of both symmetric and qualitatively different types, which provides an important reference point for the phase-controlled case analyzed later.

Fig.~\ref{fig:preliminary_phase trajectories} demonstrates a consistent qualitative agreement between the simulated and experimentally measured phase trajectories over the selected excitation frequencies. In all cases, the model governed by Eq.~(\ref{eq:motion_dimensionless_8}) reproduces the essential geometric features of the observed attractors, including single-well oscillations, inter-well excursions, and mixed responses combining both types of motion. This indicates that the identified magnetic interaction and the adopted nonlinear dissipation model capture the dominant mechanisms shaping the response of the classical system with \(p=0\).
The agreement is also visible at the level of the associated Poincar\'e sections. The numerically obtained Poincar\'e points fall within the same regions of phase space as the experimental point clouds, which confirms that the model reproduces not only the overall trajectory geometry but also the correct periodicity and phase relation with the excitation. This is particularly important because the bifurcation diagrams alone provide only a reduced representation of the dynamics, whereas the phase portraits reveal the actual structure of the underlying attractors.
At the same time, the experimental trajectories and Poincar\'e points are visibly more diffuse than their numerical counterparts. This effect is especially pronounced in regimes involving transitions between potential wells, where the dynamics passes close to separatrix-like regions and therefore becomes more sensitive to perturbations, delays, and measurement uncertainty. Additional discrepancies may be associated with unavoidable experimental imperfections, such as noise in the angle measurement, small fluctuations in the current signal, and variability of friction at low angular velocities. Nevertheless, these discrepancies remain local and do not alter the qualitative agreement between experiment and simulation.

Taken together, the results shown in Fig.~\ref{fig:preliminary_phase trajectories} provide strong support for the validity of the classical reference model. In combination with the bifurcation diagrams and Lyapunov exponent analysis, they show that the model captures both the global organization of the response and the detailed geometry of representative attractors. This provides a reliable baseline for assessing the additional dynamical effects introduced later by the state-dependent phase control.
\subsection{Parameter-space organization under phase-controlled excitation}

The regime charts presented in Fig.~\ref{fig:phase_control_charts} show that the phase-control coefficient \(p\) acts as an effective organizing parameter of the system dynamics. For all three considered excitation frequencies, the response in the \((p,A_0)\) plane is divided into extended domains of low-period oscillations, fragmented regions of higher-period solutions, and broad chaotic layers identified by positive values of the largest Lyapunov exponent. This confirms that the introduction of state-dependent phase modulation does not merely perturb the classical dynamics, but substantially reshapes the parameter-space structure of the oscillator.

A characteristic feature of all three charts is the presence of large coherent regions of regular oscillations, especially in the low-to-moderate range of \(p\), together with more fragmented structures for larger values of the phase-control coefficient. The regular domains contain embedded bands of higher-period responses, while chaotic regions appear predominantly near their boundaries and in strongly folded parts of the parameter space. This organization suggests that the transition to irregular motion is closely related to successive bifurcations of periodic solutions and to the reorganization of attractor structure under phase feedback.
Another important observation is that moderate changes in the excitation frequency \(\Omega\) do not destroy the global topology of the regime charts. Instead, they mainly shift the boundaries between the different dynamical regions. In other words, the phase-controlled system exhibits a robust qualitative regime structure with respect to \(\Omega\), while the precise thresholds of transitions between periodic and chaotic responses remain frequency dependent. This property indicates that the role of \(p\) as a control parameter is structurally persistent over the investigated frequency range.
It should also be emphasized that the charts in Fig.~\ref{fig:phase_control_charts} were obtained for fixed initial conditions. Therefore, they primarily reveal the organization of attractor types and the sensitivity of the long-term response to parameter variation, rather than providing a direct proof of multistability for each parameter pair. The actual coexistence of attractors for identical parameter values is addressed more explicitly in the subsequent analyses of selected bifurcation sections and attraction basins.

The comparison between the bifurcation sections presented in Fig.~\ref{fig:phase_control_bifurcation_Omega0664} (as well as in Fig.~\ref{fig:app_bifurcation_Omega0700_Omega0750} in Appendix \ref{app:additional_frequencies}) and computed for \(p=0\) and \(p=1.423\) demonstrates that the phase-control coefficient has a profound effect on the accessibility of regular long-term responses. For \(\Omega=0.664\) and \(\Omega=0.700\), the classical system is dominated by chaotic motion over almost the entire investigated interval of \(A_0\), with only a few isolated regular windows. In contrast, the phase-controlled system exhibits broad families of periodic solutions separated by relatively narrow chaotic intervals. This difference is confirmed quantitatively by the largest Lyapunov exponent: the fraction of scanned parameter values associated with non-positive \(\lambda_1\) increases from approximately 6\% in the classical case to more than 92\% and 97\% for \(p=1.423\), respectively.
For \(\Omega=0.750\), the classical system already contains a substantial periodic branch at higher amplitudes, and therefore the effect of phase control is less dramatic in purely qualitative terms. Nevertheless, the quantitative comparison still shows a strong increase in the share of regular responses, from 49.40\% for \(p=0\) to 99.65\% for \(p=1.423\). This indicates that the role of the phase-control coefficient should not be interpreted simply as a binary switch between chaos and regularity, but rather as a mechanism that reorganizes the bifurcation structure and shifts the balance between regular and irregular attractors over the parameter range.

From a dynamical viewpoint, these results suggest that state-dependent phase modulation strongly affects the timing of excitation in a way that promotes the persistence of periodic responses for selected amplitudes of forcing. In this sense, the phase-control coefficient may be viewed as a regulator of regime selection: for the analyzed parameter sections, it enlarges the regions in which regular attractors are accessible from the chosen initial conditions and reduces the dominance of chaotic motion. At the same time, this effect remains frequency dependent, as illustrated by the different degree of reorganization observed for \(\Omega=0.664\), \(0.700\), and \(0.750\).

A more detailed inspection of the bifurcation section for \(\Omega=0.664\) and \(p=1.423\) presented in Fig.~\ref{fig:phase_control_bifurcation_Omega0664}e indicates that, at least in the analyzed interval of \(A_0\), the onset of chaos follows the classical period-doubling scenario. The enlarged bifurcation diagram shows successive branch splittings before the appearance of a broadband irregular attractor, while the corresponding Lyapunov spectrum confirms that the largest exponent becomes positive only after this cascade is completed. This observation suggests that the phase-controlled system does not generate chaotic motion in an unstructured manner; rather, at least in selected parameter regions, chaos emerges through an ordered bifurcation sequence.
At the same time, this result should be interpreted as a representative example rather than as a universal route to chaos for the entire parameter space. The regime charts and bifurcation sections indicate that the phase-controlled pendulum exhibits several different types of transitions between regular and irregular responses. Nevertheless, the period-doubling cascade observed here provides a clear illustration of how phase feedback can reorganize the bifurcation structure and create a controlled progression from periodic motion to chaos.

The representative phase portraits and Fourier spectra shown in Fig.~\ref{fig:phasePlot_PS_control_phase} provide a geometric and spectral interpretation of the bifurcation sequence observed for \(\Omega=0.664\) and \(p=1.423\). The case \(A_0=181.2\) corresponds to a regular low-period response, while the case \(A_0=183\) shows a more complex attractor with a larger number of Poincar\'e points, consistent with the period-doubling progression identified in the bifurcation diagram. For \(A_0=185\), the dense distribution of Poincar\'e points and the broadband component of the Fourier spectrum indicate that the system has entered a chaotic regime.	
An important observation is that the dominant spectral peaks remain locked to the excitation frequency and its harmonics even in the chaotic case, while an increasing number of subharmonic components and a growing broadband background accompany the transition from regular motion to chaos. This shows that the onset of chaos does not destroy the underlying forcing structure, but rather enriches the response by introducing additional scales and irregular modulation. In this sense, the phase-controlled pendulum retains a clear imprint of the external excitation even after the regular attractor loses stability.	
The phase portraits also show that the transition to chaos is accompanied by a progressive thickening and folding of the attractor in phase space. Therefore, the combined use of bifurcation sections, Lyapunov exponents, Poincar\'e maps, and Fourier spectra provides a consistent picture of the route to chaos in this representative parameter interval.
\subsection{Robustness and physical interpretation of the intrawell period-1 branch}

One of the most distinctive regular responses observed in the phase-controlled system is the extended intrawell period-1 branch visible in the bifurcation diagrams for \(p=1.423\). Its most striking feature is the weak dependence of the oscillatory response on the excitation amplitude \(A_0\) over a broad parameter interval. Although increasing \(A_0\) corresponds to stronger magnetic forcing, the associated phase trajectories remain nearly unchanged and the corresponding Fourier spectra continue to be dominated by the forcing frequency. The nearly horizontal character of this branch in the bifurcation diagrams, together with the weak variation of the phase-space radius \(R(A_0)\), indicates that the system maintains a remarkably robust phase-locked oscillatory regime.

This observation suggests that the phase-controlled forcing does not act as a simple amplitude amplifying mechanism. Instead, the state-dependent phase modulation appears to organize the timing of energy exchange in such a way that the pendulum repeatedly follows almost the same closed orbit despite substantial changes in the forcing amplitude. In this sense, the intrawell period-1 branch can be interpreted as a self-consistent periodic response maintained by the controlled alternation of attractive and repulsive magnetic action. The fact that lower excitation frequencies correspond to slightly larger values of \(R(A_0)\) indicates that the size of this orbit is still frequency dependent, but its sensitivity to \(A_0\) remains weak throughout the investigated range.

The splitting of the original period-1 branch into two distinct period-1 attractors for larger values of \(A_0\) further enriches the picture. Importantly, this transition does not destroy the basic frequency locking of the response, since both solutions remain period-1 and retain similar spectral content. However, the coexistence of two geometrically different trajectories for the same parameter values shows that the robust single-attractor regime eventually loses uniqueness and gives way to a more complex response structure. Thus, the phase-controlled system combines two important features: persistence of a simple regular response over a wide range of forcing amplitudes, and the emergence of multistability once the excitation exceeds a critical level.

The comparison with experimental phase trajectories confirms that this regular branch is not merely a numerical artifact. Even though the experimental Poincar\'e points do not collapse to a single point and the phase trajectories are more diffuse than their numerical counterparts, the overall size, shape, and location of the measured orbits remain in good qualitative agreement with the simulations. This supports the conclusion that the intrawell period-1 branch corresponds to a physically realizable regime of the laboratory system. The observed spread of the experimental trajectories is consistent with the non-ideal character of the test stand, in particular with feedback delay, measurement noise, and small fluctuations of dissipation and current control.

The trajectory-conditioned effective potential provides an additional physical interpretation of this behavior. The corresponding three-dimensional plots show that the pendulum does not oscillate in a fixed potential landscape, but in a potential that is periodically reshaped by the magnetic interaction. During the attracting phase of the coil current, the pendulum is guided toward a local potential well, whereas during the repelling phase it is driven away from a local barrier. The sustained intrawell period-1 oscillation may therefore be interpreted as a repeated motion in a time-varying effective potential, where the controlled phase of excitation stabilizes a nearly invariant oscillatory path over a broad parameter interval.

Taken together, these results indicate that the intrawell period-1 branch is one of the clearest manifestations of the dynamical role of phase control in the studied system. Rather than merely increasing the complexity of the motion, the feedback mechanism is also capable of organizing a robust regular response and maintaining it over a wide range of excitation amplitudes. This makes the phase-controlled pendulum qualitatively different from a purely harmonically forced oscillator, for which stronger forcing would more naturally be associated with a substantial change in oscillation size or an earlier loss of regularity.
\subsection{Attractor coexistence and basin-boundary complexity near the intrawell period-1 regime}
The basins of attraction computed for the intrawell period-1 branch provide a more complete picture of its dynamical role than the bifurcation diagrams alone. For \(A_0=138.6\), the period-1 attractor already dominates the analyzed neighborhood of initial conditions, although additional period-2 and period-6 attractors are also present. For \(A_0=210.0\), this dominance becomes much more pronounced, as almost the entire investigated region converges to the same period-1 response. This indicates that, in the intermediate range of forcing amplitudes, the intrawell period-1 branch is not only stable but also locally robust in the sense of possessing a very large basin of attraction.

At the same time, the basin structure shows that this robustness is not preserved uniformly as \(A_0\) increases further. For \(A_0=243.1\), the basin of the period-1 solution remains substantial, but a competing period-3 attractor occupies a significant part of the same neighborhood of initial conditions. Thus, the increase of the excitation amplitude does not simply destabilize the period-1 motion in a monotonic way; rather, it reorganizes the local attractor landscape and reduces the dominance of the original regular branch.

This tendency becomes even more evident for \(A_0=253.0\), where two distinct period-1 attractors coexist together with a period-3 response. The corresponding basin structure reveals that the regular branch identified earlier is no longer represented by a single coherent basin, but splits into two separate period-1 regions associated with geometrically different trajectories. This means that the phase-controlled system evolves from a regime of strong local attraction toward a genuinely multistable configuration, in which several regular long-term responses compete within the same region of phase space.

The magnified basin plots and the computed box-counting dimension \(FD\approx1.87\) indicate that the boundary separating these basins has a fractal-like geometry. This conclusion is reinforced by the boundary basin entropy result \(S_{\mathrm{bb}}=0.4818\pm0.0221\), which exceeds the theoretical threshold for a smooth two-basin boundary. Together, these measures show that the complexity of the basin structure is not merely visual, but is supported by quantitative indicators of irregular boundary geometry and local basin mixing.

From a dynamical perspective, this has an important consequence: near the interwoven basin boundary, arbitrarily small perturbations of the initial condition may redirect the trajectory toward different coexisting attractors. Therefore, the phase-controlled pendulum combines two seemingly opposite properties. On the one hand, it is capable of sustaining a robust period-1 response over a broad range of forcing amplitudes. On the other hand, once the excitation exceeds a critical level, the same control mechanism gives rise to multistability and highly intricate basin geometry, which strongly increases the sensitivity of the long-term response to initial conditions.

\subsection{Energy balance of the robust period-1 response}
The energy analysis provides a physical explanation for the persistence of the intrawell period-1 branch over a broad range of excitation amplitudes. The time evolution of the mechanical energy shows that the motion is not sustained by a continuous energy supply, but rather by short, localized energy transfer events. These events occur when the pendulum passes through the angular region in which the magnetic torque is significant. As a result, the mechanical energy exhibits sharp jumps followed by slower decay intervals, reflecting the alternation between localized magnetic energy injection and dissipative losses.

The comparison between the cumulative delivered energy and the cumulative dissipated energy shows that the input energy is supplied in discrete increments, whereas dissipation increases more smoothly during the oscillation cycle. Despite this different temporal structure, both quantities remain nearly balanced over one period of the motion. This cycle-by-cycle compensation explains why the period-1 response can remain bounded and nearly unchanged even when the excitation amplitude \(A_0\) is varied. In other words, increasing the magnetic forcing amplitude does not simply lead to a proportional increase in the oscillation size; instead, the state-dependent phase control organizes the timing of energy injection so that the supplied energy compensates the losses in a repeatable manner.

The instantaneous power balance further supports this interpretation. The close agreement between \(dE_{mech}/dx\) and \(P_{deliv}-P_{loss}\) confirms that the changes in mechanical energy are accurately accounted for by the competition between magnetic power input and dissipation. This agreement also verifies the numerical consistency of the energy formulation used in the model. More importantly, it indicates that the robust period-1 orbit is maintained by a localized and phase-synchronized exchange of energy between the electromagnetic actuator and the mechanical oscillator.

Therefore, the weak dependence of the intrawell period-1 branch on \(A_0\) can be interpreted as a consequence of the controlled timing of energy transfer. The phase feedback causes the magnetic torque to act only during dynamically favorable portions of the cycle, while dissipation removes energy continuously between these localized input events. This balance produces a self-consistent periodic regime that is robust with respect to changes in forcing amplitude, at least within the parameter interval considered here.

\section{Concluding remarks}\label{sec7}

In this work, the dynamics of a strongly nonlinear magnetic pendulum subjected to phase-controlled magnetic excitation was investigated using a combined experimental and numerical approach. The key novelty lies in introducing a state-dependent phase law \(\phi_0(y)=py\), in which the phase of the excitation signal depends on the instantaneous pendulum position. In contrast to the classical formulation, this modification transforms the excitation into a feedback-type phase-modulated forcing and makes the timing of energy transfer dependent on the system state.

The proposed mathematical model, based on an experimentally identified magnetic torque, was validated against experimental data and shown to reproduce the principal dynamical features of the system. This provides a reliable basis for analyzing the effects of phase-controlled excitation.
The results demonstrate that the phase-control parameter \(p\) plays a crucial role in organizing the system dynamics. In particular, state-dependent phase modulation significantly enlarges the regions of regular motion in parameter space and reduces the dominance of chaotic responses compared with the classical case. For example, for \(\Omega=0.664\), the fraction of regular responses increases from approximately \(6\%\) in the classical case to over \(90\%\) when phase control is applied. This clearly indicates that phase control can act as an effective mechanism of dynamical regime selection.

A central finding of the study is the existence of a robust intrawell period-1 branch that persists over a wide range of excitation amplitudes \(A_0\). Despite substantial changes in forcing intensity, the corresponding oscillatory response remains almost unchanged in both geometry and frequency content. This behavior was confirmed numerically and supported by experimental observations.
The robustness of this branch can be explained by the mechanism of energy transfer. The magnetic excitation acts in short, localized intervals, producing pulse-like energy injection when the pendulum passes through a narrow angular region. Over each oscillation cycle, the supplied energy compensates dissipation, resulting in a bounded and nearly invariant periodic motion. Thus, the phase-controlled excitation regulates the timing of energy input rather than simply increasing its magnitude.

At the same time, for larger excitation amplitudes, the system exhibits multistability and increased sensitivity to initial conditions. Basin of attraction analysis revealed coexistence of multiple periodic attractors and highly intricate, fractal-like basin boundaries. These results show that the same control mechanism that stabilizes periodic motion may also lead to complex coexistence structures and sensitive dependence on initial conditions.

Overall, the study demonstrates that position-dependent phase control fundamentally alters the nature of excitation in magneto-mechanical oscillators. By introducing feedback in the phase domain, it becomes possible to regulate energy transfer, promote robust periodic responses, and control the structure of attractors in parameter space. These findings may be relevant for the design of nonlinear oscillators with tunable dynamics, including applications in vibration control, adaptive systems, and energy transfer devices.

The present work is limited to a simple linear phase-control law and a near real-time experimental implementation. Future research may focus on more advanced phase-control strategies, including nonlinear and delayed phase feedback, as well as systematic optimization of energy injection for targeted dynamical behavior.

\section*{Acknowledgments}
This work was supported by the National Science Center, Poland under the grant OPUS 26 No. 2023/51/B/ST8/00021. \textcolor{red}{For the purpose of Open Access, the authors has applied a CC-BY public copyright licence to any Author Accepted Manuscript (AAM) version arising from this submission. This is an AAM of an article published by Elsevier in European Journal of Mechanics - A/Solids on 7 November 2026, available at: \url{https://doi.org/10.1016/j.euromechsol.2026.106467} }

\section*{Declaration of generative AI and AI-assisted technologies in the manuscript preparation process} During the preparation of this work, the authors used Microsoft Copilot for translation and language editing. After using this tool, the authors reviewed and edited the output as needed and take full responsibility for the content of the publication.

\section*{Supplementary Materials}

Several supplementary GIF animations are provided in the online version of the article to enhance the understanding of the system’s dynamic behavior. Each animation illustrates a key phenomenon discussed in the main text. The animations are referenced at relevant points in the article and are described below:

\begin{itemize}
	\item \textbf{GIF 1 – Phase trajectory and Poincaré section} \\
	This animation shows the evolution of the phase trajectory and the corresponding Poincaré section for the bifurcation diagram presented in Fig.~\ref{fig:phase_control_bifurcation_Omega0664}e. It illustrates the emergence of complex attractors and transitions between periodic and chaotic regimes.
	
	\item \textbf{GIF 2 – Magnetic force vector during pendulum motion} \\
	This animation visualizes the real-time magnitude and direction of the magnetic force vector during pendulum oscillations. It demonstrates the large deflections and the dynamic changes of the force in relation to the pendulum’s position. The elastic gum element prevents full rotations of the pendulum.
	
	\item \textbf{GIF 3 – Detailed view of magnetic interaction} \\
	A slowed-down version of the pendulum's motion near the coil is presented, enabling detailed observation of how the magnetic force evolves. The animation highlights the threshold angles beyond which the magnetic torque rapidly diminishes, as discussed at the end of Section~\ref{sec5}. These thresholds are marked with green lines.
\end{itemize}

All animations are available in the supplementary section of the online version of this article and can be used to visually support the interpretation of the nonlinear effects and energy dynamics described in the study.
\appendix
\section{Numerical methods}
\label{app:numerical_methods}
All numerical integrations of the dimensionless equations of motion were performed using the \texttt{NDSolve} function in Wolfram Mathematica (version 13.0 or higher). To handle the potential stiffness of the system arising from the combination of nonlinear friction and magnetic forcing, we employed the option \texttt{Method -> {"StiffnessSwitching", "DiscontinuityProcessing" -> False}}. This setting enables an adaptive solver that automatically switches between a non-stiff (explicit Adams) method and a stiff (implicit Gear/BDF) method when the local stiffness ratio exceeds a threshold. The option \texttt{"DiscontinuityProcessing" -> False} was used to prevent the solver from attempting to detect and process discontinuities, as the system's nonlinearities are smooth and this intervention could unnecessarily interrupt integration. The integration time step was controlled by the solver's adaptive step-size algorithm, with relative and absolute error tolerances set to $10^{-8}$ and $10^{-10}$, respectively (default \texttt{PrecisionGoal} and \texttt{AccuracyGoal} in \texttt{NDSolve}). For bifurcation diagrams and Lyapunov exponents, simulations covered at least 5000 excitation periods after discarding transients, with Poincaré sections sampled at the forcing period.
\textcolor{black}{
	\section{Local linear stability analysis} \label{app:Jacobian_matrix} To determine the local stability of the lower central equilibrium, the governing equation was linearized about the equilibrium state
	 \begin{equation} \mathbf{y}^{*} = \begin{bmatrix} 0 & 0 				\end{bmatrix}^{\mathrm{T}}, 
	 \end{equation} 
	 where the state vector is defined as 
	 \begin{equation} \mathbf{y} = \begin{bmatrix} y & y' 	\end{bmatrix}^{\mathrm{T}}.
	 \end{equation} 
	   The resulting linearized system can be expressed as 
	 \begin{equation} 
	   	\mathbf{y}'=\mathbf{A}_{J}\mathbf{y},
	   	 \label{eq:linear_stability_B}
	 \end{equation}
	 where $\mathbf{A}_{J}$ is the Jacobian matrix of the corresponding autonomous system evaluated at $\mathbf{y}^{*}$. For the attractive direction of the magnetic interaction, the Jacobian matrix is 
	\begin{equation}
	   \mathbf{A}_{J}^{\mathrm{att}} = \begin{bmatrix} 0 & 1 \\ -\alpha-A_0-1 & -\beta-\sigma(\delta+\zeta) \end{bmatrix}, \label{eq:Jacobian_attraction}
	 \end{equation}
	 whereas for the repulsive direction it takes the form
	 \begin{equation}
	  	 \mathbf{A}_{J}^{\mathrm{rep}} = \begin{bmatrix} 0 & 1 \\ -\alpha+A_0-1 & -\beta-\sigma(\delta+\zeta) \end{bmatrix}.
	   \label{eq:Jacobian_repulsion}
	\end{equation}
	Here, $A_0=81.82$ denotes the magnitude of the dimensionless magnetic excitation corresponding to a coil-current magnitude of $1~\mathrm{A}$. The direction of the magnetic interaction is included explicitly through the sign preceding $A_0$ in Eqs.~\eqref{eq:Jacobian_attraction} and \eqref{eq:Jacobian_repulsion}. The local stability is determined from the eigenvalues $\mu_{1,2}$ of the corresponding Jacobian matrix, obtained from
	\begin{equation}
	   \det\left(\mathbf{A}_{J}-\mu\mathbf{I}\right)=0,
	   \label{eq:Jacobian_eigenvalues} 
	\end{equation} 
	 where $\mathbf{I}$ is the identity matrix. For a Jacobian matrix of the form
	 \begin{equation}
	 	\mathbf{A}_{J} = \begin{bmatrix} 0 & 1 \\ a_{21} & a_{22} \end{bmatrix},
	\end{equation}
	 the corresponding eigenvalues are 
		\begin{equation} \mu_{1,2} = \frac{a_{22}\pm\sqrt{a_{22}^{2}+4a_{21}}}{2}. \label{eq:general_eigenvalues}
	\end{equation} Substitution of the parameter values listed in Table~\ref{tab:parameters_1} into Eq.~\eqref{eq:Jacobian_attraction} gives
		\begin{equation} \mathbf{A}_{J}^{\mathrm{att}} = \begin{bmatrix} 0 & 1\\ -83.152 & -0.470 \end{bmatrix}. \label{eq:Jacobian_attraction_numerical}
	\end{equation} 
	The corresponding eigenvalues are
	\begin{equation}
		\mu_{1,2}^{\mathrm{att}} = -0.235 \pm 9.116\,\mathrm{i}. \label{eq:eigenvalues_attraction}
	\end{equation} 
	Since both eigenvalues have negative real parts, the lower central equilibrium is a locally asymptotically stable focus for the attractive magnetic interaction. For the repulsive magnetic interaction, substitution of the same parameter values into Eq.~\eqref{eq:Jacobian_repulsion} yields 
	\begin{equation}
		 \mathbf{A}_{J}^{\mathrm{rep}} = \begin{bmatrix} 0 & 1\\ 80.488 & -0.470 \end{bmatrix}. 
		\label{eq:Jacobian_repulsion_numerical} 
	\end{equation} 
	The corresponding eigenvalues are 
	\begin{equation} 
		\mu_{1}^{\mathrm{rep}} = 8.739, \qquad \mu_{2}^{\mathrm{rep}} = -9.210. 
		\label{eq:eigenvalues_repulsion} 
	\end{equation} 
	Since the eigenvalues are real and have opposite signs, the lower central equilibrium is an unstable saddle for the repulsive magnetic interaction. The eigenvalue analysis therefore confirms the phase-space structures presented in Fig.~\ref{fig:Stability_saddle_focus_4}. The attractive magnetic interaction preserves the lower central equilibrium as a stable focus, whereas the repulsive interaction destabilizes it and gives it a saddle-type character.}
\textcolor{black}{
\section{Supplementary bifurcation results for additional excitation frequencies}
\label{app:additional_frequencies}
This Appendix presents supplementary bifurcation diagrams and Lyapunov spectra for $\Omega=0.700$ and $\Omega=0.750$, see Fig.~\ref{fig:app_bifurcation_Omega0700_Omega0750}. These results complement the detailed analysis of the representative case $\Omega=0.664$ presented in the main text. For each excitation frequency, the classical case $p=0$ is compared with the phase-controlled case $p=1.423$ over the interval $A_0\in(130,260)$.
\begin{figure}[h!] 
	\centering 
	\begin{subfigure}{0.49\textwidth} 
	\centering \includegraphics[width=\linewidth] {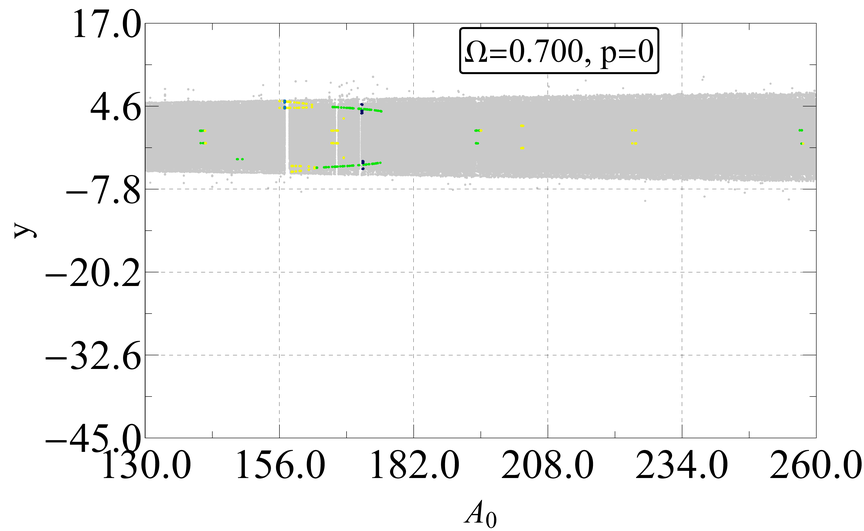} 
	\caption{} 
	\label{} 
	\end{subfigure} 
	\hfill 
	\begin{subfigure}{0.49\textwidth} 
	\centering \includegraphics[width=\linewidth] {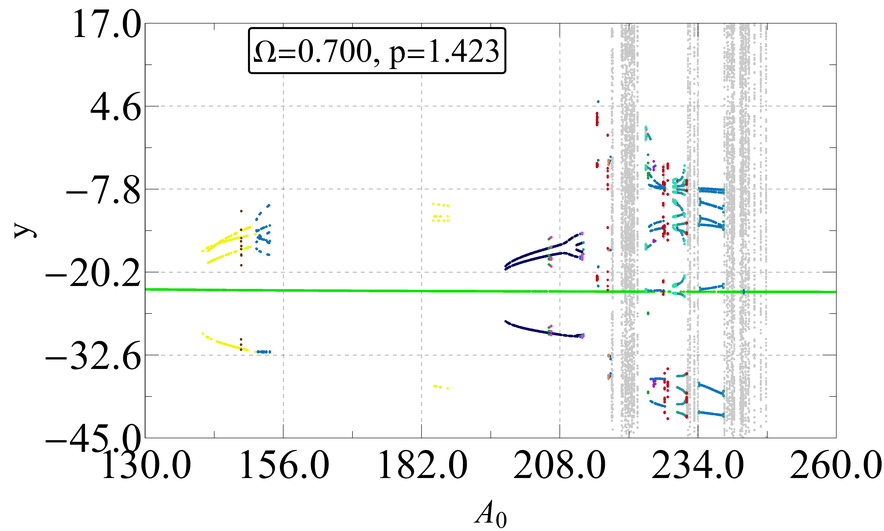} 
	\caption{} 
	\label{} 
	\end{subfigure} 
	\begin{subfigure}{0.49\textwidth} 
	\centering 
	\includegraphics[width=\linewidth] {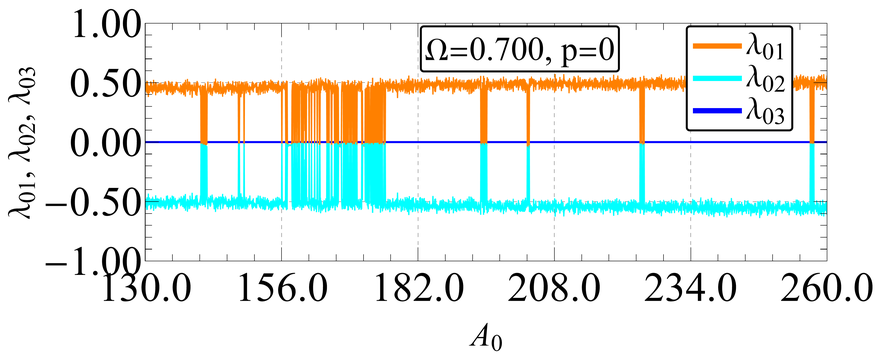} 
	\caption{} 
	\label{} 
	\end{subfigure} 
	\hfill 
	\begin{subfigure}{0.49\textwidth} 
	\centering \includegraphics[width=\linewidth] {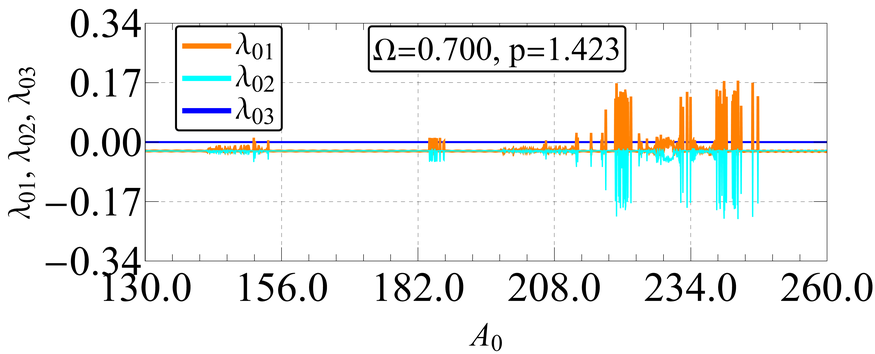} 
	\caption{} 
	\label{} 
	\end{subfigure} 
		\begin{subfigure}{0.49\textwidth} 
		\centering 
		\includegraphics[width=\linewidth] {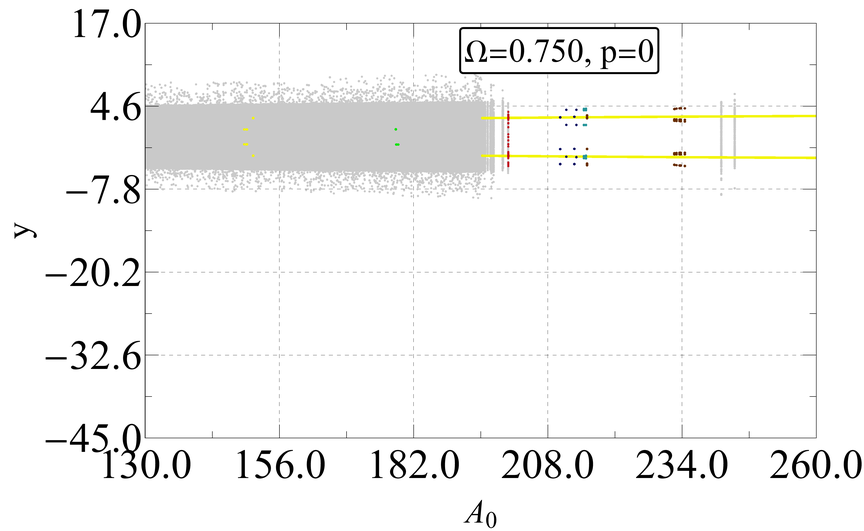} 
		\caption{} 
		\label{} 
	\end{subfigure} 
	\hfill 
	\begin{subfigure}{0.49\textwidth} 
		\centering 
		\includegraphics[width=\linewidth] {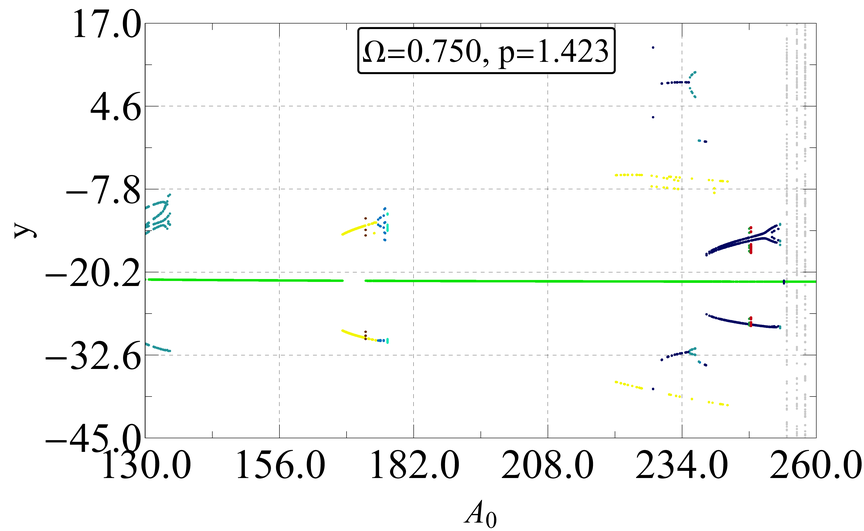} 
		\caption{} 
		\label{}
	\end{subfigure} 
	\begin{subfigure}{0.49\textwidth} 
		\centering 
		\includegraphics[width=\linewidth] {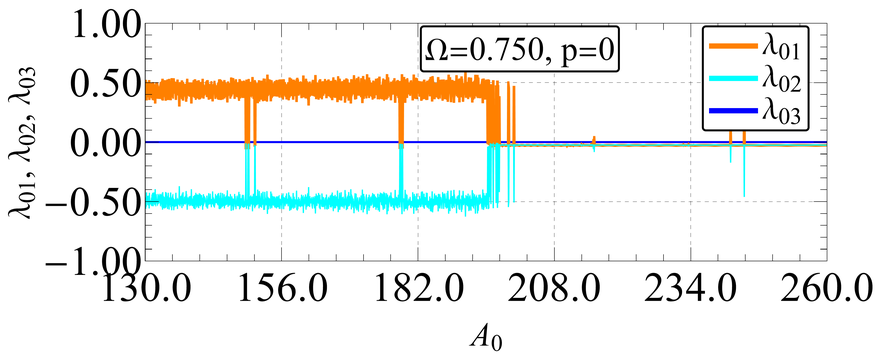} 
		\caption{} 
		\label{} 
	\end{subfigure} 
	\hfill 
	\begin{subfigure}{0.49\textwidth} 
		\centering 
		\includegraphics[width=\linewidth] {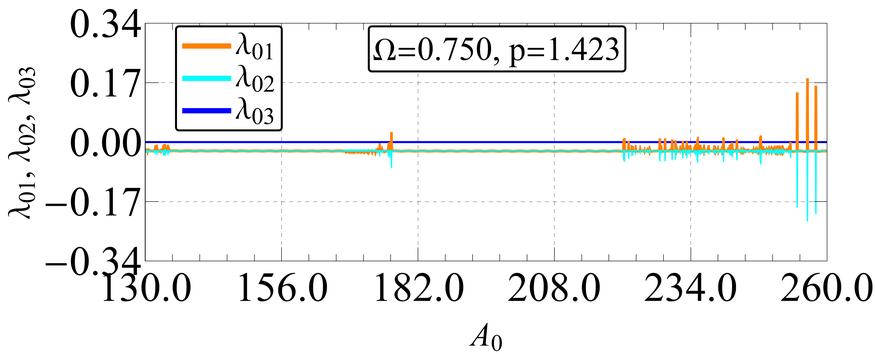} 
		\caption{} 
		\label{} 
	\end{subfigure} 
	\caption{Bifurcation diagrams and corresponding Lyapunov spectra for $\Omega=0.700$,  $\Omega=0.750$ and $A_0\in(130,260)$. Panels (a)--(b) and (e)--(f) show the bifurcation diagrams $y(A_0)$, while panels (c)--(d) and (g)--(h) show the corresponding Lyapunov spectra. The left column corresponds to $p=0$, whereas the right column corresponds to $p=1.423$.} 
	\label{fig:app_bifurcation_Omega0700_Omega0750} 
\end{figure} 
For $\Omega=0.700$, the response in the classical case $p=0$ remains predominantly chaotic over the investigated interval. By contrast, the phase-controlled case $p=1.423$ exhibits several well-defined families of periodic solutions separated by narrower intervals of irregular motion. The corresponding Lyapunov analysis shows that $97.12\%$ of the scanned parameter values satisfy $\lambda_{01}\leq0$ for $p=1.423$, compared with only $6.57\%$ for $p=0$. Thus, the phase-dependent excitation substantially increases the proportion of regular responses for this excitation frequency. 
	For $\Omega=0.750$, the difference between the classical and phase-controlled cases is less pronounced than for the two lower excitation frequencies, but it remains clearly visible. The bifurcation diagram for $p=0$ contains an extended chaotic region at lower values of $A_0$, together with a broad periodic branch at larger excitation amplitudes. For $p=1.423$, the regular responses are distributed among several separated branches. The corresponding Lyapunov spectrum indicates that $99.65\%$ of the scanned parameter values satisfy $\lambda_{01}\leq0$ for $p=1.423$, whereas the corresponding proportion for $p=0$ is $49.40\%$. These results support the conclusion that the phase-dependent excitation increases the prevalence of regular responses across the investigated excitation frequencies, although the detailed bifurcation structure remains frequency dependent.}
\section*{CRediT authorship contribution statement}
\textbf{Krystian Polczy\'nski:} Conceptualization of this study, Methodology, Software, Writing -- Original draft preparation, Investigation.
\textbf{Jan Awrejcewicz:} Supervision, Project administration, Funding acquisition.
\bibliographystyle{unsrtnat}
\bibliography{library}
\end{document}